\documentclass[11pt, aps, nofootinbib]{revtex4-1}

\usepackage{geometry}
\usepackage{latexsym}
\usepackage{graphicx}
\usepackage{amsmath}
\usepackage{amssymb}
\usepackage{color}
\usepackage{mathtools}
\usepackage[colorlinks=true,a4paper=true,backref=false,linktocpage=true,citecolor=blue,urlcolor=blue,linkcolor=blue,pdfpagemode=UseOutlines]{hyperref}
\usepackage{braket}

\newcommand{\be}{\begin{equation}}
\newcommand{\ee}{\end{equation}}
\newcommand{\bea}{\begin{eqnarray}}
\newcommand{\eea}{\end{eqnarray}}

\newcommand{\LaSapienza}{Physics Department, University of Roma ``La Sapienza'' and INFN, Sezione di Roma,\\ Piazzale Aldo Moro 5, 00185 Roma, Italy}
\newcommand{\RomatreINFN}{Istituto Nazionale di Fisica Nucleare, Sezione di Roma Tre,\\ Via della Vasca Navale 84, I-00146 Rome, Italy}
\newcommand{\SNS}{Scuola Normale Superiore, Piazza dei Cavalieri 7, I-56126, Pisa, Italy}
\newcommand{\INFNPisa}{Istituto Nazionale di Fisica Nucleare, Sezione di Pisa,\\ Largo Bruno Pontecorvo 3, I-56127 Pisa, Italy}

\begin{document}

\title{Exclusive semileptonic $B \to \pi \ell \nu_\ell$ and $B_s \to K \ell \nu_\ell$ decays through unitarity and lattice QCD}

\author{G.\,Martinelli}\affiliation{\LaSapienza}
\author{S.\,Simula}\affiliation{\RomatreINFN}
\author{L.\,Vittorio}\affiliation{\SNS}\affiliation{\INFNPisa}

\begin{abstract}
The Cabibbo-Kobayashi-Maskawa (CKM) matrix element $\vert V_{ub}\vert$ is obtained from exclusive semileptonic $B \to \pi \ell \nu_\ell$ and $B_s \to K \ell \nu_\ell$ decays adopting the unitarity-based dispersion matrix approach for the determination of the hadronic form factors (FFs) in the whole kinematical range. We use lattice computations of the relevant susceptibilities and of the FFs in the large-$q^2$ regime in order to derive their behavior in the low-$q^2$ region without assuming any specific momentum dependence and without constraining their shape using experimental data. Then, we address the extraction of $\vert V_{ub}\vert$ from the experimental data, obtaining $\vert V_{ub}\vert = (3.62 \pm 0.47) \cdot 10^{-3}$ from $B \to \pi$ and $\vert V_{ub}\vert = (3.77 \pm 0.48) \cdot 10^{-3}$ from $B_s \to K$, which after averaging yield $\vert V_{ub}\vert = (3.69 \pm 0.34) \cdot 10^{-3}$. These results are compatible with the most recent inclusive value $\vert V_{ub} \vert_{incl} = 4.13\,(26) \cdot 10^{-3}$ at the 1$\sigma$ level. We also present purely theoretical estimates of the ratio of the $\tau/\mu$ decay rates $R^{\tau/\mu}_{\pi(K)}$, the normalized forward-backward asymmetry $\bar{\mathcal{A}}_{FB}^{\ell,\pi(K)}$ and the normalized lepton polarization asymmetry $\bar{\mathcal{A}}_{polar}^{\ell,\pi(K)}$.
\end{abstract}

\maketitle

\newpage

\section{Introduction}
\label{sec:introduction}

Since many years the heavy-to-light semileptonic transitions are very intriguing processes mainly because a long-standing tension affects the inclusive and the exclusive determinations of the CKM matrix element $\vert V_{ub} \vert$. The most recent version of the FLAG report~\cite{Aoki:2021kgd} quotes for the exclusive estimate of $\vert V_{ub} \vert$ the value $\vert V_{ub} \vert_{excl} \cdot 10^3 = 3.74\,(17)$ from $B \to \pi \ell \nu_\ell$ decays, while the inclusive determination performed by HFLAV~\cite{Amhis:2016xyh} reads $\vert V_{ub} \vert_{incl} \cdot 10^3 = 4.52\,(15)\,(_{-14}^{+11})$, implying a $\sim 3 \sigma$ discrepancy between them. However, a recent measurement of the inclusive value of $\vert V_{ub} \vert$ made by Belle~\cite{Belle:2021eni} has changed the picture. In fact, the collaboration has presented the result of an average over four theoretical calculations (BLNP~\cite{Lange:2005yw}, DGE~\cite{Andersen:2005mj, Gardi:2008bb}, GGOU~\cite{Gambino:2007rp}, ADFR~\cite{Aglietti:2006yb, Aglietti:2007ik}), which reads
\be
    \label{inclBELLE}
    \vert V_{ub} \vert_{incl} \cdot 10^3 = 4.10 \pm 0.09 \pm 0.22 \pm 0.15 ~ \left[ 0.28 \right] ~ , ~
\ee
where the first two errors represent the statistical and systematic uncertainties respectively, the third one denotes the theoretical model uncertainty and the fourth one is their sum in quadrature.
The FLAG review\,\cite{Aoki:2021kgd} quotes the inclusive value $\vert V_{ub} \vert_{incl} \cdot 10^3 = 4.32\,(29)$, which does not include the Belle result\,(\ref{inclBELLE}), but takes into account in the error the spread among various theoretical calculations. The FLAG inclusive value differs from the exclusive one by $\simeq 1.7$ standard deviations.
The last PDG review\,\cite{ParticleDataGroup:2020ssz} includes both the recent Belle result and the spread among various theoretical calculations. For the exclusive and inclusive determinations of $\vert V_{ub} \vert$ the PDG\,\cite{ParticleDataGroup:2020ssz} quotes the values $\vert V_{ub} \vert_{excl} \cdot 10^3 = 3.70\,(10)_{\rm exp}\,(12)_{\rm th} = 3.70\,(16)$ from $B \to \pi \ell \nu_\ell$ decays and $\vert V_{ub} \vert_{incl} \cdot 10^3 = 4.13\,(12)_{\rm exp}\,(_{-14}^{+13})_{\rm th}\,(18)_{\rm model} = 4.13\,(26)$, which differ by $\simeq 1.4$ standard deviations.
New analyses of the exclusive $b \to u$ transitions, claiming that their exclusive determinations of $\vert V_{ub} \vert$ are consistent with the estimate (\ref{inclBELLE}) at the $1 \div 1.5\,\sigma$ level, also appeared~\cite{Leljak:2021vte, Biswas:2021qyq,Gonzalez-Solis:2021pyh}. Note that the latter results were obtained by adopting for the hadronic Form Factors (FFs) the Bourrely-Caprini-Lellouch (BCL)~\cite{Bourrely:2008za} or the Bharucha-Straub-Zwicky (BSZ)~\cite{Straub:2015ica} parameterizations or the Pad\'e approximants~\cite{Gonzalez-Solis:2018ooo}. 

In this work our aim is to re-examine the $b \to u$ transition through the Dispersive Matrix (DM) method, originally proposed in Ref.\,\cite{Lellouch:1995yv} and recently reapprised in Ref.\,\cite{DiCarlo:2021dzg}. The DM method can be applied to any semileptonic decays once lattice QCD (LQCD) computations of the relevant susceptibilities and of the FFs are available. As for the susceptibilities, we present here their computation for $b \to u$ transitions following the same strategy and the same gauge ensembles considered in the case of the $b \to c$ transition in Ref.\,\cite{Martinelli:2021frl}. The FFs, instead, are taken from the results of the RBC/UKQCD\,\cite{Flynn:2015mha} and FNAL/MILC\,\cite{Lattice:2015tia} Collaborations for the $B \to \pi \ell \nu_\ell$ decays, and from RBC/UKQCD\,\cite{Flynn:2015mha}, HPQCD\,\cite{Bouchard:2014ypa} and FNAL/MILC\,\cite{Bazavov:2019aom} Collaborations for the $B_s \to K \ell \nu_\ell$ decays.
As already done for the analysis of the exclusive $B \to D^{(*)}$ decays\,\cite{Martinelli:2021onb,Martinelli:2021myh}, we stress that only LQCD computations of the FFs for small values of the recoil will be used to determine the shape of the FFs in the whole kinematical range without making any assumption on their momentum dependence. Moreover, the experimental data are not used to constrain the shape of the FFs, but only to obtain the final exclusive determination of $\vert V_{ub} \vert$. In this way, our calculation of the FFs allows to obtain pure theoretical estimates of several quantities of phenomenological interest, namely the $\tau/\mu$ ratio of the decay rates $R^{\tau/\mu}_{\pi(K)}$, which is important for testing Lepton Flavour Universality (LFU), the normalized forward-backward asymmetry $\bar{\mathcal{A}}_{FB}^{\ell,\pi(K)}$ and the normalized lepton polarization asymmetry $\bar{\mathcal{A}}_{polar}^{\ell,\pi(K)}$.  

The paper is organized as follows. In Section\,\ref{sec:DM} we review the main properties of the DM method\,\cite{DiCarlo:2021dzg}. In Section\,\ref{sec:semileptonic} we apply our procedure to predict the FFs of interest in the whole kinematical range relevant for the semileptonic $B \to \pi$ and $B_s \to K$ decays. The non-perturbative computation of the unitarity bounds for the $b \to u$ (and as a by-product for the $c \to d$) transition is based on suitable lattice two-point correlation functions, evaluated using the gauge configurations produced by the Extended Twisted Mass Collaboration (ETMC), and it is presented in the Appendix\,\ref{sec:appA}. Then, the experimental data are used to determine $\vert V_{ub} \vert$ from a bin-per-bin analysis in the case of the $B \to \pi \ell \nu_\ell$ decays and from the total branching ratio for the $B_s \to K \ell \nu_\ell$ decays. In Section\,\ref{sec:LFU} we investigate the issue of LFU by evaluating the ratio of the $\tau/\mu$ decay rates $R^{\tau/\mu}_{\pi(K)}$ from theory. We determine also the forward-backward $\bar{\mathcal{A}}_{FB}^{\ell,\pi(K)}$ and lepton polarization $\bar{\mathcal{A}}_{polar}^{\ell,\pi(K)}$ asymmetries. Finally, in Section\,\ref{sec:conclusions} we summarize the main results of this work and sketch possible future developments in the extraction of $\vert V_{ub} \vert$ from exclusive semileptonic decays.

\section{The DM method}
\label{sec:DM}

In this Section we review the main properties of the non-perturbative DM approach to the description of the semileptonic FFs, proposed in Ref.\,\cite{DiCarlo:2021dzg} and already applied to the study of $B \to D^{(*)} \ell \nu_\ell$ decays in Refs.\,\cite{Martinelli:2021onb,Martinelli:2021myh}.

\subsection{The unitarity bounds on the FFs}

The dispersion relation for a given spin-parity channel can be written in a compact form as\,\cite{Boyd:1994tt,Boyd:1997kz,Caprini:1997mu} 
\be
\label{eq:JQ2z}
\frac{1}{2\pi i } \int_{\vert z\vert =1} \frac{dz}{z}   \vert\phi(z, q_0^2) f(z)\vert^2 \leq \chi(q_0^2)\, , 
\ee
where $f(z)$ is the FF of interest, $\phi(z, q_0^2)$ is a kinematical function (whose definition depends on the spin-parity channel), $\chi(q_0^2)$ is related to the derivative of the Fourier transform of suitable Green functions of bilinear quark operators\,\cite{Boyd:1997kz} and $q_0^2$ is an auxiliary value of the squared 4-momentum transfer. Hereafter, we will refer to the functions $\chi(q_0^2)$ as the \emph{susceptibilities}.

By introducing the inner product defined as\,\cite{Bourrely:1980gp,Lellouch:1995yv}
\be
 \label{eq:inpro}
\langle g\vert h\rangle =\frac{1}{2\pi i } \oint_{\vert z\vert=1 } \frac{dz}{z}   \bar {g}(z) h(z)\, , 
\ee
where $\bar{g}(z)$ is the  complex conjugate of the function $g(z)$, Eq.\,(\ref{eq:JQ2z}) can be also written as
\be
\label{eq:JQinpro}
0 \leq \langle \phi f \vert \phi  f\rangle \leq \chi(q_0^2)\, .
\ee
As for the $B \to D^{(*)}$ case\,\cite{Martinelli:2021onb,Martinelli:2021myh}, in this work we limit ourselves to the case $q_0^2 = 0$ and we postpone the discussion of the phenomenological implications of the choice $q_0^2 \neq 0$ to a forthcoming work.

Following Refs.\,\cite{Bourrely:1980gp,Lellouch:1995yv} we introduce the set of functions
\be
g_t(z) \equiv \frac{1}{1-\bar{z}(t) z}\, , \nonumber
\ee
where $z$ is the integration variable of Eqs.\,(\ref{eq:JQ2z})-(\ref{eq:inpro}) and $\bar{z}(t)$ is the complex conjugate of the conformal variable $z(t)$, defined as
\be
\label{eq:conformal}
z(t) = \frac{\sqrt{t_+ - t} - \sqrt{t_+ - t_-}}{\sqrt{t_+ - t} + \sqrt{t_+ - t_-}}\,
\ee
with $t \equiv q^2$ being the squared 4-momentum transfer and 
\be
\label{eq:tpm}
t_\pm \equiv (m_{B_{(s)}} \pm m_{\pi(K)})^2\, .
\ee
Using the Cauchy's theorem one has
\bea
\langle g_t|\phi f \rangle  & = & \phi(z(t),q^2)\, f\left(z(t)\right)\, , \nonumber \\[2mm]
\langle g_{t_m} | g_{t_l} \rangle  & = & \frac{1}{1- \bar{z}(t_l) z(t_m)} \, . \nonumber
\eea

The central ingredient of the DM method is the matrix\,\cite{Bourrely:1980gp,Lellouch:1995yv} 
\be
\label{eq:Delta}
\mathbf{M} \equiv \left(
\begin{array}{ccccc}
\langle\phi f | \phi f \rangle  & \langle\phi f | g_t \rangle  & \langle\phi f | g_{t_1} \rangle  &\cdots & \langle\phi f | g_{t_N}\rangle  \\
\langle g_t | \phi f \rangle  & \langle g_t |  g_t \rangle  & \langle  g_t | g_{t_1} \rangle  &\cdots & \langle g_t | g_{t_N}\rangle  \\
\langle g_{t_1} | \phi f \rangle  & \langle g_{t_1} | g_t \rangle  & \langle g_{t_1} | g_{t_1} \rangle  &\cdots & \langle g_{t_1} | g_{t_N}\rangle  \\
\vdots & \vdots & \vdots & \vdots & \vdots \\ 
\langle g_{t_N} | \phi f \rangle  & \langle g_{t_N} | g_t \rangle  & \langle g_{t_N} | g_{t_1} \rangle  &\cdots & \langle g_{t_N} | g_{t_N} \rangle  \\
\end{array} \right)  ~ , ~
\ee
where $t_1, \ldots, t_N$ are the values of the squared 4-momentum transfer at which the FF $f(z)$ is known. 
In the DM method we consider only values $f(z(t_i))$ (with $i = 1, 2, ... N$) computed nonperturbatively on the lattice. 

The important feature of the matrix $\mathbf{M}$ is that, thanks to the positivity of the inner products, its determinant is positive semidefinite, i.e.\,$\det \mathbf{M} \geq 0$.
This property is not modified when the first matrix element in Eq.\,(\ref{eq:Delta}) is replaced by the susceptibility $\chi(q_0^2)$ through the dispersion relation\,(\ref{eq:JQ2z}).
Thus, using also the fact both $z$ and $f(z)$ can assume only real values in the allowed kinematical region for semileptonic decays, the original matrix\,(\ref{eq:Delta}) can be replaced explicitly by
\be
\mathbf{M}_{\chi} = \left( 
\begin{tabular}{cccccc}
   $\chi$ & $\phi f$                            & $\phi_1 f_1$                             & $\phi_2 f_2$                           & $...$ & $\phi_N f_N$ \\[2mm] 
   $\phi f$     & $\frac{1}{1 - z^2}$     & $\frac{1}{1 - z z_1}$      & $\frac{1}{1 - z z_2}$     & $...$ & $\frac{1}{1 - z z_N}$ \\[2mm]
   $\phi_1 f_1$ & $\frac{1}{1 - z_1 z}$  & $\frac{1}{1 - z_1^2}$     & $\frac{1}{1 - z_1 z_2}$ & $...$ & $\frac{1}{1 - z_1 z_N}$ \\[2mm]
   $\phi_2 f_2$ & $\frac{1}{1 - z_2 z}$  & $\frac{1}{1 - z_2 z_1}$  & $\frac{1}{1 - z_2^2}$    & $...$ & $\frac{1}{1 - z_2 z_N}$ \\[2mm]
   $... $  & $...$                           & $...$                              & $...$                              & $...$ & $...$ \\[2mm]
   $\phi_N f_N$ & $\frac{1}{1 - z_N z}$ & $\frac{1}{1 - z_N z_1}$ & $\frac{1}{1 - z_N z_2}$ & $...$ & $\frac{1}{1 - z_N^2}$
\end{tabular}
\right) ~ , ~
\label{eq:Delta2}
\ee
where $\phi_i f_i \equiv \phi(z_i) f(z_i)$ (with $i = 1, 2, ... N$) represent the known values of $\phi(z) f(z)$ corresponding to the given set of values $z_i$.
Furthermore, in order to simplify the notation we indicate $z$ and the corresponding unknown value $\phi f$ as $z_0$ and $\phi_0 f_0 \equiv \phi(z_0) f(z_0)$, respectively, so that the index $i$ now runs from $0$ to $N$. 

By imposing the positivity of the determinant of the matrix\,(\ref{eq:Delta2}) it is possible to compute explicitly the lower and the upper unitarity bounds for the FFs of interest, namely\,\cite{DiCarlo:2021dzg}
\be
  \label{eq:bounds}
  \beta - \sqrt{\gamma} \leq f_0 \leq \beta + \sqrt{\gamma} ~ , ~
\ee 
where
\bea
      \label{eq:beta_final}
      \beta & \equiv & \frac{1}{\phi_0 d_0} \sum_{j = 1}^N \phi_j f_j d_j \frac{1 - z_j^2}{z_0 - z_j} ~ , ~ \\[2mm]
      \label{eq:gamma_final}
      \gamma & \equiv &  \frac{1}{1 - z_0^2} \frac{1}{\phi_0^2 d_0^2} \left( \chi - \chi_{DM} \right) ~ , ~ \\[2mm]
      \label{eq:chi0_final}
      \chi_{DM} & \equiv & \sum_{i, j = 1}^N \phi_i f_i \phi_j  f_j d_i d_j \frac{(1 - z_i^2) (1 - z_j^2)}{1 - z_i z_j} ~ , ~ \\[2mm]
      \label{eq:d0}
     d_0 & \equiv & \prod_{m = 1}^N \frac{1 - z_0 z_m}{z_0 - z_m}  ~ , ~ \\[2mm]
     \label{eq:di}
     d_j & \equiv & \prod_{m \neq j = 1}^N \frac{1 - z_j z_m}{z_j - z_m}  ~ . ~ 
\eea
Unitarity is satisfied only when $\gamma \geq 0$, which implies $\chi \geq \chi_{DM}$.
Since $\chi_{DM}$ does not depend on $z_0$, the above condition is either never verified or always verified for any value of $z_0$.
This means that the unitarity filter $\chi \geq \chi_{DM}$ represents a parameterization-independent test of unitarity for a given set of input values $f_j$ of the FF $f$.

We remind also an important feature of the DM approach.
When $z_0$ coincides with one of the data points, i.e.~$z_0 \to z_j$, one has $\beta \to f_j$ and $\gamma \to 0$.
In other words the DM method reproduces exactly the given set of data points.
This is at variance with what may happen using truncated BCL parametrisations, since there is no guarantee that such parametrizations can reproduce exactly the set of input data.
Thus, it is worthwhile to highlight the following important feature: the DM band given by Eqs.~(\ref{eq:bounds})-(\ref{eq:di}) is equivalent to the results of all possible fits which satisfy unitarity and at the same time reproduce exactly the input data.

\subsection{The kinematical constraint at $q^2 = 0$}

In the semileptonic $B \to \pi$ and $B_s \to K$ decays, there are two FFs the vector $f_+(q^2)$ and the scalar $f_0(q^2)$ one, which are related at zero 4-momentum transfer by the following kinematical constraint (KC) 
\begin{equation*}
f_0(0) = f_+(0).
\end{equation*}
As in Ref.\,\cite{Lellouch:1995yv}, we consider
\begin{eqnarray*}
f_{lo}^*(0)&=&\max[f_{+,lo}(0),f_{0,lo}(0)],\\
f_{up}^*(0)&=&\min[f_{+,up}(0),f_{0,up}(0)],
\end{eqnarray*}
where in terms of Eq.\,(\ref{eq:bounds}) one has $f_{+(0), lo(up)}(0) = \beta_{+(0)} \mp \sqrt{\gamma_{+(0)}}$ for $z_0 = z(t=0)$.

Putting $f(0) \equiv f_0(0) = f_+(0)$ one gets
\begin{equation}
\label{eq:rangeFFKC}
f_{lo}^*(0) \leq f(0) \leq f_{up}^*(0) \, .
\end{equation}
We now consider the FF at zero 4-momentum transfer to be uniformly distributed in the range given by Eq.\,(\ref{eq:rangeFFKC}). The resulting value is considered as a new input at $t_{N+1} = 0$. Thus, for each of the two FFs we consider a new matrix, $\mathbf{M}_{KC}$, that has one more row and one more column with respect to $\mathbf{M}$ in order to contain the common value $f(t_{N+1} = 0)$, namely
\be
\mathbf{M}_{KC} = \left( 
\begin{tabular}{ccccccc}
   $\chi$ & $\phi_0 f_0$                            & $\phi_1 f_1$                             & $\phi_2 f_2$                           & $...$ & $\phi_N f_N$ & $\phi_{N+1} f_{N+1}$ \\[2mm] 
   $\phi_0 f_0$     & $\frac{1}{1 - z_0^2}$     & $\frac{1}{1 - z_0 z_1}$      & $\frac{1}{1 - z_0 z_2}$     & $...$ & $\frac{1}{1 - z_0 z_N}$ & $\frac{1}{1 - z_0 z_{N+1}}$ \\[2mm]
   $\phi_1 f_1$ & $\frac{1}{1 - z_1 z_0}$  & $\frac{1}{1 - z_1^2}$     & $\frac{1}{1 - z_1 z_2}$ & $...$ & $\frac{1}{1 - z_1 z_N}$ & $\frac{1}{1 - z_1 z_{N+1}}$ \\[2mm]
   $\phi_2 f_2$ & $\frac{1}{1 - z_2 z_0}$  & $\frac{1}{1 - z_2 z_1}$  & $\frac{1}{1 - z_2^2}$    & $...$ & $\frac{1}{1 - z_2 z_N}$ & $\frac{1}{1 - z_2 z_{N+1}}$ \\[2mm]
   $... $  & $...$                           & $...$                              & $...$                              & $...$ & $...$ \\[2mm]
   $\phi_N f_N$ & $\frac{1}{1 - z_N z_0}$ & $\frac{1}{1 - z_N z_1}$ & $\frac{1}{1 - z_N z_2}$ & $...$ & $\frac{1}{1 - z_N^2}$ & $\frac{1}{1 - z_N z_{N+1}}$ \\[2mm]
$\phi_{N+1} f_{N+1}$ & $\frac{1}{1 - z_{N+1} z_0}$ & $\frac{1}{1 - z_{N+1} z_1}$ & $\frac{1}{1 - z_{N+1} z_2}$ & $...$ & $\frac{1}{1 - z_{N+1} z_N}$ & $\frac{1}{1 - z_{N+1}^2}$
\end{tabular}
\right) ~ . ~
\label{eq:Delta2KC}
\ee
In order to predict the DM bands for $f_{+,0}(q^2)$ in the whole kinematical range, we consider the matrix $\mathbf{M}_{KC}$ at any value of the momentum transfer and, by using the explicit forms\,(\ref{eq:bounds})-(\ref{eq:di}), we get the corresponding unitarity bounds. 

Finally, as discussed in Refs.\,\cite{DiCarlo:2021dzg,Martinelli:2021onb,Martinelli:2021myh}, we use the mean values, the uncertainties and (when available) the correlations of the LQCD computations of the FFs and the susceptibilities to construct a multivariate Gaussian distribution for generating a sample of bootstrap events to each of which the DM method is applied.

\section{Semileptonic $B \to \pi$ and $B_s \to K$ decays}
\label{sec:semileptonic}

In this Section we apply the DM method to the study of the semileptonic $B \to \pi$ and $B_s \to K$ decays. First we describe the state of the art of the LQCD computations of the relevant FFs, which are limited to large values of the 4-momentum transfer $q^2$, and then we apply the DM method to get the FFs in the whole kinematical range accessible to experiments. To this end another nonperturbative input is used, namely the values of the longitudinal and transverse vector susceptibilities, $\chi_{0^+}(0)$ and $\chi_{1^-}(0)$, whose determination based on suitable lattice two-point correlation functions for the $b \to u$ transition is illustrated in the Appendix\,\ref{sec:appA}.  Finally, we compare our theoretical results with the experimental data in order to extract $\vert V_{ub} \vert$ from the semileptonic $B \to \pi$ and $B_s \to K$ channels. 

\subsection{State of the art of the LQCD computations of the FFs}

The FFs entering semileptonic $B \to \pi$ decays have been studied by the RBC/ UKQCD\,\cite{Flynn:2015mha} and the FNAL/MILC\,\cite{Lattice:2015tia} Collaborations. 
In the case of the $B_s \to K$ transition several LQCD computations of the FFs are available, namely from the RBC/UKQCD\,\cite{Flynn:2015mha}, HPQCD\,\cite{Bouchard:2014ypa} and FNAL/MILC\,\cite{Bazavov:2019aom} Collaborations. 
For both channels the lattice computations of the FFs are available in the large-$q^2$ region, $17~\mbox{GeV}^2 \lesssim q^2 \leq q_{max}^2 \equiv (m_{B_{(s)}} - m_{\pi(K)})^2$.

The authors of Ref.\,\cite{Flynn:2015mha} provide synthetic LQCD values of the FFs (together with their statistical and systematic correlations) at three values of $q^2$ in the large-$q^2$ regime, namely $q^2 = \{ 19.0, 22.6, 25.1 \}$ GeV$^2$ for the $B \to \pi$ transition and $q^2 = \{ 17.6, 20.8, 23.4 \}$ GeV$^2$ in the case of the $B_s \to K$ transition.
These data can be directly used as inputs for our DM method. 
In the other works\,\cite{Lattice:2015tia,Bouchard:2014ypa,Bazavov:2019aom} the results of BCL fits of the FFs extrapolated to the continuum limit and to the physical pion point are available. 
Thus, from the marginalized BCL coefficients we evaluate the mean values, uncertainties and correlations of the FFs at the three values of $q^2$ given in Ref.\,\cite{Flynn:2015mha}. 
The LQCD results used as inputs for our DM method are collected in Tables\,\ref{tab:LQCDBPi} and\,\ref{tab:LQCDBsK} for the $B \to \pi$ and $B_s \to K$ decays, respectively. 
In the next future new LQCD computations of the FFs are expected to become available\,\cite{Gelzer:2019zwx,Flynn:2020nmk}.

For both $B \to \pi$ and $B_s \to K$ decays we have also combined all the LQCD determinations of the FFs corresponding to the same values of the momentum transfer. We have followed the procedure already applied in Ref.\,\cite{Martinelli:2021onb} to the $B \to D^*$ case: starting from $N$ computations of the FFs with mean values $x_i^{(k)}$ and uncertainties $\sigma_i^{(k)}$ ($k=1,\cdots,N$) corresponding to a given value $q_i^2$ of the squared 4-momentum transfer, the \emph{combined} LQCD average $x_i$ and uncertainty $\sigma_i$ are given by (see Ref.\,\cite{Carrasco:2014cwa})
\begin{eqnarray}
\label{eq28Carr}
x_i &=& \sum_{k=1}^N \omega^{(k)} x_i^{(k)},\\
\label{eq28Carrb}
\sigma_i^2 &=& \sum_{k=1}^N \omega^{(k)} (\sigma_i^{(k)})^2 + \sum_{k=1}^N \omega^{(k)} (x_i^{(k)} - x_i)^2 ~ , ~
\end{eqnarray}
where $\omega^{(k)}$ represents the weight associated to the $k$-th calculation ($\sum_{k = 1}^N \omega^{(k)} = 1$).
Since the uncertainties of the various lattice computations are comparable, in what follows we assume the same weight for all the computations, i.e.~we consider the simple choice $\omega^{(k)} = 1 / N$. 
The results of Eqs.\,(\ref{eq28Carr})-(\ref{eq28Carrb}) are shown in the last columns of the Tables \ref{tab:LQCDBPi} and \ref{tab:LQCDBsK} for both the $B \to \pi$ and the $B_s \to K$ cases, respectively. Moreover, the covariance matrix $C$ of the combined data can be easily evaluated in terms of the covariance matrices $C^{(k)}$ of each single LQCD computation as 
\begin{equation}
\label{eq28Carrc}
C_{i j}\equiv \frac{1}{N} \sum_{k=1}^N C_{i j}^{(k)} + \frac{1}{N} \sum_{k=1}^N (x_i^{(k)} - x_i) (x_j^{(k)} - x_j) ~ , ~
\end{equation}
where the indices $i$ and $j$ run over the number of values of the 4-momentum transfer at which the LQCD computations of the FFs have been performed, namely in this work $i, j = 1, 2, 3$.

\begin{table}[htb!]
\renewcommand{\arraystretch}{1.1}
\begin{center}
\begin{tabular}{|c||c|c|c|}
\hline
& ~ RBC/UKQCD ~ & ~ FNAL/MILC ~ & ~ Combined ~ \\
\hline \hline
$f^{\pi}_+(19.0$ GeV$^2$)& 1.21(10)(9) & 1.17(8) & 1.19(11)\\
$f^{\pi}_+(22.6$ GeV$^2$)& 2.27(13)(14) & 2.24(12)&2.25(16)\\
$f^{\pi}_+(25.1$ GeV$^2$)& 4.11(51)(29) & 4.46(23)& 4.29(48)\\
\hline \hline
$f^{\pi}_0(19.0$ GeV$^2$) & 0.46(3)(5) & 0.46(3)& 0.46(5)\\
$f^{\pi}_0(22.6$ GeV$^2$) & 0.68(3)(6) & 0.65(3)& 0.66(5)\\
$f^{\pi}_0(25.1$ GeV$^2$) & 0.92(3)(6) & 0.86(3)& 0.89(6)\\
\hline
\end{tabular}
\caption{\it Mean values and uncertainties of the LQCD computations of the FFs $f_{+,0}^\pi(q^2)$ obtained at three selected values of $q^2$ from the results of the RBC/UKQCD\,\cite{Flynn:2015mha} and FNAL/MILC\,\cite{Lattice:2015tia} Collaborations. For the RBC/UKQCD computations the first error is statistical while the second one is systematic. The last column contains the results of the combination procedure given in Eqs.\,(\ref{eq28Carr})-(\ref{eq28Carrb}) with $\omega^{(k)} = 1 / N$.}
\label{tab:LQCDBPi}
\end{center}
\renewcommand{\arraystretch}{1.0}
\end{table}

\begin{table}[htb!]
\renewcommand{\arraystretch}{1.1}
\begin{center}
\begin{tabular}{|c||c|c|c|c|}
\hline
& ~ RBC/UKQCD ~ & ~ HPQCD ~ & ~ FNAL/MILC ~ & ~ Combined ~ \\
\hline \hline
$f^K_+(17.6$ GeV$^2$) & 0.99(4)(5)  & 1.04(5) & 1.01(4)& 1.01(6)\\
$f^K_+(20.8$ GeV$^2$) & 1.64(6)(7)   & 1.68(7) & 1.68(5)& 1.67(8)\\
$f^K_+(23.4$ GeV$^2$) & 2.77(9)(11)   & 2.94(13) & 2.91(9)& 2.87(15)\\
\hline \hline
$f^K_0(17.6$ GeV$^2$) & 0.48(2)(3)  & 0.53(3) & 0.44(2)& 0.48(4)\\
$f^K_0(20.8$ GeV$^2$) & 0.63(2)(4)  & 0.64(3) & 0.59(1)& 0.62(4)\\
$f^K_0(23.4$ GeV$^2$) & 0.81(2)(5)  & 0.79(4) & 0.76(2)& 0.79(5)\\
\hline
\end{tabular}
\caption{\it Mean values and uncertainties of the LQCD computations of the FFs $f_{+,0}^K(q^2)$ obtained at three selected values of $q^2$ from the results of the RBC/UKQCD\,\cite{Flynn:2015mha}, HPQCD\,\cite{Bouchard:2014ypa} and FNAL/MILC\,\cite{Bazavov:2019aom} Collaborations. For the RBC/UKQCD computations the first error is statistical while the second one is systematic. The last column contains the results of the combination procedure given in Eqs.\,(\ref{eq28Carr})-(\ref{eq28Carrb}) with $\omega^{(k)} = 1 / N$.}
\label{tab:LQCDBsK}
\end{center}
\renewcommand{\arraystretch}{1.0}
\end{table}

\subsection{Theoretical expression of the differential decay width}

For the semileptonic $B \to \pi \ell \nu_\ell$ and the $B_s \to K \ell \nu_\ell$ decays the vector $f_+^{\pi(K)}(q^2)$ and scalar $f_0^{\pi(K)}(q^2)$ FFs are related to the matrix elements of the weak vector current $V^{\mu} \equiv \bar{b} \gamma^{\mu} u$ by
\be
\label{FINALmatrelem}
 \Braket{\pi(K) | V^{\mu} |B_{(s)}} = f_+^{\pi(K)}(q^2)\left(P^{\mu} - \frac{P \cdot q}{q^2}q^{\mu}\right) +f_0^{\pi(K)}(q^2) \frac{P \cdot q}{q^2} q^{\mu} ~ , ~
\ee
where $P^{\mu} = p_{B_{(s)}}^{\mu} + p_{\pi(K)}^{\mu}$, $q^{\mu} = p_{B_{(s)}}^{\mu} - p_{\pi(K)}^{\mu}$ and $P \cdot q = m_{B_{(s)}}^2 - m_{\pi(K)}^2$. We remind that the two FFs in Eq.\,(\ref{FINALmatrelem}) are constrained at zero momentum transfer by the kinematical relation $f_+^{\pi(K)}(0) = f_0^{\pi(K)}(0)$.

A direct computation of the two-fold differential decay width within the Standard Model gives the final expression
\begin{eqnarray}
\label{finaldiff333pre}
&& \frac{d^2\Gamma(B_{(s)}\to \pi(K) \ell \nu_\ell)}{dq^2 d \cos\theta_{\ell}} = \frac{G_F^2 \vert V_{ub} \vert^2}{128\pi^3 m_{B_{(s)}}^2} \left(1-\frac{m_{\ell}^2} {q^2}\right)^2 \nonumber \\[2mm]
&& \hspace{1.5cm} \cdot \Big\{4 m_{B_{(s)}}^2 \vert \vec{p}_{\pi(K)}\vert^3 \left(\sin^2\theta_{\ell}+\frac{m_{\ell}^2}{2q^2} \cos^2 \theta_{\ell}\right) \vert f_{+}^{\pi(K)} (q^2) \vert^2 \nonumber \\[2mm]
&& \hspace{1.5cm} + \frac{4 m_{\ell}^2}{q^2} (m_{B_{(s)}}^2 - m_{\pi(K)}^2) m_{B_{(s)}} \vert \vec{p}_{\pi(K)} \vert^2 \cos\theta_{\ell} \, \Re\left(f_+^{\pi(K)}(q^2) f_0^{*\pi(K)}(q^2)\right) \nonumber \\
&& \hspace{1.5cm} + \frac{m_{\ell}^2}{q^2} ( m_{B_{(s)}}^2 - m_{\pi(K)}^2)^2  \vert \vec{p}_{\pi(K)} \vert  \vert f_{0}^{\pi(K)} (q^2) \vert^2 \Big\} ~ , ~
\end{eqnarray}
where $G_F$ is the Fermi constant, $\vec{p}_{\pi(K)}$ the 3-momentum of the $\pi(K)$ meson in the $B_{(s)}$-meson rest frame, $m_\ell$ the mass of the produced lepton and $\theta_\ell$ represents the angle between the final charged lepton and the $B_{(s)}$-meson momenta in the rest frame of the final state leptons. By integrating out the dependence on the angle $\theta_\ell$ one gets
\begin{eqnarray}
\label{finaldiff333}
    \frac{d\Gamma(B_{(s)}\to \pi(K) \ell \nu_\ell)}{dq^2} & = & \frac{G_F^2 \vert V_{ub} \vert^2}{24\pi^3} \left(1-\frac{m_\ell^2}{q^2}\right)^2 
            \left[\vert \vec{p}_{\pi(K)} \vert^3 \left(1+\frac{m_\ell^2}{2q^2}\right) \vert f_{+}^{\pi(K)} (q^2) \vert^2 \right. \nonumber \\[2mm]
            & + & \left. m_{B_{(s)}}^2 \vert \vec{p}_{\pi(K)} \vert \left( 1-r_{\pi(K)}^2 \right)^2 \frac{3m_\ell^2}{8q^2} \vert f_{0}^{\pi(K)} (q^2) \vert^2\right] ~ , ~
\end{eqnarray}
where explicitly
\begin{equation}
     \vert \vec{p}_{\pi(K)} \vert = m_{\pi(K)} \sqrt{ \left( \frac{1 + r_{\pi(K)}^2 - q^2 / m_{B_{(s)}}^2}{2 r_{\pi(K)}} \right)^2 - 1}
\end{equation}
with $r_{\pi(K)} \equiv m_{\pi(K)} / m_{B_{(s)}}$.

\subsection{Application of the DM method to the description of the FFs}
\label{sec:FFs_DM}

The kinematical functions $\phi_0$ and $\phi_+$ corresponding to the scalar and vector FFs of the $B_{(s)} \to \pi(K)$ decays are given by\,\cite{Boyd:1997kz}
\begin{eqnarray}
\phi_0(z, 0) & = & \sqrt{\frac{2n_I}{3}} \sqrt{\frac{3 t_+ t_-}{4 \pi}} \frac{1}{t_+ - t_-} \frac{1+z}{(1-z)^{5/2}} 
                             \left(  \sqrt{\frac{t_+}{t_+-t_-}} + \frac{1+z}{1-z}\right)^{-4}\, , \nonumber  \\
\phi_+(z, 0) & = & \sqrt{\frac{2n_I}{3}} \sqrt{\frac{1}{\pi (t_+-t_-)}} \frac{(1+z)^2}{(1-z)^{9/2}} \left( \sqrt{\frac{t_+}{t_+-t_-}} + \frac{1+z}{1-z}\right)^{-5}\, .  
\label{eq:kinfun}
\end{eqnarray}
where $z \equiv z(t = q^2)$ is defined in Eq.\,(\ref{eq:conformal}) and $n_I$ is an isospin Clebsh-Gordan factor equal to $n_I = 3/2$ for the $B \to \pi$ decays and to $n_I = 1$ for the $B_s \to K$ case. 
In order to take into account the $B^*$ pole in the transverse channel, the transverse kinematical function $\phi_+$ is modified as
\begin{equation}
\phi_+(z, 0) \to \phi_+(z, 0) \cdot \frac{z - z(m_{B^*}^2)}{1 - z \, \bar{z}(m_{B^*}^2)} ~
\label{eq:modify} 
\end{equation}
with $m_{B^*} = 5.325$ GeV from the PDG\,\cite{ParticleDataGroup:2020ssz}.

The evaluation of the unitarity bound $\chi(0) \geq \chi_{DM}$ (see Eq.\,(\ref{eq:chi0_final})) requires the knowledge of the susceptibilities $\chi(0)$ appearing in the DM matrix\,(\ref{eq:Delta2}). For the $b \to u$ transition we have been computed them nonperturbatively using suitable two-point lattice correlators, as described in the Appendix~\ref{sec:appA}. The nonperturbative values for the susceptibilities relevant for the scalar $f_0(q^2)$ and vector $f_+(q^2)$ FFs are respectively 
\bea
    \label{eq:chi0+}
    \chi_{0^+}(0) & = & (2.04 \pm 0.20) \cdot 10^{-2} ~ , ~ \\[2mm]
    \label{eq:chi1-}
    \chi_{1^-}(0) & = & (4.45 \pm 1.16) \cdot 10^{-4} ~ \mbox{GeV}^{-2} ~ 
\eea 
after subtraction of the contribution of the $B^*$-meson bound state (see Appendix~\ref{sec:appA}).

\begin{figure}[htb!]
\centering{\includegraphics[scale=0.40]{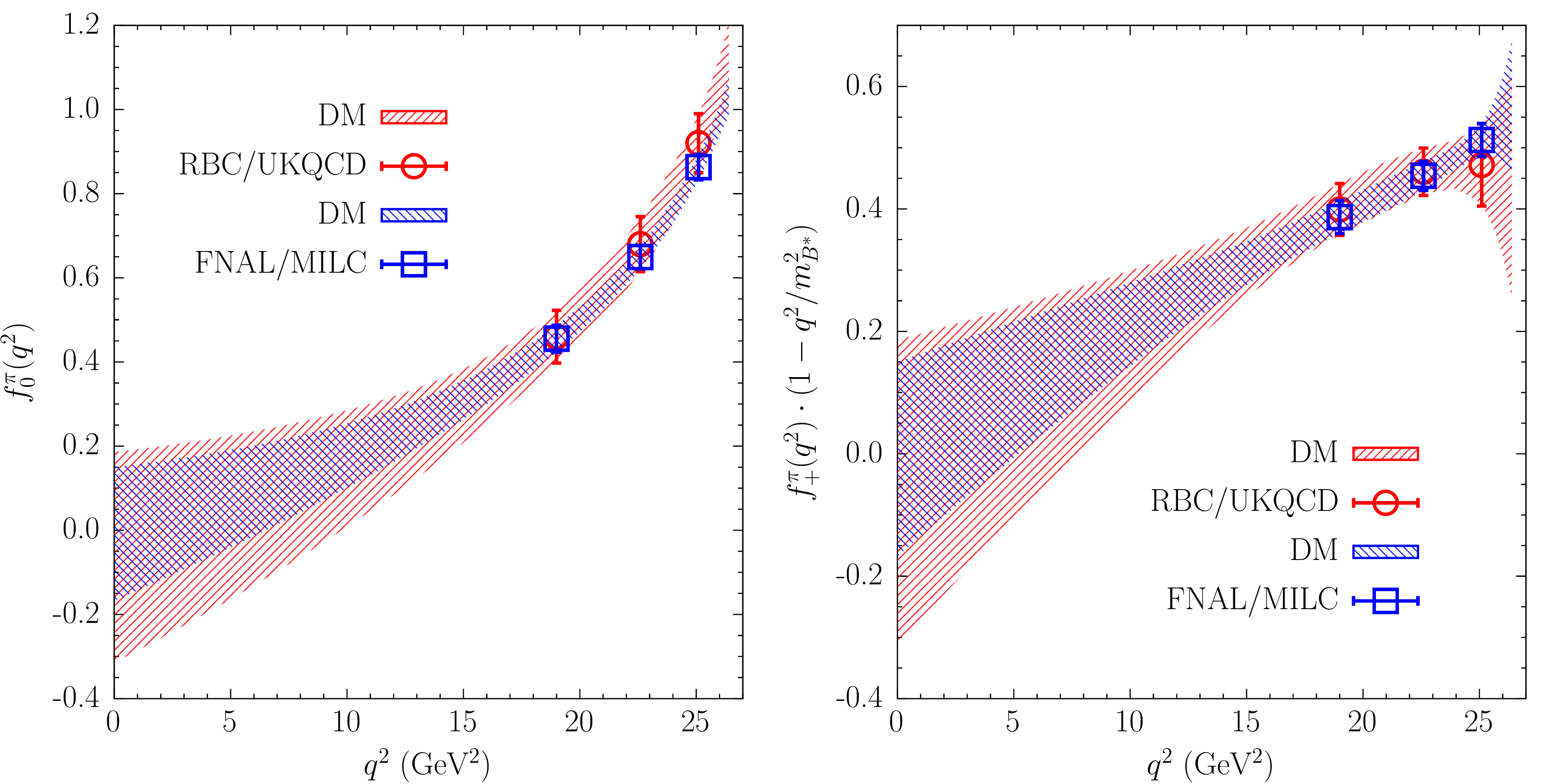}}
\caption{\it \small The scalar $f_0^\pi(q^2)$ (left panel) and vector $f_+^\pi(q^2)$ (right panel) FFs entering the semileptonic $B \to \pi \ell \nu_\ell$ decays computed by the DM method as a function of the 4-momentum transfer $q^2$ using the LQCD inputs from RBC/UKQCD\,\cite{Flynn:2015mha} and FNAL/MILC\,\cite{Lattice:2015tia} Collaborations (see Table\,\ref{tab:LQCDBPi}). For both FFs the red and blue bands correspond to the DM results obtained at $1 \sigma$ level using the RBC/UKQCD data (red circles) and FNAL/MILC (blue squares) data, respectively. In the right panel the vector FF is multiplied by the factor $(1 - q^2 / m_{B^*}^2)$ with $m_{B^*} = 5.325$ GeV.\hspace*{\fill}}
\label{FFMMBpi}
\end{figure}

\begin{figure}[htb!]
\centering{\includegraphics[scale=0.40]{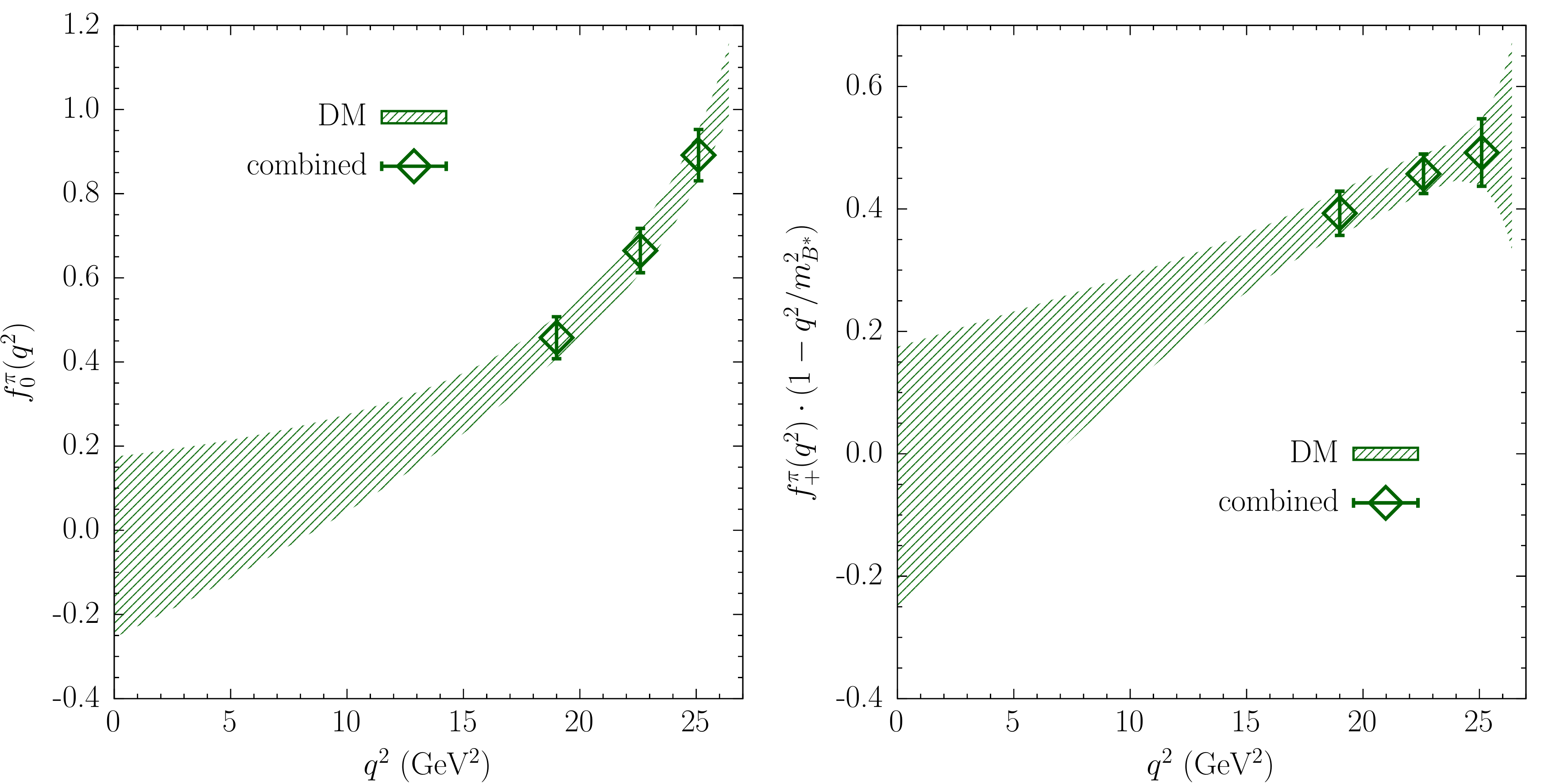}}
\caption{\it \small The bands of the scalar $f_0^\pi(q^2)$ (left panel) and vector $f_+^\pi(q^2)$ (right panel) FFs entering the semileptonic $B \to \pi \ell \nu_\ell$ decays computed by the DM method at $1 \sigma$ level using as lattice inputs the combined LQCD data of Table\,\ref{tab:LQCDBPi}, shown as green diamonds. In the right panel the vector FF is multiplied by the factor $(1 - q^2 / m_{B^*}^2)$ with $m_{B^*} = 5.325$ GeV.\hspace*{\fill}}
\label{FFMMBpiCOMB}
\end{figure}

We now apply the DM method to the $B \to \pi$ decay using as inputs the lattice data of Table\,\ref{tab:LQCDBPi} corresponding to the three sets labelled RBC/UKQCD, FNAL/MILC and combined. A total of $5 \cdot 10^4$ events are generated using the multivariate Gaussian distribution including the correlations among the LQCD data.
It turns out that the unitarity bounds for both  $f_0^\pi$ and $f_+^\pi$ as well as the KC $f_0^\pi(0) = f_+^\pi(0) \equiv f^\pi(0)$ are satisfied by $98 \div 100 \%$ of the events and, therefore, neither the skeptical nor the iterative procedures described in Refs.\,\cite{DiCarlo:2021dzg,Martinelli:2021onb,Martinelli:2021myh} need to be applied.
In Figs.\,\ref{FFMMBpi} and \,\ref{FFMMBpiCOMB} we show the resulting bands of the two FFs. 
The extrapolation to $q^2 = 0$, which is crucial in order to analyze the experimental data, reads
\begin{eqnarray*}
\label{fplusBpizmax}
f^{\pi}(q^2 = 0)\vert_{\rm{RBC/UKQCD}} & = & -0.06 \pm 0.25 ~ , ~ \\[2mm] 
f^{\pi}(q^2 = 0)\vert_{\rm{FNAL/MILC}} & = & -0.01 \pm 0.16 ~ , ~ \\[2mm]
f^{\pi}(q^2 = 0)\vert_{\rm{combined}} & = & -0.04 \pm 0.22 ~ . ~
\end{eqnarray*}
The above results exhibit large uncertainties due to the long extrapolation from the high-$q^2$ region of the input data down to $q^2 = 0$.
We stress again that our results do not depend on any parameterization of the shape of the FFs. This is at variance with what happens with the BCL parameterizations of Refs.\,\cite{Flynn:2015mha,Lattice:2015tia}, where the extrapolated mean values and uncertainties of the FFs at $q^2 = 0$ are plagued by instabilities with respect to the order of the truncation of the expansion. 

Our results for the FFs at $q^2 = 0$ are consistent within $1.4 \div 1.8$ standard deviations with the recent estimate obtained in Ref.\,\cite{Leljak:2021vte} using Light Cone Sum Rules (LCSR), namely
\begin{equation*}
f^{\pi}(q^2=0)_{\rm{LCSR}} = 0.28 \pm 0.03.
\end{equation*}

As for the semileptonic $B_s \to K$ decays few differences have to be considered with respect to the $B \to \pi$ case besides the obvious changes in the masses of the mesons involved. First, in the kinematical functions\,(\ref{eq:kinfun}) the isospin factor $n_I$ is now equal to unity instead of $3/2$ as in the $B \to \pi$ case. This is due to the fact that in the $B_s \to  K$ decays only the strange quark can be the spectator quark of the transition.
Second, following Refs.\,\cite{Flynn:2015mha,Bouchard:2014ypa,Bazavov:2019aom} a modification like the one in the Eq.\,(\ref{eq:modify}) has to be applied also to $\phi_{0}(z, 0)$ due to the presence of a scalar resonance $B^*(0^+)$ with a mass close to $5.68$ GeV, expected from the lattice results of Ref.\,\cite{Gregory:2010gm}, lying below the pair production threshold located at $M_{B_s} + M_{K} \simeq 5.86$ GeV.
For the susceptibilities $\chi_{0^+}(0)$ and $\chi_{1^-}(0)$ we adopt conservatively the same values of the $B \to \pi$ case.

\begin{figure}[htb!]
\centering{\includegraphics[scale=0.40]{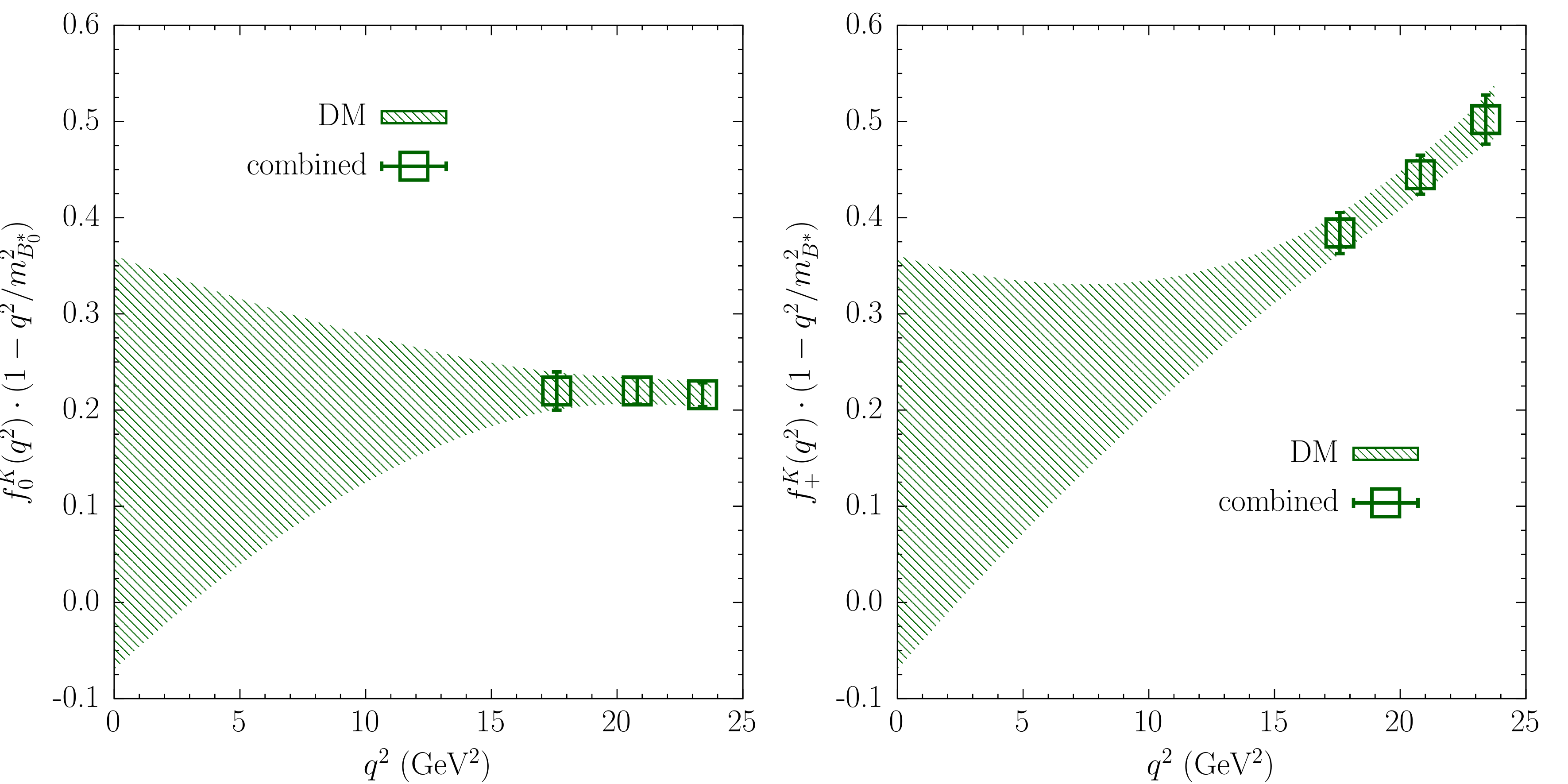}}
\caption{\it \small The bands of the scalar $f_0^K(q^2)$ (left panel) and vector $f_+^K(q^2)$ (right panel) FFs entering the semileptonic $B_s \to K \ell \nu_\ell$ decays computed by the DM method at $1 \sigma$ level using as inputs the combined LQCD data of Table\,\ref{tab:LQCDBsK}, shown as green diamonds. In the left panel the scalar FF is multiplied by the factor $(1 - q^2 / m_{B_0^*}^2)$ with $m_{B_0^*} = 5.68$ GeV, while in the right panel the vector FF is multiplied by the factor $(1 - q^2 / m_{B^*}^2)$ with $m_{B^*} = 5.325$ GeV.\hspace*{\fill}}
\label{FFMMBsKCOMB}
\end{figure}

We apply the DM method using as inputs the various sets of LQCD data of Table\,\ref{tab:LQCDBsK}.  A total of $5 \cdot 10^4$ events are generated using the multivariate Gaussian distributions including the correlations among the LQCD computations. As in the $B \to \pi$ case, the unitarity bounds for both $f_0^K$ and $f_+^K$ as well as the KC $f_0^K(0) = f_+^K(0) \equiv f^K(0)$ are satisfied by $98 \div 100 \%$ of the events.
The DM bands for the FFs corresponding to the use of the combined LQCD data of Table\,\ref{tab:LQCDBsK} are shown in Fig.\,\ref{FFMMBsKCOMB}.
Note the impact of the KC at $q^2 = 0$ on the extrapolation of the FFs in the low-$q^2$ region.

The extrapolation of the FFs to $q^2 = 0$ reads
\begin{eqnarray*}
\label{fplusBsKzmax}
f^K(q^2=0)\vert_{\rm{RBC/UKQCD}} & = & 0.08 \pm 0.15 ~ , ~ \\[2mm]
f^K(q^2=0)\vert_{\rm{HPQCD}} & = & 0.28 \pm 0.21 ~ , ~ \\[2mm]
f^K(q^2=0)\vert_{\rm{FNAL/MILC}} & = & 0.07 \pm 0.11 ~ , ~ \\[2mm]
f^K(q^2=0)\vert_{\rm{combined}} & = & 0.15 \pm 0.21 ~ . ~
\end{eqnarray*}

The above results can be compared with the recent LCSR estimate of Ref.\,\cite{Khodjamirian:2017fxg}, which is
\begin{equation*}
f^K(q^2=0)_{\rm{LCSR}} = 0.336 \pm 0.023 ~ . ~
\end{equation*}
It can be seen that the results based on the RBC/UKQCD and FNAL/MILC data differ respectively by $1.7$ and $2.4$ standard deviations from the LCSR estimate, while the results based on the HPQCD data and the combined LQCD ones are in agreement thanks to larger mean values and uncertainties.

The DM results presented so far indicate clearly that for both the $B \to \pi$ and $B_s \to K$ channels the extension of direct LQCD computations of the FFs toward values of $q^2$ lower than $\sim 17$ GeV$^2$ is crucial for improving the precision of their extrapolation to $q^2 = 0$ without resorting to the use of the experimental data. 

\subsection{New estimate of $\vert V_{ub} \vert$}

In order to obtain $\vert V_{ub} \vert$ we use our results for the FFs in the whole kinematical range and the experimental data. For the semileptonic $B \to \pi$ decays the BaBar and the Belle Collaborations\,\cite{delAmoSanchez:2010af, Ha:2010rf, Lees:2012vv, Sibidanov:2013rkk} have measured the \emph{differential} branching ratios (BRs) in different bins of the 4-momentum transfer $q^2$. Instead, for the $B_s \to K$ decays only the ratio of the \emph{total} branching fractions of  the semileptonic $B_s \to K$ and $B_s \to D_s$ decays is available at present \cite{Aaij:2020nvo}.

\subsubsection{$\vert V_{ub} \vert$ from $B \to \pi \ell \nu_\ell$ decays}

For the extraction of the CKM matrix element we follow the procedure used in Refs.\,\cite{Riggio:2017zwh, Martinelli:2021onb} in the case of several semileptonic heavy-meson decays characterized by the production of a final pseudoscalar meson. In what follows, we will distinguish the two different channels that have been measured by the experiments, $i.e.$~$B^0 \to \pi^- \ell^+ \nu$ and $B^+ \to \pi^0 \ell^+ \nu$ with $\ell = e, \mu$. Starting from the Eq.\,(\ref{finaldiff333}), for the generic $i$-th bin in $q^2$ we have\,\cite{Sibidanov:2013rkk}
\begin{equation}
\label{VubFINAL}
\vert V_{ub} \vert _i= \sqrt{\frac{C_v}{\tau_{B^v}} \cdot \frac{\Delta\mathcal{B}\vert_i^{exp} }{\Delta \zeta_i}},
\end{equation}
where $\Delta\mathcal{B}\vert_i^{exp}$ is the experimental branching fraction and $\Delta \zeta_i$ the corresponding theoretical decay width (without $\vert V_{ub} \vert$ therein) in the given bin.
Since $\ket{\pi^0} \equiv \left( \ket{u \bar{u}}-\ket{d \bar{d}} \right) / \sqrt{2}$, the isospin coefficient $C_v$ is equal to 2 for the $B^+ \to \pi^0 \ell^+ \nu$ decays and to 1 for the $B^0 \to \pi^- \ell^+ \nu$ transitions. Finally, $\tau_{B^v}$ is the lifetime of the decaying $B$-meson.

Our procedure can be summarised as follows:

\begin{itemize}

\item using the mean values and the covariance matrices available for the of the FFs and of the susceptibilities computed in LQCD we generate a multivariate Gaussian distribution of events of input data to each of which the DM method is applied for obtaining the subset of events passing the unitarity filter and satisfying the KC at $q^2 = 0$ (see Section\,\ref{sec:FFs_DM});

\item for each of the surviving events we evaluate the vector FF $f_+(q^2)$ at several values of $q^2$, which allow to perform the partial integration needed to calculate the theoretical differential decay width $\Delta \zeta_i$ in each of the experimental $q^2$-bins (in the massless lepton limit);

\item from the resulting distribution of values of $\Delta \zeta_i$ we evaluate the mean values $\langle \Delta \zeta_i \rangle$ for each bin and the corresponding covariance matrix; 

\item through multivariate Gaussian distributions we generate $N_{boot}$ events of the measured differential branching fraction $\Delta\mathcal{B}\vert_i^{exp}$ for each bin in $q^2$ and experiment\,\cite{delAmoSanchez:2010af, Ha:2010rf, Lees:2012vv, Sibidanov:2013rkk} and, independently, $N_{boot}$ events of the theoretical decay widths $\Delta \zeta_i$;

\item we compute $N_{boot}$ values $\vert V_{ub} \vert_i$ for each $q^2$-bin and each experiment through Eq.\,(\ref{VubFINAL});

\item using the distributions of values of $\vert V_{ub} \vert_i$ we calculate the corresponding mean values $\langle \vert V_{ub} \vert_i \rangle$ and covariance matrix $\mathbf{C}_{ij}$ among the bins for each experiment;

\item we evaluate the CKM matrix element $\vert V_{ub} \vert$ for the $n$-th experiment ($n = 1, \ldots, 6$ for the semileptonic $B \to \pi$ decays) as the best constant fit over all the bins of the given experiment, i.e.~through the following formulae
\begin{equation}
\label{muVubfinal}
\vert V_{ub} \vert_n = \frac{\sum_{i,j} (\mathbf{C}^{-1})_{ij}\langle  \vert V_{ub} \vert_i \rangle}{\sum_{i,j} (\mathbf{C}^{-1})_{ij}},\,\,\,\,\,\,\,\,\,\,\,\,\sigma^2_{\vert V_{ub} \vert_{n}} = \frac{1}{\sum_{i,j} (\mathbf{C}^{-1})_{ij}},
\end{equation} 
where the indices $i,j$ run over all the $q^2$-bins of the $n$-th experiment.

\end{itemize}

\begin{figure}[htb!]
\centering{\includegraphics[scale=0.40]{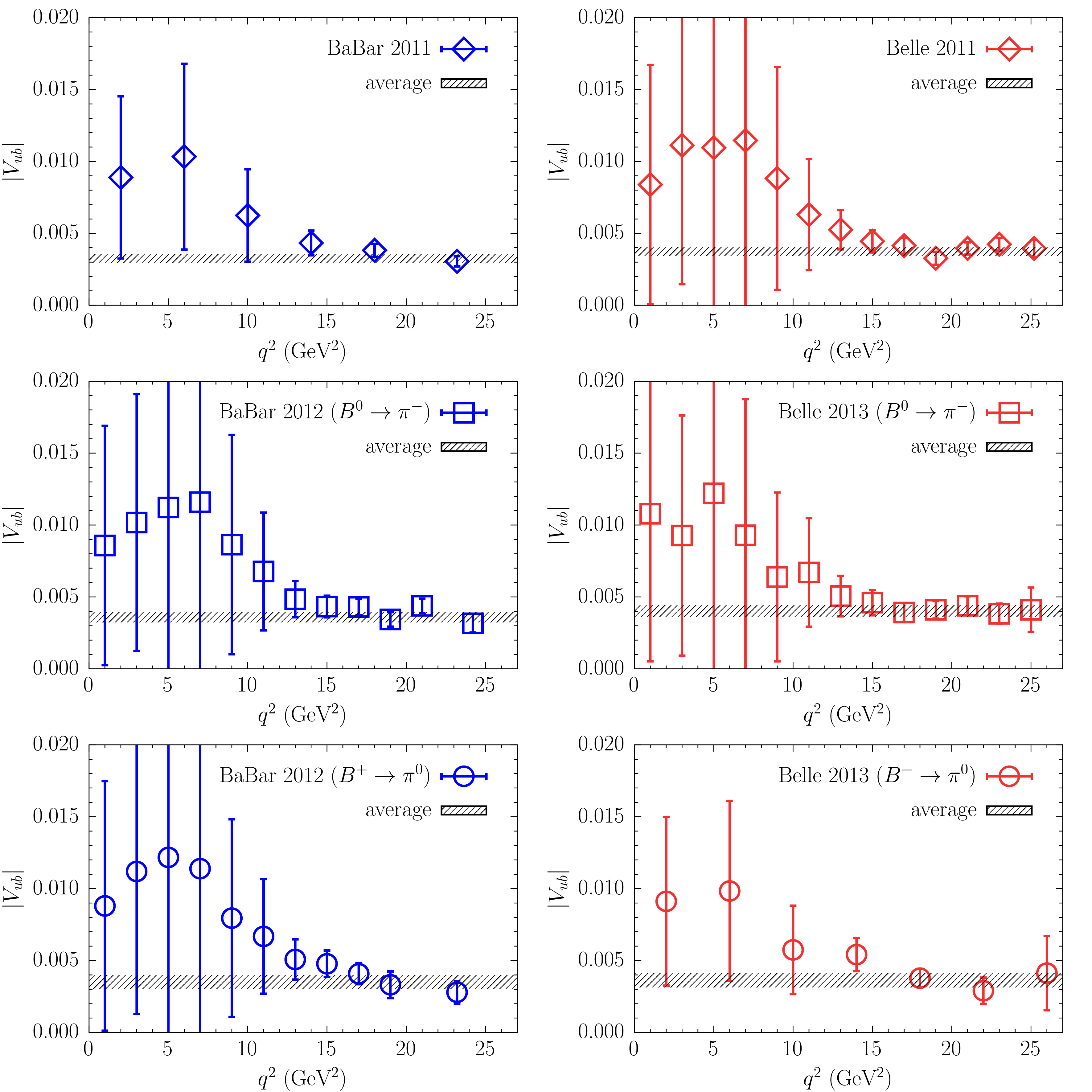}}
\caption{\it \small Bin-per-bin estimates of $\vert V_{ub} \vert$ obtained using Eq.\,(\ref{VubFINAL}) for each of the six experiments of Refs.\,\cite{delAmoSanchez:2010af, Ha:2010rf, Lees:2012vv, Sibidanov:2013rkk} specified in the insets of the panels as a function of $q^2$. The theoretical DM bands of the FFs correspond to the use of the combined LQCD data of Table\,\ref{tab:LQCDBPi} as inputs. The black dashed bands represent the correlated weighted averages\,(\ref{muVubfinal}) for each experiment, shown in Table\,\ref{tab:Vubmean}.}
\label{VubBpifig}
\end{figure}

In Fig.\,\ref{VubBpifig} we show our results for $\vert V_{ub} \vert$ for each of the semileptonic $B \to \pi$ experiments, together with the mean values\,(\ref{muVubfinal}), adopting the DM results for the FFs obtained using as inputs the combined LQCD data of Table\,\ref{tab:LQCDBPi}.
For each experiment the corresponding correlated mean values\,(\ref{muVubfinal}) are collected in Table\,\ref{tab:Vubmean}.
\begin{table}[htb!]
\renewcommand{\arraystretch}{1.1}
\begin{center}
\begin{tabular}{||c||c|c|c||}
\hline
experiment & ~ BaBar 2011 ~ & ~ BaBar 2012 ($B^0 \to \pi^-$) ~ & ~ BaBar 2012 ($B^+ \to \pi^0$) ~ \\
\hline
$\vert V_{ub} \vert \cdot 10^3$ & $3.25 \pm 0.33$ & $3.58 \pm 0.34$ & $3.51 \pm 0.46$ \\
\hline
\end{tabular}
\end{center}
\vspace*{-0.25cm}
\begin{center}
\begin{tabular}{||c||c|c|c||}
\hline
experiment & ~~ Belle 2011 ~~ & ~~ Belle 2013 ($B^0 \to \pi^-$) ~~ & ~~ Belle 2013 ($B^+ \to \pi^0$) ~~ \\
\hline
$\vert V_{ub} \vert \cdot 10^3$ & $3.74 \pm 0.32$ & $4.03 \pm 0.41$ & $3.64 \pm 0.51$ \\
\hline
\end{tabular}
\vspace*{0.25cm}
\caption{\it \small The correlated weighted averages\,(\ref{muVubfinal}) for each of the six experiments of Refs.\,\cite{delAmoSanchez:2010af, Ha:2010rf, Lees:2012vv, Sibidanov:2013rkk}. The theoretical DM bands of the FFs correspond to the use of the combined LQCD data of Table\,\ref{tab:LQCDBPi} as inputs.}
\label{tab:Vubmean}
\end{center}
\renewcommand{\arraystretch}{1.0}
\end{table}

As shown in Figs.\,\ref{FFMMBpi} and \ref{FFMMBpiCOMB}, the form factor $f_+(q^2)$, which is the only one contributing to the decay rate in the limit of massless leptons, may become numerically very small (in absolute value) below $q^2 \approx 10$ GeV$^2$.
Since the theoretical decay rate appears in the denominator of Eq.\,(\ref{VubFINAL}), the resulting values of $|V_{ub}|$ for the bins corresponding to $q^2 \lesssim 10$ GeV$^2$ exhibit a tendency to larger values. However, the uncertainties are quite large for those bins (due to the present uncertainties of the input lattice data for $f_+(q^2)$ and to the long extrapolation to low values of $q^2$) and, therefore, for each experiment the average\,(\ref{muVubfinal}) is dominated by the contributions of the large-$q^2$ bins.
Direct lattice calculations at smaller values of $q^2$ will allow in the future to clarify this point.

Our final results for $\vert V_{ub} \vert$, evaluated making use of the averaging procedure given by Eqs.\,(\ref{eq28Carr})-(\ref{eq28Carrb}), read
\bea
\label{VubLASTRBC}
\vert V_{ub} \vert^{B\pi}_{\rm{RBC/UKQCD}} \cdot 10^{3} & = & 3.52 \pm 0.49 ~ , ~ \nonumber \\[2mm]
\label{VubLASTMILC}
\vert V_{ub} \vert^{B\pi}_{\rm{FNAL/MILC}} \cdot 10^{3} & = & 3.76 \pm 0.41 ~ , ~ \\[2mm]
\label{VubLASTCOMB}
\vert V_{ub} \vert^{B\pi}_{\rm{combined}} \cdot 10^{3} & = & 3.62 \pm 0.47 ~ , ~ \nonumber 
\eea
which are consistent with the latest exclusive determination $\vert V_{ub} \vert_{excl} \cdot 10^3 = 3.70\,(16)$ from PDG\,\cite{ParticleDataGroup:2020ssz}. Our uncertainties are larger than the PDG one, because we do not mix the theoretical calculations of the FFs with the experimental data to constrain the shape of the FFs in order to avoid possible biases.
We are currently investigating strategies to improve the precision of the determination of $\vert V_{ub} \vert$ within our DM approach.

\subsubsection{$\vert V_{ub} \vert$ from $B_s \to K \ell \nu_\ell$ decays}

The LHCb Collaboration has observed for the first time the semileptonic $B_s \to K \ell \nu_\ell$ decays\,\cite{Aaij:2020nvo} and measured the ratio of the branching fractions of the $B_s^0 \to K^- \mu^+ \nu_{\mu}$ and the $B_s^0 \to D_s^- \mu^+ \nu_{\mu}$ processes,
\begin{equation*}
    \label{RBF}
    R_{BF} \equiv \frac{\mathcal{B}(B_s^0 \to K^- \mu^+ \nu_{\mu})}{\mathcal{B}(B_s^0 \to D_s^- \mu^+ \nu_{\mu})} ~ , ~
\end{equation*}
in two different $q^2$-bins, namely
\bea
    \label{RBFlow}
    R_{BF}(\mbox{low}) & = & (1.66 \pm 0.08 \pm 0.07 \pm 0.05) \cdot 10^{-3} 
                                              \qquad \mbox{for ~} q^2 \leq 7\,\mbox{GeV}^2 ~ , ~ \\[2mm]
    \label{RBFhigh}
    R_{BF}(\mbox{high}) & = & (3.25 \pm 0.21_{-0.17}^{+0.16} \pm 0.09) \cdot 10^{-3} 
                                                \qquad \mbox{for ~} q^2 \geq 7\,\mbox{GeV}^2 ~ , ~ 
\eea
where the first error is statistical, the second one is systematic and the third one is due to the uncertainty on the $D_s^- \to K^+ K^- \pi^-$ branching fraction.

In order to obtain an exclusive estimate of $\vert V_{ub} \vert$ we make use of the life time of the $B_s$-meson, $\tau_{B_s^0} = (1.516 \pm 0.006) \cdot 10^{-12}$ s\,\cite{ParticleDataGroup:2020ssz}, and of the experimental value of the branching ratio $\mathcal{B}(B_s^0 \to D_s^- \mu^+ \nu_{\mu})$ measured by the LHCb Collaboration\,\cite{Aaij:2020hsi}
\begin{equation*}
    \mathcal{B}(B_s^0 \to D_s^- \mu^+ \nu_{\mu})= (2.49 \pm 0.12 \pm 0.14 \pm 0.16) \cdot 10^{-2} ~ , ~
\end{equation*}
where the first error is statistical, the second one is systematic and the third one is due to limited knowledge of the normalization branching fractions.

Then, we use the FFs obtained with our DM method to compute the differential decay width $d\Gamma / dq^2 $ according to the formula (except $\vert V_{ub} \vert^2$) given in Eq.\,(\ref{finaldiff333}). 
Our results for $\vert V_{ub} \vert$ are collected in Table\,\ref{tab:BsK_corr}.
\begin{table}[htb!]
\renewcommand{\arraystretch}{1.2}	 
\begin{center}	
\begin{tabular}{||c||c|c|c||c||}
\hline
$q^2$-bin & RBC/UKQCD & FNAL/MILC& HPQCD & combined \\
\hline \hline
low  & $6.70 \pm 3.26$ & $6.43 \pm 2.03$ & $3.57 \pm 1.94$ & $5.31 \pm 3.02$ \\ \hline
high & $4.20 \pm 0.56$ & $4.10 \pm 0.38$ & $3.54 \pm 0.43$ & $3.94 \pm 0.59$ \\ \hline \hline
\end{tabular}
\end{center}
\renewcommand{\arraystretch}{1.0}
\caption{\it \small Values of $\vert V_{ub} \vert \cdot 10^3$ extracted from the $B_s \to K \ell \nu_\ell$ decays measured at LHCb in the low ($q^2 \leq 7$ GeV$^2$) and high ($q^2 \geq 7$ GeV$^2$) $q^2$-bins using the DM bands for the theoretical FFs.}
\label{tab:BsK_corr}
\end{table}

Assuming (conservatively) that the systematic errors of the two experimental results\,(\ref{RBFlow})-(\ref{RBFhigh}) are $100 \%$ correlated (which corresponds to a correlation coefficient equal to $0.486$ in the experimental, statistical plus systematic covariance matrix), the weighted averages of the two bins, carried out following Eqs.\,(\ref{eq28Carr})-(\ref{eq28Carrb}) for each set of FFs, read
\bea
\label{VubBsKRBC}
\vert V_{ub} \vert^{B_sK}_{\rm{RBC/UKQCD}} \cdot 10^{3} & = & 3.93 \pm 0.46 ~ , ~ \nonumber \\[2mm]
\label{VubBsKMILC}
\vert V_{ub} \vert^{B_sK}_{\rm{FNAL/MILC}} \cdot 10^{3} & = & 3.93 \pm 0.35 ~ , ~ \\[2mm]
\label{VubBsKHPQCD}
\vert V_{ub} \vert^{B_sK}_{\rm{HPQCD}} \cdot 10^{3} & = & 3.54 \pm 0.35 ~ , ~ \nonumber \\[2mm]
\label{VubBsKCOMB}
\vert V_{ub} \vert^{B_sK}_{\rm{combined}} \cdot 10^{3} & = & 3.77 \pm 0.48 ~ , ~ \nonumber 
\eea
which are consistent with our results\,(\ref{VubLASTMILC}), obtained from the analysis of the $B \to \pi \ell \nu_\ell$ decays, and with the latest exclusive determination $\vert V_{ub} \vert_{excl} \cdot 10^3 = 3.70\,(16)$ from PDG\,\cite{ParticleDataGroup:2020ssz}. We remind that the PDG uncertainty results from analyses in which theoretical calculations of the FFs and experimental data are mixed in order to constrain the shape of the FFs.

\section{Theoretical estimate of $R^{\tau/\mu}_{\pi(K)}$, $\bar{\mathcal{A}}_{FB}^{\ell,\pi(K)}$ and $\bar{\mathcal{A}}_{polar}^{\ell,\pi(K)}$}
\label{sec:LFU}

In this Section we give pure theoretical estimates of various quantities of phenomenological interest, which are independent of $\vert V_{ub} \vert$, namely the ratio of the $\tau/\mu$ decay rates $R^{\tau/\mu}_{\pi(K)}$, the normalized forward-backward asymmetry $\bar{\mathcal{A}}_{FB}^{\ell,\pi(K)}$ and the normalized lepton polarization asymmetry $\bar{\mathcal{A}}_{polar}^{\ell,\pi(K)}$.

The $\tau/\mu$ ratio $R^{\tau/\mu}_{\pi(K)}$ is defined as 
\begin{equation}
\label{Rpidef}
R^{\tau/\mu}_{\pi(K)} \equiv \frac{\Gamma(B_{(s)} \to \pi (K) \tau \nu_\tau)}{\Gamma(B_{(s)} \to \pi (K) \mu \nu_\mu)} ~ , ~
\end{equation}
where
\begin{eqnarray*}
\label{Rpi(K)}
\Gamma(B_{(s)} \to \pi (K) \ell \nu_\ell) & = & \frac{G_F^2 \vert V_{ub} \vert^2}{24\pi^3} \int_{m_\ell^2}^{(m_{B_{(s)}} - m_{\pi(K)})^2} dq^2 \left[ \vert \vec{p}_{\pi(K)} \vert^3 L_+\left( \frac{m_\ell^2}{q^2} \right) \vert f_+^{\pi(K)}(q^2) \vert^2 \right. \\[2mm]
& + & \left. m_{B_{(s)}}^2 \left( 1-r_{\pi(K)}^2 \right)^2 \vert \vec{p}_{\pi(K)} \vert L_0\left( \frac{m_\ell^2}{q^2} \right) \vert f_0^{\pi(K)}(q^2) \vert^2  \right] ~ 
\end{eqnarray*}
with $m_\ell$ being the lepton mass ($\ell = \tau, \mu$) and
\begin{equation*}
    L_+(x) = (1 - x)^2 \left( 1 + \frac{x}{2} \right) ~, ~ \qquad L_0(x) = (1 - x)^2 \frac{3x}{8} ~ . ~ 
\end{equation*}

The forward-backward asymmetry $\mathcal{A}_{FB}^{\ell,\pi(K)}$ is defined as
\begin{equation*}
\mathcal{A}_{FB}^{\ell,\pi(K)}(q^2) \equiv \int_0^1 \frac{d^2\Gamma}{dq^2 d\cos \theta_l}   d\cos \theta_l - \int_{-1}^0 \frac{d^2\Gamma}{dq^2 d\cos \theta_l}   d\cos \theta_l ~ , ~
\end{equation*}
where from Eq.\,(\ref{finaldiff333pre}) one has
\begin{eqnarray}
    \mathcal{A}_{FB}^{\ell,\pi(K)}(q^2) & = & \frac{G_F^2 \vert V_{ub} \vert^2}{32 \pi^3 m_{B_{(s)}}} \left(1-\frac{m_\ell^2}{q^2}\right)^2 \vert 
        \vec{p}_{\pi(K)} \vert^2 \frac{m_\ell^2}{q^2} (m_{B_{(s)}}^2-m_{\pi(K)}^2) \nonumber \\[2mm]
        & \cdot & \Re[f_+^{\pi(K)}(q^2) f_0^{*\pi(K)}(q^2)] ~ . ~ \nonumber
\end{eqnarray}
Then, the normalized forward-backward asymmetry $\bar{\mathcal{A}}_{FB}^{\ell,\pi(K)}$ is given by
\begin{equation}
\label{normAFB}
\bar{\mathcal{A}}_{FB}^{\ell,\pi(K)} \equiv \frac{\int dq^2\, \mathcal{A}_{FB}^{\ell,\pi(K)}(q^2)}{\int dq^2\, d\Gamma^{\pi(K)}/dq^2} ~ . ~
\end{equation}

We compute also the lepton polarization asymmetry $\mathcal{A}_{polar}^{\ell,\pi(K)}$ defined as 
\begin{equation*}
\mathcal{A}_{polar}^{\ell,\pi(K)}(q^2) \equiv \frac{d\Gamma_-^{\pi(K)}}{dq^2} - \frac{d\Gamma_+^{\pi(K)}}{dq^2} ~ , ~
\end{equation*}
where\,\cite{Meissner:2013pba}
\begin{eqnarray*}
\frac{d\Gamma_-^{\pi(K)}}{dq^2} &=& \frac{G_F^2 \vert V_{ub} \vert^2}{24 \pi^3} \left(1-\frac{m_\ell^2}{q^2}\right)^2 \vert \vec{p}_{\pi(K)} \vert^3 \vert f_+^{\pi(K)}(q^2) \vert^2 ~ , ~ \\[2mm]
\frac{d\Gamma_+^{\pi(K)}}{dq^2} &=& \frac{G_F^2 \vert V_{ub} \vert^2}{24 \pi^3} \left(1-\frac{m_\ell^2}{q^2}\right)^2 \frac{m_\ell^2}{q^2} \vert \vec{p}_{\pi(K)} \vert \\[2mm]
& \cdot & \left[ \frac{3}{8} \frac{\left(m_{B_{(s)}}^2-m_{\pi(K)}^2\right)^2}{m_{B_{(s)}}^2} \vert f_0^{\pi(K)}(q^2) \vert^2 + \frac{1}{2} \vert \vec{p}_{\pi(K)} \vert^2 \vert f_+^{\pi(K)}(q^2) \vert^2 \right] ~ . ~
\end{eqnarray*}
The normalized  lepton polarization asymmetry $\bar{\mathcal{A}}_{polar}^{\ell,\pi(K)}$ is given by
\begin{equation}
\label{normApol}
\bar{\mathcal{A}}_{polar}^{\ell,\pi(K)} \equiv \frac{\int dq^2\, \mathcal{A}_{polar}^{\ell,\pi(K)}(q^2)}{\int dq^2\, d\Gamma^{\pi(K)}/dq^2}.
\end{equation}

In Tables\,\ref{tab:phenoBpi} and \ref{tab:phenoBsK} we collect our theoretical estimates of the quantities\,(\ref{Rpidef}-\ref{normApol}) for each set of LQCD computations of the FFs in the case of the $B \to \pi$ and $B_s \to K$ decays, respectively. 
Within the uncertainties our results are consistent with recent estimates\,\cite{Rajeev:2018txm,Leljak:2021vte,Biswas:2021cyd} based on the BCL or BSZ parameterizations of the FFs.

\begin{table}[htb!]
\renewcommand{\arraystretch}{1.1}
\begin{center}
{\small
\begin{tabular}{|c||c||c||c|}
\hline
& ~ RBC/UKQCD ~ & ~ FNAL/MILC ~ & ~ combined ~\\
\hline
\hline
$R^{\tau/\mu}_{\pi}$ & 0.767(145) & 0.838(75) & 0.793(118) \\
$\bar{\mathcal{A}}_{FB}^{\mu,\pi}$ & 0.0043(39) & 0.0018(14) & 0.0034(31)\\
$\bar{\mathcal{A}}_{FB}^{\tau,\pi}$ & 0.219(25) & 0.221(19) & 0.220(24) \\
$\bar{\mathcal{A}}_{polar}^{\mu,\pi}$ & 0.985(11) & 0.991(4) & 0.988(9)\\
$\bar{\mathcal{A}}_{polar}^{\tau,\pi}$ & 0.294(87) & 0.309(82) & 0.301(86)\\
\hline
\end{tabular}
}
\caption{\it The theoretical values of the quantities\,(\ref{Rpidef})-(\ref{normApol}) in the case of the semileptonic $B \to \pi \ell \nu_\ell$ decays with $\ell = \mu, \tau$ adopting the RBC/UKQCD, the FNAL/MILC and the combined LQCD data of Table\,\ref{tab:LQCDBPi} as inputs for our DM method.\hspace*{\fill}}
\label{tab:phenoBpi}
\end{center}
\end{table}

\begin{table}[htb!]
\renewcommand{\arraystretch}{1.1}
\begin{center}
{\small
\begin{tabular}{|c||c||c||c||c|}
\hline
& ~ RBC/UKQCD ~ & ~ FNAL/MILC ~ & ~ HPQCD ~ & ~ combined ~\\
\hline
\hline
$R^{\tau/\mu}_{K}$ & 0.845(122) & 0.816(64) & 0.680(134) & 0.755(138)\\
$\bar{\mathcal{A}}_{FB}^{\mu,K}$ & 0.0032(18) & 0.0024(12) & 0.0059(29)& 0.0046(28)\\
$\bar{\mathcal{A}}_{FB}^{\tau,K}$ & 0.257(14) & 0.246(14) & 0.278(19) & 0.262(23)\\
$\bar{\mathcal{A}}_{polar}^{\mu,K}$ & 0.990(5) & 0.992(4) & 0.982(8) & 0.986(7)\\
$\bar{\mathcal{A}}_{polar}^{\tau,K}$ & 0.172(54) & 0.254(64) & 0.112(79) & 0.172(91)\\
\hline
\end{tabular}
}
\caption{\it The same as in Table\,\ref{tab:phenoBpi}, but in the case of the semileptonic $B_{s} \to K$ decays adopting the RBC/UKQCD, the FNAL/MILC, the HPQCD and the combined LQCD data of Table\,\ref{tab:LQCDBsK} as inputs for our DM method.\hspace*{\fill}}
\label{tab:phenoBsK}
\end{center}
\end{table}

As for the experimental side, only one measurement of $R^{\tau/\mu}_{\pi}$ by Belle is presently available, namely\,\cite{Hamer:2015jsa}
\begin{equation}
\label{RBelle}
R_{\pi}^{\tau/\mu}\vert_{exp} = 1.05 \pm 0.51 ~ , ~
\end{equation}
which still has a large uncertainty compared to our theoretical ones.
Note that the uncertainty on the above ratio expected by Belle II at 50 ab$^{-1}$ of luminosity\,\cite{Kou:2018nap} is $\delta R_{\pi}^{\tau/\mu} \simeq 0.09$, which will be comparable to our present theoretical uncertainties. 

Since it is likely that experimental measurements of $R^{\tau/\mu}_{\pi(K)}$, $\bar{\mathcal{A}}_{FB}^{\ell,\pi(K)}$ and $\bar{\mathcal{A}}_{polar}^{\ell,\pi(K)}$ will be carried out in limited regions of the phase space, we provide in Appendix\,\ref{sec:appB} our theoretical estimated of these quantities in three selected $q^2$ regions.

\section{Conclusions}
\label{sec:conclusions}

In this work we have analysed the available lattice and experimental data concerning the semileptonic $B \to \pi$ and $B_s \to K$ decays. We have obtained new exclusive estimates of the CKM matrix element $\vert V_{ub} \vert$ in a rigorous model-independent way in order to shed a new light onto the tension between its inclusive and exclusive determinations. This has been achieved by evaluating the semileptonic FFs according to the non-perturbative and model-independent DM method proposed in Ref.\,\cite{DiCarlo:2021dzg} and by computing for the first time non-perturbatively the susceptibilities relevant for the unitarity bounds in the $b \to u$ transition.

Our results for $\vert V_{ub} \vert$ can be summarized as  
\begin{itemize}
\item from the semileptonic $B \to \pi$ decays
\begin{equation*}
    \vert V_{ub} \vert \cdot 10^3 = 3.62 \pm 0.47 ~ , ~
\end{equation*}
\item from the semileptonic $B_s \to K$ processes
\begin{equation*}
    \vert V_{ub} \vert \cdot 10^3 = 3.77 \pm 0.48 ~ . ~
\end{equation*}
\end{itemize}
They are compatible with each other and also consistent within $1 \sigma$ level with the latest exclusive and inclusive determinations of $\vert V_{ub} \vert$, $\vert V_{ub} \vert_{excl} \cdot 10^3 = 3.70\,(16)$ and $\vert V_{ub} \vert_{incl} \cdot 10^3 = 4.13\,(26)$, taken from PDG\,\cite{ParticleDataGroup:2020ssz}.

Then, by averaging the above results corresponding to the $B \to \pi$ and $B_s \to K$ channels our final estimate of $\vert V_{ub} \vert$ reads
\begin{equation*}
    \vert V_{ub} \vert \cdot 10^3 = 3.69 \pm 0.34 ~ . ~
\end{equation*}

We have also investigated the issue of LFU by computing the $\tau/\mu$ ratio $R^{\tau/\mu}_{\pi(K)}$ given in Eq.\,(\ref{Rpidef}). 
Our results read
\begin{equation*}
R_\pi^{\tau/\mu} = 0.793 \pm 0.118 ~ , ~ \qquad \quad R_K^{\tau/\mu} = 0.755 \pm 0.138 ~ . ~
\end{equation*}
Our estimate of $R^{\tau/\mu}_{\pi}$ is compatible with the Belle measurement\,(\ref{RBelle}) within the present large experimental uncertainty\,\cite{Hamer:2015jsa}.

We have computed also the normalized forward-backward asymmetry $\bar{\mathcal{A}}_{FB}^{\ell,\pi(K)}$ and the normalized lepton polarization asymmetry $\bar{\mathcal{A}}_{polar}^{\ell,\pi(K)}$ given in Eqs.~\.(\ref{normAFB})-(\ref{normApol}). For $\ell =\mu, \tau$ we have got
\begin{eqnarray*}
\bar{\mathcal{A}}_{FB}^{\mu,\pi} & = & 0.0034 \pm 0.0031 ~ , ~ \qquad ~ \bar{\mathcal{A}}_{FB}^{\mu,K} = 0.0046 \pm 0.0028 ~ , ~ \\[2mm]
\bar{\mathcal{A}}_{FB}^{\tau,\pi} & = & 0.220 \pm 0.024 ~ , ~ \qquad \quad ~ \bar{\mathcal{A}}_{FB}^{\tau,K} = 0.262 \pm 0.023 ~ , ~ \\[2mm]
\bar{\mathcal{A}}_{polar}^{\mu,\pi} & = & 0.988 \pm 0.009 ~ , ~ \qquad \quad \bar{\mathcal{A}}_{polar}^{\mu,K} = 0.986 \pm 0.007~ , ~ \\[2mm]
\bar{\mathcal{A}}_{polar}^{\tau,\pi} & = & 0.301 \pm 0.086 ~ , ~ \qquad \quad \bar{\mathcal{A}}_{polar}^{\tau,K} = 0.172 \pm 0.091 ~ . ~ 
\end{eqnarray*}

We stress that other exclusive estimates of $\vert V_{ub} \vert$ can be obtained by investigating the semileptonic $B \to \rho$ and $B \to \omega$ decays. In these cases the analysis is more involved due to the vector nature of the final $\rho$ and $\omega$ mesons. Nevertheless, once LQCD computations of the FFs of interest for these processes will be available, our DM method can be applied, as already demonstrated in Ref.\,\cite{Martinelli:2021onb} for the case of the semileptonic $B \to D^* \ell \nu_\ell$ decays.

\begin{acknowledgments}
We warmly thanks Fabio Ferrari for fruitful discussions concerning the LHCb experiment of Ref.\,\cite{Aaij:2020nvo}.
We acknowledge PRACE for awarding us access to Marconi at CINECA (Italy) under the grant of the PRACE project PRA067.  We also acknowledge use of CPU time provided by CINECA under the specific initiative INFN-LQCD123.
G.M.~and S.S.~thank the Italian Ministry of Research (MIUR) for partial support under the contract PRIN 2015. 
S.S.~is supported by MIUR also under the grant PRIN 20172LNEEZ.
\end{acknowledgments}

\appendix

\section{Lattice computation of the susceptibilities for the $b \to u$ transition}
\label{sec:appA}

In this Appendix we describe the non-perturbative computation of the unitarity bounds for the $b \to u$ (and as a by-product for the $c \to d$) transition based on suitable lattice two-point correlation functions. We strictly follow the procedure adopted already in Ref.\,\cite{Martinelli:2021frl} for our determination of the susceptibilities in the case of the $b \to c$ transition.
The above procedure includes the ETMC ratio method\,\cite{ETM:2009sed,ETM:2016nbo} for reaching the physical $b$-quark point.

\subsection{Basic definitions}

Let us first recall the basic definitions of the susceptibilities $\chi(Q^2)$ we are interested in this work, namely
\bea
     \label{eq:chiVL}
       \chi_{0^+}(Q^2) & \equiv & \frac{\partial}{\partial Q^2} \left[ Q^2 \Pi_{0^+}(Q^2) \right] = \int_0^\infty dt ~ t^2 j_0(Qt) ~ C_{0^+}(t) ~ , \\[2mm]
       \label{eq:chiVT}
        \chi_{1^-}(Q^2) & \equiv & - \frac{1}{2} \frac{\partial^2}{\partial^2 Q^2} \left[ Q^2 \Pi_{1^-}(Q^2) \right] = 
                                                   \frac{1}{4} \int_0^\infty dt ~ t^4 \frac{j_1(Qt)}{Qt} ~ C_{1^-}(t) ~ , \\[2mm]
       \label{eq:chiAL}
       \chi_{0^-}(Q^2) & \equiv & \frac{\partial}{\partial Q^2} \left[ Q^2 \Pi_{0^-}(Q^2) \right] = \int_0^\infty dt ~ t^2 j_0(Qt) ~ C_{0^-}(t)~ , \\[2mm]
       \label{eq:chiAT}
       \chi_{1^+}(Q^2) & \equiv & - \frac{1}{2} \frac{\partial^2}{\partial^2 Q^2} \left[ Q^2 \Pi_{1^+}(Q^2) \right] = 
                                                   \frac{1}{4} \int_0^\infty dt ~ t^4 \frac{j_1(Qt)}{Qt} ~ C_{1^+}(t) ~ ,
\eea
where the quantities $\Pi_j(Q^2)$ with $j = \{0^+, 1^-, 0^-, 1^+ \}$ are the vacuum polarization functions corresponding to definite spin-parity channels (see for more details Refs.\,\cite{DiCarlo:2021dzg,Martinelli:2021frl}), $Q$  is an Euclidean 4-momentum, $j_0(x) = \mbox{sin}(x) / x$ and $j_1(x) = [\mbox{sin}(x) / x -  \mbox{cos}(x)] / x$ are spherical Bessel functions and the Euclidean correlators $C_j(t)$ are given by
\bea
    \label{eq:CVL}
    C_{0^+}(t) & = & \int d^3x  \langle 0 | T\left[ \bar{b}(x) \gamma_0 u(x) ~ \bar{u}(0) \gamma_0 b(0) \right] | 0 \rangle ~ , \\[2mm]
    \label{eq:CVT}
    C_{1^-}(t) & = & \frac{1}{3} \sum_{j=1}^3 \int d^3x  \langle 0 | T\left[ \bar{b}(x) \gamma_j u(x) ~ \bar{u}(0) \gamma_j b(0) \right] | 0 \rangle ~ , \\[2mm]
   \label{eq:CAL}
    C_{0^-}(t) & = & \int d^3x  \langle 0 | T\left[ \bar{b}(x) \gamma_0 \gamma_5 u(x) ~ \bar{u}(0) \gamma_0 \gamma_5 b(0) \right] | 0 \rangle ~ , \\[2mm]
    \label{eq:CAT}
    C_{1^+}(t) & = & \frac{1}{3} \sum_{j=1}^3 \int d^3x \langle 0 | T\left[ \bar{b}(x) \gamma_j \gamma_5 u(x) ~ 
                               \bar{u}(0) \gamma_j \gamma_5 b(0) \right] | 0 \rangle ~ 
\eea
with $u(x)$ representing the light-quark $u$ field.
Note that the longitudinal (first) derivatives (\ref{eq:chiVL}) and (\ref{eq:chiAL}) are dimensionless, while the transverse (second) ones (\ref{eq:chiVT}) and (\ref{eq:chiAT}) have the dimension of $[E]^{-2}$, where $E$ is an energy.

As shown in Ref.~\cite{DiCarlo:2021dzg}, Eqs.~(\ref{eq:chiVL})-(\ref{eq:chiAT}) are obtained in the Euclidean region $Q^2 \geq 0$, but they can be easily generalized also to the case $Q^2 < 0$.
In the Euclidean region $Q^2 \geq 0$ a good convergence of the perturbative calculation of the above derivatives is expected to occur far from the kinematical regions where resonances can contribute.
In the case of the $b \to u$ weak transition this means down to $Q^2 = 0$~\cite{Lellouch:1995yv} and, indeed, this is the value of $Q^2$ that has been generally employed in the evaluation of the dispersive bounds for heavy-to-light \cite{Lellouch:1995yv,Bourrely:2008za} and also for heavy-to-heavy \cite{Boyd:1997kz,Caprini:1997mu,Bigi:2016mdz,Bigi:2017njr,Bigi:2017jbd} semileptonic form factors. 
By contrast, with a non-perturbative determination of the two-point correlation functions we can use the most convenient value of $Q^2$ at disposal, namely the value which will allow the most stringent bounds on the semileptonic form factors. 
In this work we will limit ourselves to the usual choice $Q^2 = 0$, which will allow the comparison with perturbative results, and we will leave the investigation of the choice $Q^2 \neq 0$ to a future, separate work.

At $Q^2 = 0$ the derivatives of the longitudinal and transverse polarization functions correspond to the second and fourth moments of the longitudinal and transverse Euclidean correlators, respectively.  However, the evaluation of the longitudinal susceptibilities~(\ref{eq:chiVL}) and~(\ref{eq:chiAL}) is plagued by contact terms related to the product of two current operators, since only the second moment of the longitudinal correlators are involved. In Ref.~\cite{DiCarlo:2021dzg} it has been shown that the use of the Ward Identities (WIs), which should be satisfied by the vector and axial-vector quark currents, allow to avoid the effects of the contact terms. Indeed, thanks to the WIs all the susceptibilities at $Q^2 = 0$ can be written as the fourth moment of suitable Euclidean correlators, namely
 \bea
       \label{eq:chiVL_WI0}
       \chi_{0^+}(Q^2 = 0) & = &\frac{1}{12} (m_b - m_u)^2 \int_0^\infty dt ~ t^4 ~ C_S(t) ~ , ~ \\[2mm]
       \label{eq:chiVT0}
       \chi_{1^-}(Q^2=0) & = & \frac{1}{12} \int_0^\infty dt ~ t^4 ~ C_{1^-}(t) ~ , ~ \\[2mm]
       \label{eq:chiAL_WI0}
       \chi_{0^-}(Q^2 = 0) & = & \frac{1}{12} (m_b + m_u)^2 \int_0^\infty dt ~ t^4 ~ C_P(t) ~ , ~ \\[2mm]
       \label{eq:chiAT0}
       \chi_{1^+}(Q^2=0) & = & \frac{1}{12} \int_0^\infty dt ~ t^4 ~ C_{1^+}(t) ~ , ~ 
\eea
where $C_S(t)$ and $C_P(t)$ are the scalar and pseudoscalar Euclidean correlators
\bea
   \label{eq:CS}
    C_S(t) & = & \int d^3x  \langle 0 | T\left[ \bar{b}(x) u(x) ~ \bar{u}(0) b(0) \right] | 0 \rangle ~ , \\[2mm]
    \label{eq:CP}
    C_P(t) & = & \int d^3x  \langle 0 | T\left[ \bar{b}(x) \gamma_5 u(x) ~ \bar{u}(0) \gamma_5b(0) \right] | 0 \rangle ~ .
\eea

\subsection{Lattice correlators}

The gauge ensembles used in this work have been generated by ETMC with $N_f = 2 + 1 + 1$ dynamical quarks, which include in the sea, besides two light mass-degenerate quarks ($m_u = m_d = m_{ud}$), also the strange and the charm quarks with masses close to their physical values~\cite{Baron:2010bv,ETM:2010cqp}. 
They are the same adopted in the case of the study of the $b \to c$ transition and details can be found in the Appendix A of Ref.\,\cite{Martinelli:2021frl}.

Here, we mention that the simulations have been carried out at three values of the lattice spacing ($a \simeq 0.062, 0.082, 0.089$ fm) and with pion masses in the range $\simeq 210 - 450$ MeV.
The physical up/down, strange and charm quark masses have been determined in Ref.\,\cite{Carrasco:2014cwa} obtaining $m_{ud}^{phys} = 3.72 \pm 0.17$ MeV, $m_s^{phys} = 99.6 \pm 4.3$ MeV and $m_c^{phys} = 1.176 \pm 0.039$ GeV in the $\overline{\mathrm{MS}}$ scheme at a renormalization scale of 2 GeV.
In Ref.~\cite{ETM:2016nbo} the physical b-quark mass has been determined adopting the ETMC ratio method~\cite{ETM:2009sed}, obtaining $m_b^{phys}(m_b^{phys}) = 4.26 \pm 0.10$ GeV which corresponds to $m_b^{phys} = 5.198 \pm 0.122$ GeV in the $\overline{\mathrm{MS}}(2~\mbox{GeV})$ scheme.

Using the ETMC gauge ensembles the computation of the susceptibilities\,(\ref{eq:chiVL_WI0})-(\ref{eq:chiAT0}) require the evaluation of the following two-point correlation functions 
\bea
   \label{eq:CS12}
   C_S(t) & = & \widetilde{Z}_S^2 ~ \int d^3x \langle 0 | T\left[ \bar{q}_1(x) q_2(x) ~  \bar{q}_2(0) q_1(0) \right] | 0 \rangle ~ , \\[2mm]
    \label{eq:CVT12}
    C_{1^-}(t) & = & \widetilde{Z}_V^2 ~ \frac{1}{3} \sum_{j=1}^3 \int d^3x  \langle 0 | T\left[ \bar{q}_1(x) \gamma_j q_2(x) ~ \bar{q}_2(0) \gamma_j q_1(0) \right] | 0 \rangle ~ , \\[2mm]
    \label{eq:CP12}
    C_P(t) & = & \widetilde{Z}_P^2 ~ \int d^3x  \langle 0 | T\left[ \bar{q}_1(x) \gamma_5 q_2(x) ~ \bar{q}_2(0) \gamma_5 q_1(0) \right] | 0 \rangle ~ , ~ \\[2mm]
    \label{eq:CAT12}
    C_{1^+}(t) & = & \widetilde{Z}_A^2 ~ \frac{1}{3} \sum_{j=1}^3 \int d^3x \langle 0 | T\left[ \bar{q}_1(x) \gamma_j \gamma_5 q_2(x) ~ \bar{q}_2(0) \gamma_j \gamma_5 q_1(0) \right] | 0 \rangle ~ , ~
\eea
where $q_1$ and $q_2$ are the two valence quarks involved in the weak transition with bare masses $a \mu_1$ and $a \mu_2$ given in Table~VII of Ref.\,\cite{Martinelli:2021frl}, while the multiplicative factor $\widetilde{Z}_O$ ($O = \{ S, V, P, A \}$) is an appropriate renormalization constant (RC), which will be specified in a while.
Indeed, we consider either opposite or equal values for the Wilson parameters $r_1$ and $r_2$ of the two valence quarks, namely either the case $r_1 = - r_2$ or the case $r_1 = r_2$. 
Since our twisted-mass setup is at its maximal twist, in the case $r_1 = - r_2$ we have $\widetilde{Z}_O = \{ Z_P, Z_A, Z_S, Z_V \}$, while in the case $r_1 = r_2$ we have $\widetilde{Z}_O = \{ Z_S, Z_V, Z_P, Z_A \}$, where the RCs of the various bilinear operators have been determined in Ref.\,\cite{Carrasco:2014cwa} (using the RI$^\prime$-MOM scheme for $Z_P$, $Z_A$ and $Z_S$, and the vector WI for $Z_V$).

Once renormalized, the correlation functions (\ref{eq:CS12}-\ref{eq:CAT12}) and, consequently, also the susceptibilities\,(\ref{eq:chiVL_WI0})-(\ref{eq:chiAT0}) corresponding to either opposite or equal values of the Wilson parameters $r_1$ and $r_2$ differ only by effects of order ${\cal{O}}(a^2)$.
For the sake of simplicity, in what follows we will denote by $\chi_j$ with $j = \{ 0^+, 1^-, 0^-, 1^+ \}$ the susceptibilities evaluated at $Q^2 = 0$.

For each ETMC gauge ensemble the susceptibilities $\chi_j$ have been evaluated for many combinations of the two valence quark masses $m_1 = a \mu_1 / (Z_P a)$ and $m_2 = a \mu_2 / (Z_P a)$, namely for 14 values in the light, strange, charm and heavier-than-charm sectors in the case of $a \mu_1$, while the values of $a \mu_2$ have been chosen in the light, strange and charm regions for a total of 7 values (see Table~VII of Ref.\,\cite{Martinelli:2021frl}).

\subsection{The $h \to u$ transition}

In this work we limit ourselves to the quark mass combinations $a \mu_1 = a \mu_h \geq a \mu_c$ and $a \mu_2 = a \mu_{ud}$, which in our isosymmetric QCD setup correspond to $h \to u$ transitions.

The values of the simulated susceptibilities $\chi_{0^\pm(1^\pm)}$ are smoothly interpolated at a series of values of the heavy-quark mass $m_h = a \mu_h / (Z_P a)$, dictated by the analysis of Ref.~\cite{ETM:2016nbo}, namely
\be
     \label{eq:mh_n}
     m_h(n) = \lambda^{n-1} ~ m_c^{phys} \qquad \mbox{for}~n = 1, 2, ...
\ee
with $\lambda \equiv [m_b^{phys} / m_c^{phys}]^{1/10} = [5.198 / 1.176]^{1/10} \simeq 1.1602$, and starting from $m_h(1) = m_c^{phys}$.
The value of $\lambda$, which is the same as the one adopted in Ref.~\cite{ETM:2016nbo}, is such that $m_h(n = 11) = m_b^{phys}$.
Correspondingly, the uncertainty $\delta m_h(n)$ is given by
\be
     \delta m_h(n) = \epsilon^{n-1} \delta m_c^{phys} \qquad \mbox{for}~n = 1, 2, ...
\ee
with $\epsilon \equiv [\delta m_b^{phys} / \delta m_c^{phys}]^{1/10} = [0.122 / 0.039]^{1/10} \simeq 1.1208$. 
Given the number of simulated values of $m_h > m_c^{phys}$, the susceptibilities $\chi_j$ are interpolated at the series of values (\ref{eq:mh_n}) up to $n = 9$, which corresponds to $m_h(9) \simeq 3.9~\mbox{GeV} \simeq 0.75~m_b^{phys}$.

Following Ref.\,\cite{Martinelli:2021frl} the analysis is split into the eight branches originally introduced in Ref.\,\cite{Carrasco:2014cwa}. They differ in: ~ i) the continuum extrapolation adopting for the matching of the lattice scale either the Sommer parameter $r_0$ or the mass of a fictitious P-meson made up of two valence strange(charm)-like quarks; ~ ii) the chiral extrapolation performed with fitting functions chosen to be either a polynomial expansion or a Chiral Perturbation Theory Ansatz in the light-quark mass; and ~ iii) the choice between the methods M1 and M2, which differ by ${\cal{O}}(a^2)$ effects, used to determine the RCs in the RI$^\prime$-MOM scheme. For each branch the central values and the errors of the input parameters are evaluated using a bootstrap sample with ${\cal{O}}(100)$ events (see Tables VIII and IX of Ref.\,\cite{Martinelli:2021frl}). Unless otherwise stated, the results that will be shown in the Figures of this Appendix correspond to the average of the first four branches of the bootstrap analysis.

\begin{figure}[htb!]
\begin{center}
\includegraphics[scale=0.75]{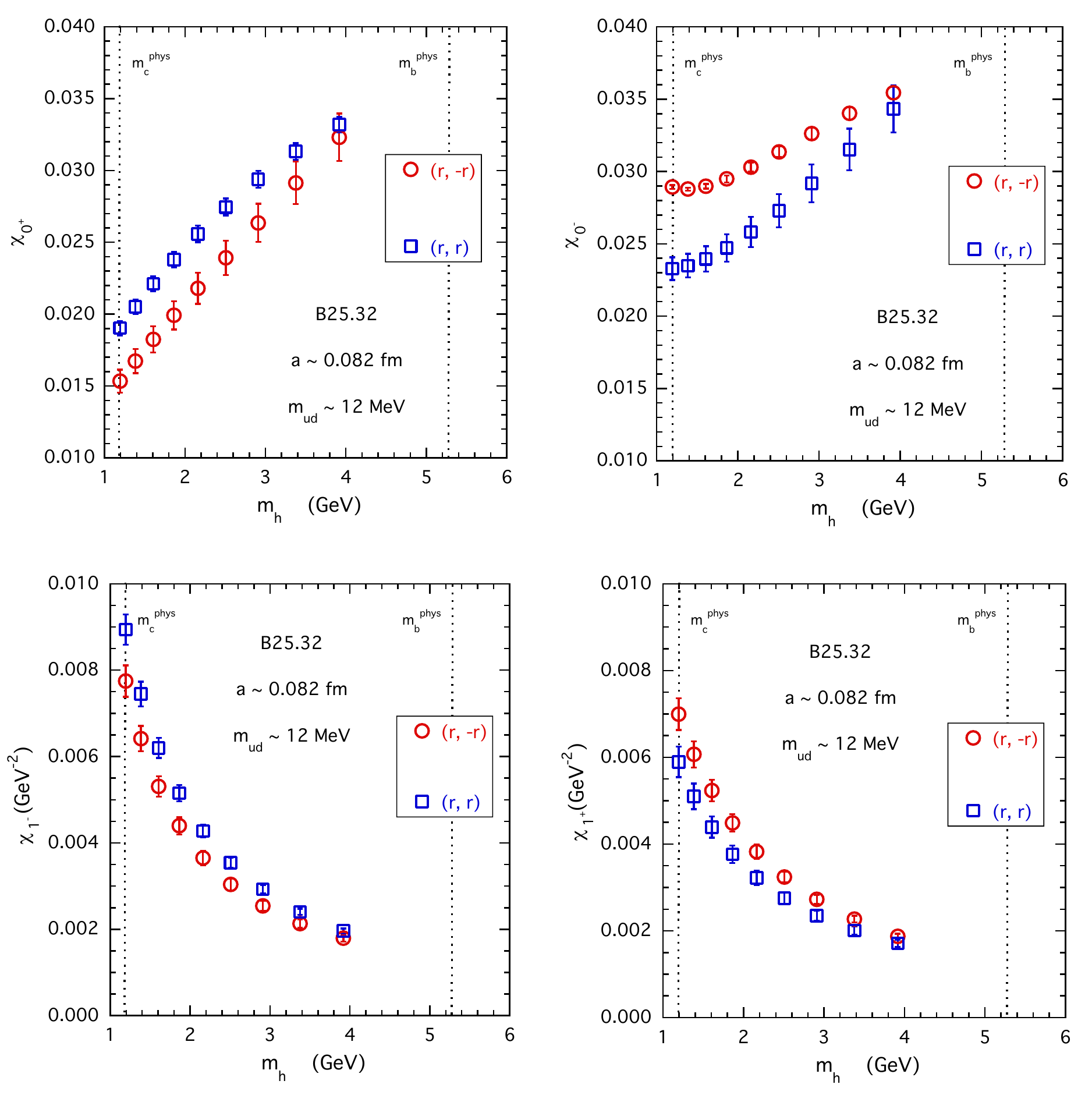}
\end{center}
\vspace{-1cm}
\caption{\it \small The susceptibilities $\chi_{0^+}$ (left top panel), $\chi_{0^-}$ (right top panel), $\chi_{1^-}$ (left bottom panel) and $\chi_{1^+}$ (right bottom panel) corresponding to the gauge ensemble B25.32 for the $h \to u$ transitions as a function of the heavy-quark mass $m_h$ given by Eq.~(\ref{eq:mh_n}) up to $n = 9$, i.e.~up to $m_h(9) \simeq 3.9$ GeV. The red circles correspond to the choice of opposite values $(r, -r)$ of the two valence-quark Wilson parameters, while the blue squares refer to the case of equal values $(r, r)$.}
\label{fig:VLALVTAT}
\end{figure}

Using the gauge ensemble B25.32 as a representative case, our results for the vector and axial-vector, longitudinal and transverse susceptibilities $\chi_{0^\pm(1^\pm)}$ are shown in Fig.~\ref{fig:VLALVTAT} at either opposite or equal values of the valence-quark Wilson parameters, which will be denoted hereafter by $(r, -r)$ and $(r, r)$.
It can be seen that the difference between the susceptibilities corresponding to the two different $r$-combinations does not exceed $\sim 25 \%$.

\subsection{The ETMC ratios}

According to the ETMC ratio method of Ref.\,\cite{ETM:2009sed} we now consider the ratios of the lattice susceptibilities $\chi_j = \chi_j[m_h(n); a^2, m_{ud}]$ interpolated for each ETMC gauge ensemble at subsequent values of the heavy-quark mass $m_h(n)$ given by Eq.~(\ref{eq:mh_n}), namely 
\be
    \label{eq:ETMC_ratios}
    R_j(n; a^2, m_{ud}) \equiv \frac{\chi_j[m_h(n); a^2, m_{ud}]}{\chi_j[m_h(n-1); a^2, m_{ud}]} ~ \frac{\rho_j[m_h(n)]}{\rho_j[m_h(n-1)]} ~ , ~
\ee
where $n = 2, 3, ... \, 9$ for $j = \{0^+, 1^-, 0^-, 1^+ \}$.

In Eq.~(\ref{eq:ETMC_ratios}) the factor $\rho_j(m_h)$ is introduced to guarantee that in the heavy-quark limit $m_h \to \infty$ (i.e., $n \to \infty$) one has $R_j \to 1$.
Using the perturbative results of Ref.~\cite{Boyd:1997kz} the above condition is satisfied by
\bea
    \label{eq:rhoL}
    \rho_{0^+} (m_h) = \rho_{0^-}(m_h) & = & 1 ~ , ~ \\[2mm]
    \label{eq:rhoT}
    \rho_{1^-}(m_h) = \rho_{1^+}(m_h) & = & (m_h^{pole})^2 ~ , ~ 
\eea
where $m_h^{pole}$ is the pole heavy-quark mass.
The latter one can be constructed from the $\overline{MS}(2~\rm{GeV})$ mass $m_h$ in two steps.
First, the PT scale is evolved from $\mu = 2$ GeV to the value $\mu = m_h$ using $\rm N^3LO$ perturbation theory \cite{Chetyrkin:1999pq} with four quark flavors ($n_\ell = 4$) and $\Lambda_{QCD}^{Nf = 4} = 294\,(12)$ MeV \cite{Aoki:2019cca}, obtaining in this way $m_h(m_h)$.
Then, at order ${\cal{O}}(\alpha_s^2)$ the pole quark mass $m_h^{pole}$ is given in terms of the $\overline{MS}$ mass $m_h(m_h)$ by 
\bea
      m_h^{pole} & = & m_h(m_h) \left\{ 1 + \frac{4}{3} \frac{\alpha_s(m_h)}{\pi} + \left( \frac{\alpha_s(m_h)}{\pi} \right)^2 \right. \nonumber \\
                         & \cdot & \left. \left[ \frac{\beta_0}{24} (8 \pi^2 + 71) + \frac{35}{24} + \frac{\pi^2}{9} \mbox{ln}(2) - 
                                         \frac{7 \pi^2}{12} -\frac{\zeta_3}{6} \right] + {\cal{O}}(\alpha_s^3) \right\} ~ , 
       \label{eq:mh_pole}
\eea
where $\beta_0 = (33 - 2 n_\ell) / 12$ and $\zeta_3 \simeq 1.20206$.
The relation between $m_h^{pole}$ and $m_h(m_h)$ is known up to order ${\cal{O}}(\alpha_s^3)$ (see Refs.~\cite{Chetyrkin:1999qi,Melnikov:2000qh}), but the ratios of the transverse factors (\ref{eq:rhoT}) appearing in Eq.~(\ref{eq:ETMC_ratios}) turn out to be almost insensitive to such high-order corrections.

\begin{figure}[htb!]
\begin{center}
\includegraphics[scale=0.75]{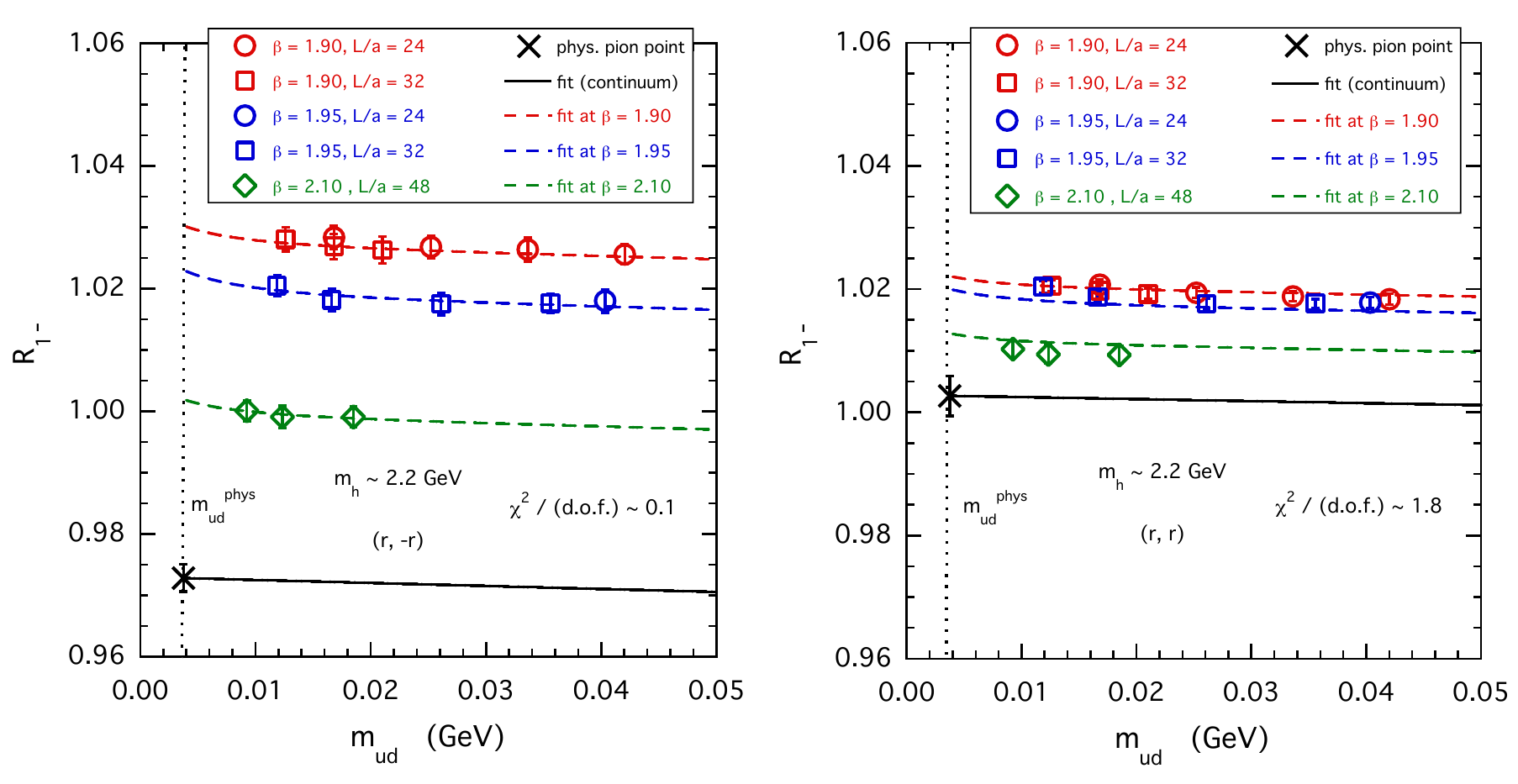}
\end{center}
\vspace{-1cm}
\caption{\it \small Light-quark mass dependence of the ratio of the susceptibilities corresponding to Eq.~(\ref{eq:ETMC_ratios}) for $j = 1^-$ and $n = 5$ for the two combinations $(r, -r)$ (left panel) and $(r,r)$ (right panel) of the Wilson $r$-parameters. The solid lines represent the results of the fitting function (\ref{eq:ratiofit_a2}) evaluated in the continuum and infinite volume limits, while the dashed ones correspond to the fitting function evaluated at each value of $\beta$ and for the largest value of $L /a$. The crosses represent the value of the ratio extrapolated at the physical pion point ($m_{ud} = m_{ud}^{phys}$) and in the continuum and infinite volume limits.}
\label{fig:RVT_a2}
\end{figure}

Thanks to the large correlation between the numerator and the denominator in Eq.~(\ref{eq:ETMC_ratios}) the statistical uncertainty of the ETMC ratios $R_j(n; a^2, m_{ud})$ is much smaller than those of the separate susceptibilities and it may reach the permille level, as it is shown in Fig.~\ref{fig:RVT_a2} in the case $j = 1^-$ and $n=5$ as an illustrative example.

The light-quark mass dependence of the ratios\,(\ref{eq:ETMC_ratios}) turns out to be very mild and, therefore, for each value of the heavy-quark mass $m_h(n)$ we fit the lattice data by adopting a simple linear Ansatz both in the light-quark mass $m_{ud}$  and in the values of the squared lattice spacing $a^2$ (since in our lattice setup the susceptibilities are ${\cal{O}}(a)$-improved) with an additional phenomenological term aimed at describing finite volume effects (FVEs), namely
\bea
    \label{eq:ratiofit_a2}
    R_j(n; a^2, m_{ud}) & = & R_j(n) \left[ 1 + A_1 \left( m_{ud} - m_{ud}^{phys} \right) + D_1 ~ \frac{a^2}{r_0^2} \right] ~ \\[2mm]
                                    & \cdot & \left( 1 + F_1 \frac{\overline{M}^2}{(4\pi f)^2} \frac{e^{-\overline{M} L}}{(\overline{M}L)^p} \right) ~ , ~ \nonumber
\eea
where  $r_0$ is the Sommer parameter, $\overline{M}^2 \equiv 2 B m_{ud}$ and $R_j(n)$ stands for $R_j(n; 0, m_{ud}^{phys})$. The values of $r_0$ and of the low-energy constants $B$ and $f$ have been determined for our lattice setup in Ref.~\cite{Carrasco:2014cwa}\footnote{As for the FVEs, they appear to be generally small. Nevertheless, we have tried several values of the power $p$ finding that the optimal choice is $p = 1.5$, which is the value adopted in what follows.}. For sake of simplicity, in Eq.\,(\ref{eq:ratiofit_a2}) we have dropped in the notation of the coefficients $A_1$, $D_1$ and $F_1$ their dependence on the specific channel $j$ as well as on the specific value of the heavy-quark mass $m_h$. The fitting procedure\,(\ref{eq:ratiofit_a2}) is applied for each of the four channels $j = \{ 0^-, 0^+, 1^-, 1^+ \}$, for eight values of $n$ ($n = 2, 3, ..., 9$) and for the two $r$-combinations. 
For each of the $64$ fits the number of data points is $15$ and the number of free parameters is $4$.

The results obtained with the fitting function (\ref{eq:ratiofit_a2}) are shown in Fig.~\ref{fig:RVT_a2} in the case of the ratio $R_{1^-}$ for $n = 5$.
The quality of the fitting procedure may be quite good in several cases, as shown in the left panel of Fig.~\ref{fig:RVT_a2} where the value of $\chi^2/(d.o.f.)$ is significantly less than $1$, but it may be also quite poor, as shown in the right panel of  Fig.~\ref{fig:RVT_a2} where the value of $\chi^2/(d.o.f.)$ is significantly larger than $1$.
In the latter cases discretization effects beyond the order ${\cal{O}}(a^2)$ seem to be required.
Moreover, in Eq.~(\ref{eq:ratiofit_a2}) the coefficient $R_j(n)$ represents the value of the ETMC ratio extrapolated to the physical pion point and to the continuum and infinite volume limits.
However, the susceptibilities corresponding to the two combinations $(r, -r)$ and $(r,r)$ of the Wilson $r$-parameters should differ only by discretization effects (at least of order ${\cal{O}}(a^2)$ in our maximally twisted setup).
This means that the value of $R_j(n)$ should be independent of the choice of the Wilson $r$-parameters.
The conclusion is that the Anstaz (\ref{eq:ratiofit_a2}) is not sufficient for describing the lattice data, since discretization effects beyond the order ${\cal{O}}(a^2)$ should be taken into account.

\begin{figure}[htb!]
\begin{center}
\includegraphics[scale=0.75]{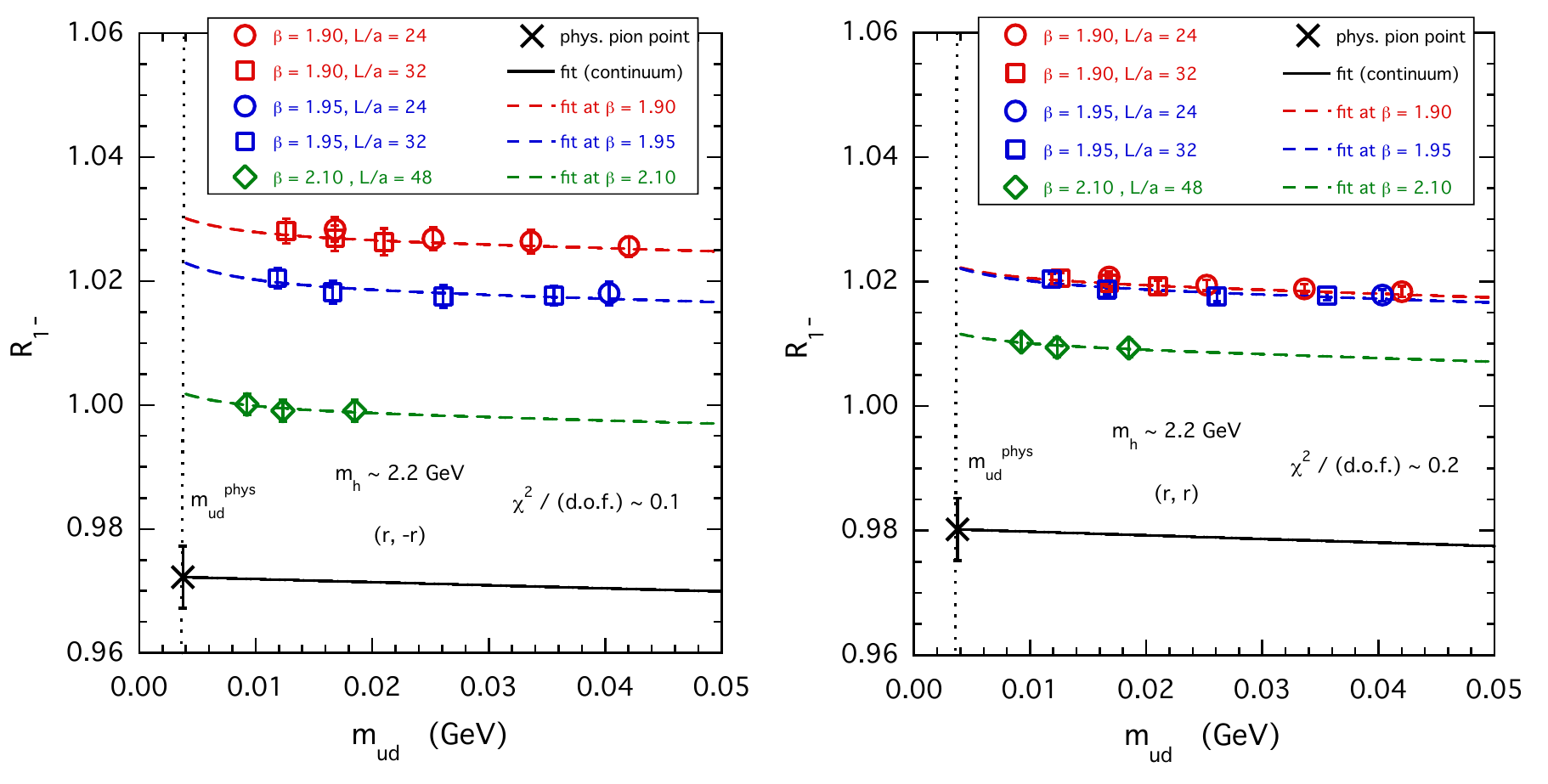}
\end{center}
\vspace{-1cm}
\caption{\it \small The same as in Fig.\,\ref{fig:RVT_a2}, but here the solid and dashed lines represent the results of the fitting function (\ref{eq:ratiofit_a4}) evaluated in the continuum and infinite volume limits and at each value of $\beta$ (for the largest value of $L /a$).}
\label{fig:RVT_a4}
\end{figure}

Following Ref.\,\cite{Martinelli:2021frl} a possible option is to add a term proportional to $a^4$, namely
\bea
    \label{eq:ratiofit_a4}
    R_j(n; a^2, m_{ud}) & = & R_j(n) \left[ 1 + A_1 \left( m_{ud} - m_{ud}^{phys} \right) + D_1 ~ \frac{a^2}{r_0^2} + D_2 ~ \frac{a^4}{r_0^4} \right] ~ \\[2mm]
                                    & \cdot & \left( 1 + F_1 \frac{\overline{M}^2}{(4\pi f)^2} \frac{e^{-\overline{M} L}}{(\overline{M}L)^p} \right) ~ . ~ \nonumber   
\eea
Since our lattice setup includes only three values of the lattice spacing, it would be reasonable to expect that Eq.\,(\ref{eq:ratiofit_a4}) would require the use of a (gaussian) prior on the two parameters $D_1$ and $D_2$. However, at variance with the case of the $b \to c$ transition analyzed in Ref.\,\cite{Martinelli:2021frl} there is no need to introduce a prior for describing the discretization effects for the ETMC ratios of the $b \to u$ transition.
This is clearly illustrated in Fig.\,\ref{fig:RVT_a4}, where the introduction of a discretization term proportional to $a^4$ is greatly beneficial for obtaining fits with good quality for both $r$-combinations.

It turns out that for $m_h(n) \lesssim 2.5$ GeV the ratios $R_j(n)$ corresponding to the two combinations $(r, -r)$ and $(r,r)$ of the Wilson $r$-parameters and extrapolated to the physical pion point and to the continuum and infinite volume limits agree within the errors, while for $m_h(n) \gtrsim 2.5$ GeV the agreement deteriorates and holds only within $\sim 2.5$ standard deviations.
Consequently, we enforce that the extrapolated values $R_j(n)$ must be independent of the specific $r$-combination by performing the following combined extrapolation 
\bea
    \label{eq:combinedfit}
    R_j^{(r, \pm r)}(n; a^2, m_{ud}) & = & R_j(n) \left[ 1 + A_1 \left( m_{ud} - m_{ud}^{phys} \right) + D_1^{(r, \pm r)} ~ \frac{a^2}{r_0^2} + D_2^{(r, \pm r)} ~ \frac{a^4}{r_0^4} \right] 
                                                                \qquad  \\[2mm]
                                                      & \cdot & \left( 1 + F_1 \frac{\overline{M}^2}{(4\pi f)^2} \frac{e^{-\overline{M} L}}{(\overline{M}L)^p} \right) ~ , ~ \nonumber  
\eea
where now only the coefficients $D_1^{(r, \pm r)}$ and $D_2^{(r, \pm r)}$ depend explicitly on the $r$-combination.

\begin{figure}[htb!]
\begin{center}
\includegraphics[scale=0.75]{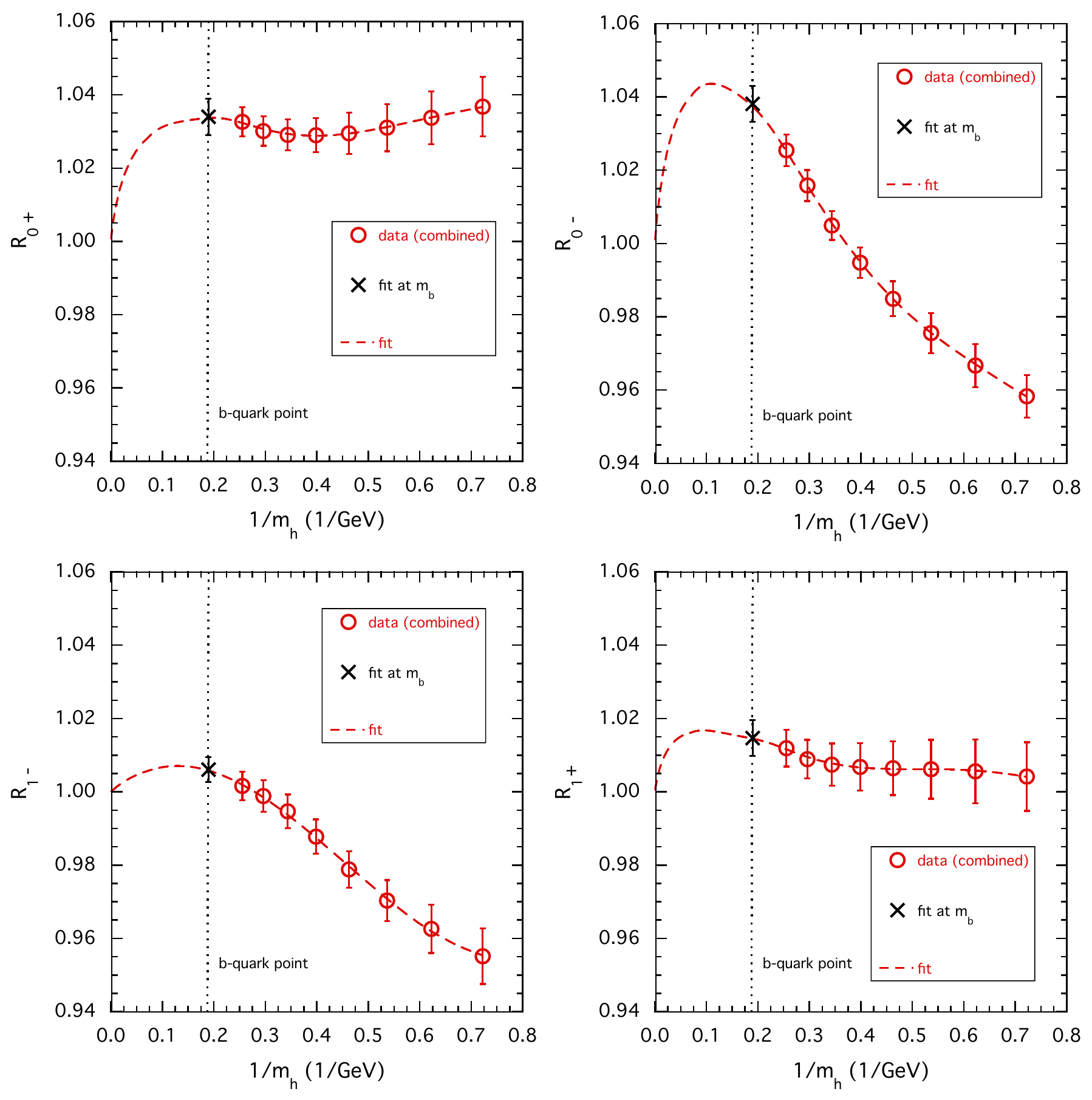}
\end{center}
\vspace{-1cm}
\caption{\it \small The susceptibility ratios $R_{0^+}$ (left top panel), $R_{0^-}$ (right top panel), $R_{1^-}$ (left bottom panel) and $R_{1^+}$ (right bottom panel) after extrapolation to the physical pion point and to the continuum and infinite volume limits based on the combined fit~(\ref{eq:combinedfit}) of the data corresponding to the two combinations $(r, -r)$ and $(r,r)$ of the Wilson $r$-parameters, versus the inverse heavy-quark mass $1 / m_h$. The dashed lines are the results of the fitting procedure\,(\ref{eq:ratio_fit}), described in the next subsection, and the crosses represent the values of the ratios $R_j$ at the physical $b$-quark point, shown as a vertical dotted line.}
\label{fig:ratios_combined}
\end{figure}

The quality of the combined fitting procedure\,(\ref{eq:combinedfit}) is always good for all channels and heavy-quark masses ($\chi^2 / ({\rm d.o.f.}) \lesssim0.7$ with $30$ data points and $7$ free parameters for each of the $32$ fits). 
The results obtained for $R_j(n)$ are shown in Fig.~\ref{fig:ratios_combined}.

\subsection{Extrapolation to the $b$-quark point}

The important feature of the ETMC ratio method is that the extrapolation to the physical $b$-quark point of the ratios $R_j$ can be carried out taking advantage of the fact that by construction
\be
     \mbox{lim}_{n \to \infty} ~ R_j(n) = 1 ~ . ~
\ee

Thus, we fit the lattice data for the ratios $R_j(n)$ adopting the following Ansatz
 \be
    \label{eq:ratio_fit}
    R_j(n) = 1 + \sum_{k=1}^M \left[ A_k + A_k^s \frac{\alpha_s(m_h(n))}{\pi} \right] \left( \frac{1}{m_h(n)} \right)^k  ~ , ~ 
\ee
which contains $2M$ parameters to be determined by a $\chi^2$-minimization procedure\footnote{We remind that, for sake of simplicity, we have dropped in the notation of all the parameters their dependence on the specific channel $j$.}.
We have considered either $M = 2$ or $M = 3$ in Eq.\,(\ref{eq:ratio_fit}), i.e.~either 4 or 6 free parameters, obtaining very similar results.
The quality of the fitting procedure\,(\ref{eq:ratio_fit}) with $M = 3$ is shown by the dashed lines in Fig.~\ref{fig:ratios_combined}, where the values corresponding to the physical $b$-quark point are represented by the crosses and are obtained after a quite short extrapolation from the lattice data.

The susceptibilities $\chi_j(m_b^{phys})$ at the physical $b$-quark point for $j = \{0^+, 1^-, 0^-, 1^+ \}$ can be expressed in terms of the corresponding ones  $\chi_j(m_c^{phys})$ at the physical $c$-quark point as
\be
   \label{eq:chi_mb}
    \chi_j(m_b^{phys}) = \chi_j(m_c^{phys}) \cdot \frac{\rho_j(m_c^{phys})}{\rho_j(m_b^{phys})} \cdot \prod_{n=2}^{11}  R_j(n) ~ , ~
\ee
where the functions $\rho_j$ are given by Eqs.\,(\ref{eq:rhoL})-(\ref{eq:rhoT}) and the product over $n$ include the lattice data up to $n = 9$ and, then, the results of the fitting function\,(\ref{eq:ratio_fit}) only for $n = 10$ and $n = 11$ (namely, after a quite short extrapolation up to the physical $b$-quark point).

\begin{figure}[htb!]
\begin{center}
\includegraphics[scale=0.75]{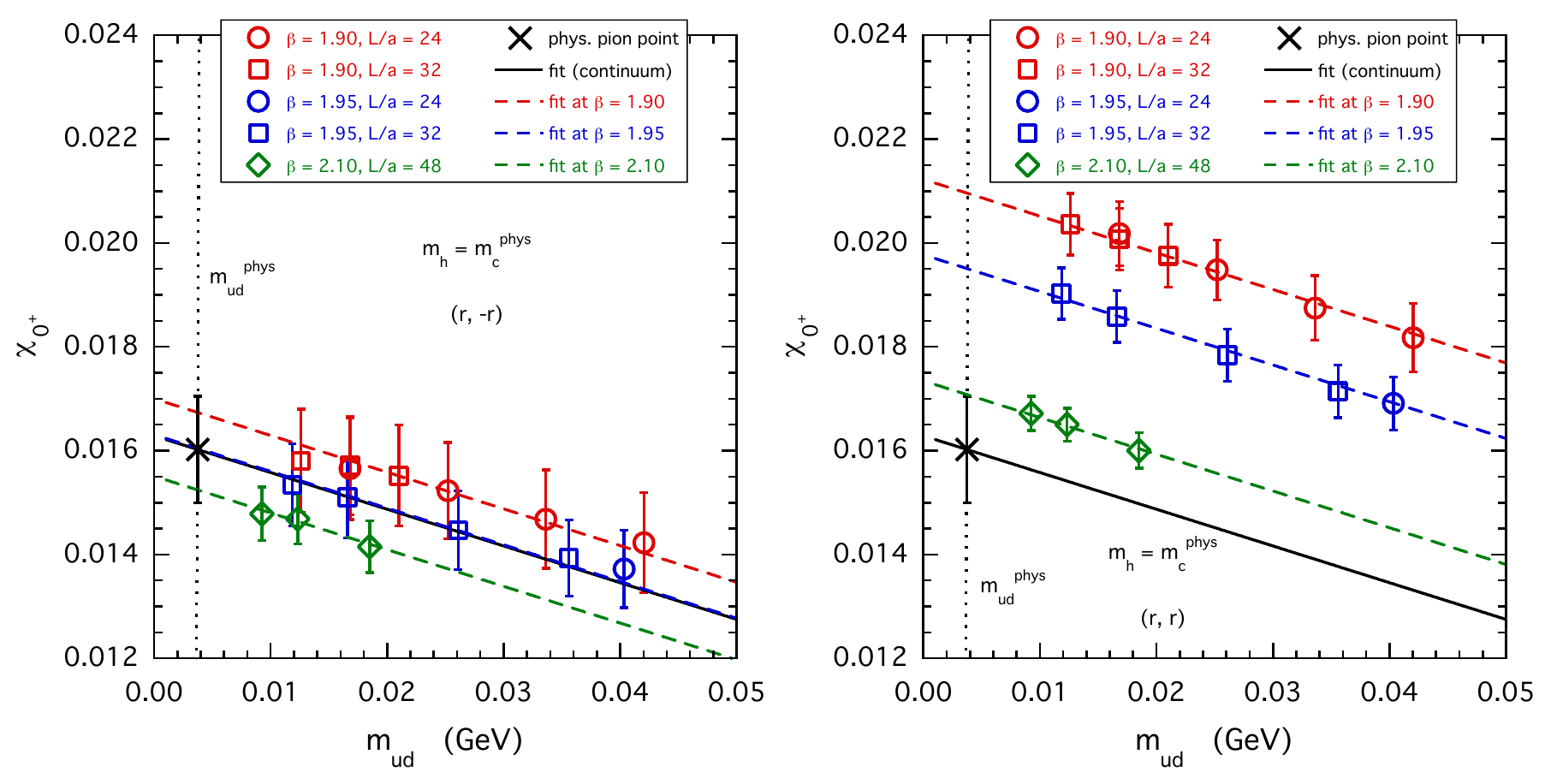}
\end{center}
\vspace{-1cm}
\caption{\it \small Light-quark mass dependence of the longitudinal vector susceptibility $\chi_{0^+}(m_c^{phys})$ for the two combinations $(r, -r)$ (left panel) and $(r,r)$ (right panel) of the Wilson $r$-parameters. The solid lines represent the results of the combined fit evaluated in the continuum and infinite volume limits, while the dashed ones correspond to the fitting function evaluated at each value of $\beta$ and for the largest value of $L /a$ (see text). The crosses represent the value of the susceptibility extrapolated to the physical pion point ($m_{ud} = m_{ud}^{phys}$) and to the continuum and infinite volume limits.}
\label{fig:VL_trig}
\end{figure}

 The ingredients that remain to be determined are the susceptibilities $\chi_j(m_c^{phys})$ evaluated at the physical $c$-quark point, which represent upper limits to the dispersive bounds for the semileptonic FFs related to the $c \to d$ transition.
The extrapolation to the physical pion point and to the continuum and infinite volume limits is performed using a fitting function similar to the one given in the r.h.s.~of Eq.\,(\ref{eq:combinedfit}), i.e.~a combined fit of the lattice data corresponding to the two $r$-combinations.
As an illustrative example, the results obtained in the case of the longitudinal vector susceptibility $\chi_{0^+}(m_c^{phys})$ are shown in Fig.~\ref{fig:VL_trig} .

After averaging over all the branches of our bootstrap analysis the final, nonperturbative values for the susceptibilities $\chi_j(m_c^{phys})$ and $\chi_j(m_b^{phys})$ are collected in Table\,\ref{tab:chi_final}.
\begin{table}[htb!]
\renewcommand{\arraystretch}{1.10}
\begin{center}
\begin{tabular}{||c||c||c||}
\hline 
~ channel $j$ ~ & $\chi_j(m_c^{phys})$ & $\chi_j(m_b^{phys})$ \\ \hline
$0^+$  & ~ $(1.50 \pm 0.13) \cdot 10^{-2}$ ~ & ~ $(2.04 \pm 0.20) \cdot 10^{-2}$ ~\\ \hline
$1^-$   & ~ $(4.81 \pm 1.14) \cdot 10^{-3} ~ \mbox{GeV}^{-2}$ ~ & ~ $(4.88 \pm 1.16) \cdot 10^{-4} ~ \mbox{GeV}^{-2}$ ~ \\ \hline
$0^-$   & ~ $(2.36 \pm 0.15) \cdot 10^{-2}$ ~ & ~ $(2.34 \pm 0.13) \cdot 10^{-2}$ ~ \\ \hline
$1^+$  & ~ $(3.61 \pm 0.81) \cdot 10^{-3} ~ \mbox{GeV}^{-2}$ ~ & ~ $(4.65 \pm 1.02) \cdot 10^{-4} ~ \mbox{GeV}^{-2}$ ~ \\ \hline
\end{tabular}
\end{center}
\renewcommand{\arraystretch}{1.0}
\caption{\it \small Values of the longitudinal and transverse, vector and axial-vector susceptibilities evaluated in this work at the physical $c$- and $b$-quark points after extrapolation to the physical pion point and to the continuum and infinite volume limits.}
\label{tab:chi_final}
\end{table}
Our non-perturbative result for $\chi_{1^-}(m_b^{phys})$ is consistent with the corresponding estimate $\chi_{1^-}(m_b^{phys}) = 5.01 \cdot 10^{-4}$ GeV$^{-2}$ made in Ref.\,\cite{Bourrely:2008za} using perturbative QCD with the addition of small contributions from quark and gluon condensates.

\subsection{Subtraction of bound-state contributions}

The results of the previous subsection represent upper limits to the dispersive bounds on the form factors relevant in the semileptonic $c \to d \ell \nu_\ell$ and $b \to u \ell \nu_\ell$ transitions, respectively. 
Such limits can be improved by removing the contributions of the bound states lying below the pair production threshold.

In the case of the semileptonic $B \to \pi \ell \nu_\ell$ decays the above situation occurs in the longitudinal axial-vector and the transverse vector channels due to the presence of the $B$- and $B^*$-meson ground states, respectively. 
Their contributions, $\chi_{0^-}^{(gs)}(m_b^{phys})$ and $\chi_{1^-}^{(gs)}(m_b^{phys})$, are explicitly given by
\bea
      \label{eq:chiAL_grs}
      \chi_{0^-}^{(gs)}(m_b^{phys}) & = & \frac{f_B^2}{M_B^2} ~ , ~  \\[2mm] 
      \label{eq:chiVT_grs}
      \chi_{1^-}^{(gs)}(m_b^{phys}) & = & \frac{f_{B^*}^2}{M_{B^*}^4} ~ , ~     
\eea
where $f_{B(B^*)}$ and $M_{B(B^*)}$ are respectively the $B(B^*)$-meson decay constant and mass.
Adopting the experimental value of the $B$-meson mass\,\cite{ParticleDataGroup:2020ssz} and the lattice values $f_B = 0.193\,(6)$ GeV, $f_{B^*} = 0.1859\,(72)$ GeV and $M_{B^*} = 5.3205\,(76)$ GeV, obtained in Refs.\,\cite{ETM:2016nbo,Lubicz:2017asp} using the same ETMC gauge ensembles of this work, one gets 
\bea
      \label{eq:chiALgrs}
      \chi_{0^-}^{(gs)}(m_b^{phys}) & = & (0.134 \pm 0.008) \cdot 10^{-2} ~ , ~ \\[2mm]
      \label{eq:chiVTgrs}
     \chi_{1^-}^{(gs)}(m_b^{phys}) & = & (0.431 \pm 0.033) \cdot 10^{-4} ~ \mbox{GeV}^{-2} ~ , ~
\eea
leading to the following upper bound for the longitudinal axial-vector and transverse vector channels at the physical $b$-quark point
\bea
     \label{eq:chiAL_final}
     \chi_{0^-}(m_b^{phys}) & = & (2.20 \pm 0.13) \cdot 10^{-2} ~ , ~ \\[2mm]
     \label{eq:chiVT_final}
     \chi_{1^-}(m_b^{phys}) & = & (4.45 \pm 1.16) \cdot 10^{-4} ~ \mbox{GeV}^{-2} ~ . ~
\eea
The values of the susceptibilities $\chi_{0^+}(m_b^{phys})$ and $\chi_{1^-}(m_b^{phys})$ relevant in this work are given in Eqs.\,(\ref{eq:chi0+})-(\ref{eq:chi1-}).

\section{Theoretical estimate of $R^{\tau/\mu}_{\pi(K)}$, $\bar{\mathcal{A}}_{FB}^{\ell,\pi(K)}$ and $\bar{\mathcal{A}}_{polar}^{\ell,\pi(K)}$ in selected $q^2$-bins}
\label{sec:appB}

In this Appendix we collect our theoretical predictions for the ratio of the $\tau/\mu$ decay rates $R^{\tau/\mu}_{\pi(K)}$, the normalized forward-backward asymmetry $\bar{\mathcal{A}}_{FB}^{\ell,\pi(K)}$ and the normalized lepton polarization asymmetry $\bar{\mathcal{A}}_{polar}^{\ell,\pi(K)}$ evaluated in limited regions of the phase space.
In other words we evaluate the quantities\,(\ref{Rpidef})-(\ref{normApol}) restricting the integration over $q^2$ in both the numerator and the denominator to limited kinematical regions. 

We have selected three different $q^2$-bins, namely for the $B_{(s)} \to \pi(K)$ decays
\begin{itemize}
\item low-$q^2$ region: from $q^2 = 0$ to $q^2 = 9$ GeV$^2$ (8 GeV$^2$);
\item intermediate-$q^2$ region: from $q^2 = 9$ GeV$^2$ (8 GeV$^2$) to $q^2 = 18$ GeV$^2$ (16 GeV$^2$);
\item high-$q^2$ region:  from $q^2 = 18$ GeV$^2$ (16 GeV$^2$) to $q^2 = t_- = 26.4$ GeV$^2$ ($23.7$ GeV$^2$).
\end{itemize}
Note that in the low-$q^2$ region, when the $\tau$-lepton is involved, the minimum value of $q^2$ is equal to $m_\tau^2$.

The results, based on the combined LQCD data of Tables\,\ref{tab:LQCDBPi} and \ref{tab:LQCDBsK} used as inputs for our DM method, are collected in Tables\,\ref{tab:phenoBpi_bins} and \ref{tab:phenoBsK_bins} for the $B \to \pi$ and $B_s \to K$ decays, respectively.
\begin{table}[htb!]
\renewcommand{\arraystretch}{1.1}
\begin{center}
{\small
\begin{tabular}{|c||c||c||c|}
\hline
& ~ low-$q^2$ ~ & ~ intermediate-$q^2$ ~ & ~ high-$q^2$ ~\\
\hline
\hline
$R^{\tau/\mu}_{\pi}$ & 0.250(120) & 0.852(69) & 1.152(57) \\
$\bar{\mathcal{A}}_{FB}^{\mu,\pi}$ & 0.0092(105) & 0.00113(27) & 0.00115(9)\\
$\bar{\mathcal{A}}_{FB}^{\tau,\pi}$ & 0.223(133) & 0.220(32) & 0.208(6) \\
$\bar{\mathcal{A}}_{polar}^{\mu,\pi}$ & 3.32(1.13) & 1.16(29) & 1.06(30)\\
$\bar{\mathcal{A}}_{polar}^{\tau,\pi}$ & 14.6(13.4) & 1.16(29) & 1.06(30)\\
\hline
\end{tabular}
}
\caption{\it The theoretical values of the ratio of the $\tau/\mu$ decay rates $R^{\tau/\mu}_{\pi(K)}$, the normalized forward-backward asymmetry $\bar{\mathcal{A}}_{FB}^{\ell,\pi(K)}$ and the normalized lepton polarization asymmetry $\bar{\mathcal{A}}_{polar}^{\ell,\pi(K)}$  evaluated in the three selected $q^2$-bins in the case of the semileptonic $B \to \pi \ell \nu_\ell$ decays with $\ell = \mu, \tau$ adopting the combined LQCD data of Table\,\ref{tab:LQCDBPi} as inputs for our DM method.\hspace*{\fill}}
\label{tab:phenoBpi_bins}
\end{center}
\end{table}
\begin{table}[htb!]
\renewcommand{\arraystretch}{1.1}
\begin{center}
{\small
\begin{tabular}{|c||c||c||c|}
\hline
& ~ low-$q^2$ ~ & ~ intermediate-$q^2$ ~ & ~ high-$q^2$ ~\\
\hline
\hline
$R^{\tau/\mu}_{K}$ & 0.249(85) & 0.889(72) & 1.163(54) \\
$\bar{\mathcal{A}}_{FB}^{\mu,K}$ & 0.0105(73) & 0.00159(29) & 0.00132(9)\\
$\bar{\mathcal{A}}_{FB}^{\tau,K}$ & 0.341(82) & 0.268(23) & 0.225(5) \\
$\bar{\mathcal{A}}_{polar}^{\mu,K}$ & 4.15(1.98) & 0.88(27) & 1.04(29)\\
$\bar{\mathcal{A}}_{polar}^{\tau,K}$ & 11.4(16.7) &  0.88(27) & 1.04(29)\\
\hline
\end{tabular}
}
\caption{\it The same as in Table\,\ref{tab:phenoBpi_bins}, but in the case of the semileptonic $B_s \to K \ell \nu_\ell$ decays with $\ell = \mu, \tau$ adopting the combined LQCD data of Table\,\ref{tab:LQCDBsK} as inputs for our DM method.\hspace*{\fill}}
\label{tab:phenoBsK_bins}
\end{center}
\end{table}
It can be seen that large (and even quite large) uncertainties affect the theoretical predictions of some of the quantities in the low-$q^2$ bin.
This is related to the large uncertainties of the hadronic form factors at low values of $q^2$ (see Figs.\,\ref{FFMMBpiCOMB}-\ref{FFMMBsKCOMB}), which are a consequence of the present uncertainties of the input lattice data and of the long extrapolation to low values of $q^2$.
Direct lattice calculations at smaller values of $q^2$ will allow in the future to reach a more significative precision also in the low-$q^2$ bin.

\bibliography{BPi}

\begin{thebibliography}{58}%
\makeatletter
\providecommand \@ifxundefined [1]{%
 \@ifx{#1\undefined}
}%
\providecommand \@ifnum [1]{%
 \ifnum #1\expandafter \@firstoftwo
 \else \expandafter \@secondoftwo
 \fi
}%
\providecommand \@ifx [1]{%
 \ifx #1\expandafter \@firstoftwo
 \else \expandafter \@secondoftwo
 \fi
}%
\providecommand \natexlab [1]{#1}%
\providecommand \enquote  [1]{``#1''}%
\providecommand \bibnamefont  [1]{#1}%
\providecommand \bibfnamefont [1]{#1}%
\providecommand \citenamefont [1]{#1}%
\providecommand \href@noop [0]{\@secondoftwo}%
\providecommand \href [0]{\begingroup \@sanitize@url \@href}%
\providecommand \@href[1]{\@@startlink{#1}\@@href}%
\providecommand \@@href[1]{\endgroup#1\@@endlink}%
\providecommand \@sanitize@url [0]{\catcode `\\12\catcode `\$12\catcode
  `\&12\catcode `\#12\catcode `\^12\catcode `\_12\catcode `\%12\relax}%
\providecommand \@@startlink[1]{}%
\providecommand \@@endlink[0]{}%
\providecommand \url  [0]{\begingroup\@sanitize@url \@url }%
\providecommand \@url [1]{\endgroup\@href {#1}{\urlprefix }}%
\providecommand \urlprefix  [0]{URL }%
\providecommand \Eprint [0]{\href }%
\providecommand \doibase [0]{http://dx.doi.org/}%
\providecommand \selectlanguage [0]{\@gobble}%
\providecommand \bibinfo  [0]{\@secondoftwo}%
\providecommand \bibfield  [0]{\@secondoftwo}%
\providecommand \translation [1]{[#1]}%
\providecommand \BibitemOpen [0]{}%
\providecommand \bibitemStop [0]{}%
\providecommand \bibitemNoStop [0]{.\EOS\space}%
\providecommand \EOS [0]{\spacefactor3000\relax}%
\providecommand \BibitemShut  [1]{\csname bibitem#1\endcsname}%
\let\auto@bib@innerbib\@empty
\bibitem [{\citenamefont {Aoki}\ \emph {et~al.}(2021)\citenamefont {Aoki} \emph
  {et~al.}}]{Aoki:2021kgd}%
  \BibitemOpen
  \bibfield  {author} {\bibinfo {author} {\bibfnamefont {Y.}~\bibnamefont
  {Aoki}} \emph {et~al.},\ }\href@noop {} {\  (\bibinfo {year} {2021})},\
  \Eprint {http://arxiv.org/abs/2111.09849} {arXiv:2111.09849 [hep-lat]}
  \BibitemShut {NoStop}%
\bibitem [{\citenamefont {Amhis}\ \emph {et~al.}(2017)\citenamefont {Amhis}
  \emph {et~al.}}]{Amhis:2016xyh}%
  \BibitemOpen
  \bibfield  {author} {\bibinfo {author} {\bibfnamefont {Y.}~\bibnamefont
  {Amhis}} \emph {et~al.} (\bibinfo {collaboration} {HFLAV}),\ }\href {\doibase
  10.1140/epjc/s10052-017-5058-4} {\bibfield  {journal} {\bibinfo  {journal}
  {Eur. Phys. J. C}\ }\textbf {\bibinfo {volume} {77}},\ \bibinfo {pages} {895}
  (\bibinfo {year} {2017})},\ \Eprint {http://arxiv.org/abs/1612.07233}
  {arXiv:1612.07233 [hep-ex]} \BibitemShut {NoStop}%
\bibitem [{\citenamefont {Cao}\ \emph {et~al.}(2021)\citenamefont {Cao} \emph
  {et~al.}}]{Belle:2021eni}%
  \BibitemOpen
  \bibfield  {author} {\bibinfo {author} {\bibfnamefont {L.}~\bibnamefont
  {Cao}} \emph {et~al.} (\bibinfo {collaboration} {Belle}),\ }\href {\doibase
  10.1103/PhysRevD.104.012008} {\bibfield  {journal} {\bibinfo  {journal}
  {Phys. Rev. D}\ }\textbf {\bibinfo {volume} {104}},\ \bibinfo {pages}
  {012008} (\bibinfo {year} {2021})},\ \Eprint
  {http://arxiv.org/abs/2102.00020} {arXiv:2102.00020 [hep-ex]} \BibitemShut
  {NoStop}%
\bibitem [{\citenamefont {Lange}\ \emph {et~al.}(2005)\citenamefont {Lange},
  \citenamefont {Neubert},\ and\ \citenamefont {Paz}}]{Lange:2005yw}%
  \BibitemOpen
  \bibfield  {author} {\bibinfo {author} {\bibfnamefont {B.~O.}\ \bibnamefont
  {Lange}}, \bibinfo {author} {\bibfnamefont {M.}~\bibnamefont {Neubert}}, \
  and\ \bibinfo {author} {\bibfnamefont {G.}~\bibnamefont {Paz}},\ }\href
  {\doibase 10.1103/PhysRevD.72.073006} {\bibfield  {journal} {\bibinfo
  {journal} {Phys. Rev. D}\ }\textbf {\bibinfo {volume} {72}},\ \bibinfo
  {pages} {073006} (\bibinfo {year} {2005})},\ \Eprint
  {http://arxiv.org/abs/hep-ph/0504071} {arXiv:hep-ph/0504071} \BibitemShut
  {NoStop}%
\bibitem [{\citenamefont {Andersen}\ and\ \citenamefont
  {Gardi}(2006)}]{Andersen:2005mj}%
  \BibitemOpen
  \bibfield  {author} {\bibinfo {author} {\bibfnamefont {J.~R.}\ \bibnamefont
  {Andersen}}\ and\ \bibinfo {author} {\bibfnamefont {E.}~\bibnamefont
  {Gardi}},\ }\href {\doibase 10.1088/1126-6708/2006/01/097} {\bibfield
  {journal} {\bibinfo  {journal} {JHEP}\ }\textbf {\bibinfo {volume} {01}},\
  \bibinfo {pages} {097} (\bibinfo {year} {2006})},\ \Eprint
  {http://arxiv.org/abs/hep-ph/0509360} {arXiv:hep-ph/0509360} \BibitemShut
  {NoStop}%
\bibitem [{\citenamefont {Gardi}(2008)}]{Gardi:2008bb}%
  \BibitemOpen
  \bibfield  {author} {\bibinfo {author} {\bibfnamefont {E.}~\bibnamefont
  {Gardi}},\ }\href@noop {} {\bibfield  {journal} {\bibinfo  {journal}
  {Frascati Phys. Ser.}\ }\textbf {\bibinfo {volume} {47}},\ \bibinfo {pages}
  {381} (\bibinfo {year} {2008})},\ \Eprint {http://arxiv.org/abs/0806.4524}
  {arXiv:0806.4524 [hep-ph]} \BibitemShut {NoStop}%
\bibitem [{\citenamefont {Gambino}\ \emph {et~al.}(2007)\citenamefont
  {Gambino}, \citenamefont {Giordano}, \citenamefont {Ossola},\ and\
  \citenamefont {Uraltsev}}]{Gambino:2007rp}%
  \BibitemOpen
  \bibfield  {author} {\bibinfo {author} {\bibfnamefont {P.}~\bibnamefont
  {Gambino}}, \bibinfo {author} {\bibfnamefont {P.}~\bibnamefont {Giordano}},
  \bibinfo {author} {\bibfnamefont {G.}~\bibnamefont {Ossola}}, \ and\ \bibinfo
  {author} {\bibfnamefont {N.}~\bibnamefont {Uraltsev}},\ }\href {\doibase
  10.1088/1126-6708/2007/10/058} {\bibfield  {journal} {\bibinfo  {journal}
  {JHEP}\ }\textbf {\bibinfo {volume} {10}},\ \bibinfo {pages} {058} (\bibinfo
  {year} {2007})},\ \Eprint {http://arxiv.org/abs/0707.2493} {arXiv:0707.2493
  [hep-ph]} \BibitemShut {NoStop}%
\bibitem [{\citenamefont {Aglietti}\ \emph {et~al.}(2007)\citenamefont
  {Aglietti}, \citenamefont {Ferrera},\ and\ \citenamefont
  {Ricciardi}}]{Aglietti:2006yb}%
  \BibitemOpen
  \bibfield  {author} {\bibinfo {author} {\bibfnamefont {U.}~\bibnamefont
  {Aglietti}}, \bibinfo {author} {\bibfnamefont {G.}~\bibnamefont {Ferrera}}, \
  and\ \bibinfo {author} {\bibfnamefont {G.}~\bibnamefont {Ricciardi}},\ }\href
  {\doibase 10.1016/j.nuclphysb.2007.01.014} {\bibfield  {journal} {\bibinfo
  {journal} {Nucl. Phys. B}\ }\textbf {\bibinfo {volume} {768}},\ \bibinfo
  {pages} {85} (\bibinfo {year} {2007})},\ \Eprint
  {http://arxiv.org/abs/hep-ph/0608047} {arXiv:hep-ph/0608047} \BibitemShut
  {NoStop}%
\bibitem [{\citenamefont {Aglietti}\ \emph {et~al.}(2009)\citenamefont
  {Aglietti}, \citenamefont {Di~Lodovico}, \citenamefont {Ferrera},\ and\
  \citenamefont {Ricciardi}}]{Aglietti:2007ik}%
  \BibitemOpen
  \bibfield  {author} {\bibinfo {author} {\bibfnamefont {U.}~\bibnamefont
  {Aglietti}}, \bibinfo {author} {\bibfnamefont {F.}~\bibnamefont
  {Di~Lodovico}}, \bibinfo {author} {\bibfnamefont {G.}~\bibnamefont
  {Ferrera}}, \ and\ \bibinfo {author} {\bibfnamefont {G.}~\bibnamefont
  {Ricciardi}},\ }\href {\doibase 10.1140/epjc/s10052-008-0817-x} {\bibfield
  {journal} {\bibinfo  {journal} {Eur. Phys. J. C}\ }\textbf {\bibinfo {volume}
  {59}},\ \bibinfo {pages} {831} (\bibinfo {year} {2009})},\ \Eprint
  {http://arxiv.org/abs/0711.0860} {arXiv:0711.0860 [hep-ph]} \BibitemShut
  {NoStop}%
\bibitem [{\citenamefont {Zyla}\ \emph {et~al.}(2020)\citenamefont {Zyla} \emph
  {et~al.}}]{ParticleDataGroup:2020ssz}%
  \BibitemOpen
  \bibfield  {author} {\bibinfo {author} {\bibfnamefont {P.~A.}\ \bibnamefont
  {Zyla}} \emph {et~al.} (\bibinfo {collaboration} {Particle Data Group}),\
  }\href {\doibase 10.1093/ptep/ptaa104} {\bibfield  {journal} {\bibinfo
  {journal} {PTEP}\ }\textbf {\bibinfo {volume} {2020}},\ \bibinfo {pages}
  {083C01} (\bibinfo {year} {2020})}\BibitemShut {NoStop}%
\bibitem [{\citenamefont {Leljak}\ \emph {et~al.}(2021)\citenamefont {Leljak},
  \citenamefont {Meli\'c},\ and\ \citenamefont {van Dyk}}]{Leljak:2021vte}%
  \BibitemOpen
  \bibfield  {author} {\bibinfo {author} {\bibfnamefont {D.}~\bibnamefont
  {Leljak}}, \bibinfo {author} {\bibfnamefont {B.}~\bibnamefont {Meli\'c}}, \
  and\ \bibinfo {author} {\bibfnamefont {D.}~\bibnamefont {van Dyk}},\ }\href
  {\doibase 10.1007/JHEP07(2021)036} {\bibfield  {journal} {\bibinfo  {journal}
  {JHEP}\ }\textbf {\bibinfo {volume} {07}},\ \bibinfo {pages} {036} (\bibinfo
  {year} {2021})},\ \Eprint {http://arxiv.org/abs/2102.07233} {arXiv:2102.07233
  [hep-ph]} \BibitemShut {NoStop}%
\bibitem [{\citenamefont {Biswas}\ \emph {et~al.}(2021)\citenamefont {Biswas},
  \citenamefont {Nandi}, \citenamefont {Patra},\ and\ \citenamefont
  {Ray}}]{Biswas:2021qyq}%
  \BibitemOpen
  \bibfield  {author} {\bibinfo {author} {\bibfnamefont {A.}~\bibnamefont
  {Biswas}}, \bibinfo {author} {\bibfnamefont {S.}~\bibnamefont {Nandi}},
  \bibinfo {author} {\bibfnamefont {S.~K.}\ \bibnamefont {Patra}}, \ and\
  \bibinfo {author} {\bibfnamefont {I.}~\bibnamefont {Ray}},\ }\href {\doibase
  10.1007/JHEP07(2021)082} {\bibfield  {journal} {\bibinfo  {journal} {JHEP}\
  }\textbf {\bibinfo {volume} {07}},\ \bibinfo {pages} {082} (\bibinfo {year}
  {2021})},\ \Eprint {http://arxiv.org/abs/2103.01809} {arXiv:2103.01809
  [hep-ph]} \BibitemShut {NoStop}%
\bibitem [{\citenamefont {Gonz\`alez-Sol\'\i{}s}\ \emph
  {et~al.}(2021)\citenamefont {Gonz\`alez-Sol\'\i{}s}, \citenamefont
  {Masjuan},\ and\ \citenamefont {Rojas}}]{Gonzalez-Solis:2021pyh}%
  \BibitemOpen
  \bibfield  {author} {\bibinfo {author} {\bibfnamefont {S.}~\bibnamefont
  {Gonz\`alez-Sol\'\i{}s}}, \bibinfo {author} {\bibfnamefont {P.}~\bibnamefont
  {Masjuan}}, \ and\ \bibinfo {author} {\bibfnamefont {C.}~\bibnamefont
  {Rojas}},\ }\href {\doibase 10.1103/PhysRevD.104.114041} {\bibfield
  {journal} {\bibinfo  {journal} {Phys. Rev. D}\ }\textbf {\bibinfo {volume}
  {104}},\ \bibinfo {pages} {114041} (\bibinfo {year} {2021})},\ \Eprint
  {http://arxiv.org/abs/2110.06153} {arXiv:2110.06153 [hep-ph]} \BibitemShut
  {NoStop}%
\bibitem [{\citenamefont {Bourrely}\ \emph {et~al.}(2009)\citenamefont
  {Bourrely}, \citenamefont {Caprini},\ and\ \citenamefont
  {Lellouch}}]{Bourrely:2008za}%
  \BibitemOpen
  \bibfield  {author} {\bibinfo {author} {\bibfnamefont {C.}~\bibnamefont
  {Bourrely}}, \bibinfo {author} {\bibfnamefont {I.}~\bibnamefont {Caprini}}, \
  and\ \bibinfo {author} {\bibfnamefont {L.}~\bibnamefont {Lellouch}},\ }\href
  {\doibase 10.1103/PhysRevD.82.099902} {\bibfield  {journal} {\bibinfo
  {journal} {Phys. Rev. D}\ }\textbf {\bibinfo {volume} {79}},\ \bibinfo
  {pages} {013008} (\bibinfo {year} {2009})},\ \bibinfo {note} {[Erratum:
  Phys.Rev.D 82, 099902 (2010)]},\ \Eprint {http://arxiv.org/abs/0807.2722}
  {arXiv:0807.2722 [hep-ph]} \BibitemShut {NoStop}%
\bibitem [{\citenamefont {Bharucha}\ \emph {et~al.}(2016)\citenamefont
  {Bharucha}, \citenamefont {Straub},\ and\ \citenamefont
  {Zwicky}}]{Straub:2015ica}%
  \BibitemOpen
  \bibfield  {author} {\bibinfo {author} {\bibfnamefont {A.}~\bibnamefont
  {Bharucha}}, \bibinfo {author} {\bibfnamefont {D.~M.}\ \bibnamefont
  {Straub}}, \ and\ \bibinfo {author} {\bibfnamefont {R.}~\bibnamefont
  {Zwicky}},\ }\href {\doibase 10.1007/JHEP08(2016)098} {\bibfield  {journal}
  {\bibinfo  {journal} {JHEP}\ }\textbf {\bibinfo {volume} {08}},\ \bibinfo
  {pages} {098} (\bibinfo {year} {2016})},\ \Eprint
  {http://arxiv.org/abs/1503.05534} {arXiv:1503.05534 [hep-ph]} \BibitemShut
  {NoStop}%
\bibitem [{\citenamefont {Gonz\`alez-Sol\'\i{}s}\ and\ \citenamefont
  {Masjuan}(2018)}]{Gonzalez-Solis:2018ooo}%
  \BibitemOpen
  \bibfield  {author} {\bibinfo {author} {\bibfnamefont {S.}~\bibnamefont
  {Gonz\`alez-Sol\'\i{}s}}\ and\ \bibinfo {author} {\bibfnamefont
  {P.}~\bibnamefont {Masjuan}},\ }\href {\doibase 10.1103/PhysRevD.98.034027}
  {\bibfield  {journal} {\bibinfo  {journal} {Phys. Rev. D}\ }\textbf {\bibinfo
  {volume} {98}},\ \bibinfo {pages} {034027} (\bibinfo {year} {2018})},\
  \Eprint {http://arxiv.org/abs/1805.11262} {arXiv:1805.11262 [hep-ph]}
  \BibitemShut {NoStop}%
\bibitem [{\citenamefont {Lellouch}(1996)}]{Lellouch:1995yv}%
  \BibitemOpen
  \bibfield  {author} {\bibinfo {author} {\bibfnamefont {L.}~\bibnamefont
  {Lellouch}},\ }\href {\doibase 10.1016/0550-3213(96)00443-9} {\bibfield
  {journal} {\bibinfo  {journal} {Nucl. Phys. B}\ }\textbf {\bibinfo {volume}
  {479}},\ \bibinfo {pages} {353} (\bibinfo {year} {1996})},\ \Eprint
  {http://arxiv.org/abs/hep-ph/9509358} {arXiv:hep-ph/9509358} \BibitemShut
  {NoStop}%
\bibitem [{\citenamefont {Di~Carlo}\ \emph {et~al.}(2021)\citenamefont
  {Di~Carlo}, \citenamefont {Martinelli}, \citenamefont {Naviglio},
  \citenamefont {Sanfilippo}, \citenamefont {Simula},\ and\ \citenamefont
  {Vittorio}}]{DiCarlo:2021dzg}%
  \BibitemOpen
  \bibfield  {author} {\bibinfo {author} {\bibfnamefont {M.}~\bibnamefont
  {Di~Carlo}}, \bibinfo {author} {\bibfnamefont {G.}~\bibnamefont
  {Martinelli}}, \bibinfo {author} {\bibfnamefont {M.}~\bibnamefont
  {Naviglio}}, \bibinfo {author} {\bibfnamefont {F.}~\bibnamefont
  {Sanfilippo}}, \bibinfo {author} {\bibfnamefont {S.}~\bibnamefont {Simula}},
  \ and\ \bibinfo {author} {\bibfnamefont {L.}~\bibnamefont {Vittorio}},\
  }\href {\doibase 10.1103/PhysRevD.104.054502} {\bibfield  {journal} {\bibinfo
   {journal} {Phys. Rev. D}\ }\textbf {\bibinfo {volume} {104}},\ \bibinfo
  {pages} {054502} (\bibinfo {year} {2021})},\ \Eprint
  {http://arxiv.org/abs/2105.02497} {arXiv:2105.02497 [hep-lat]} \BibitemShut
  {NoStop}%
\bibitem [{\citenamefont {Martinelli}\ \emph
  {et~al.}(2021{\natexlab{a}})\citenamefont {Martinelli}, \citenamefont
  {Simula},\ and\ \citenamefont {Vittorio}}]{Martinelli:2021frl}%
  \BibitemOpen
  \bibfield  {author} {\bibinfo {author} {\bibfnamefont {G.}~\bibnamefont
  {Martinelli}}, \bibinfo {author} {\bibfnamefont {S.}~\bibnamefont {Simula}},
  \ and\ \bibinfo {author} {\bibfnamefont {L.}~\bibnamefont {Vittorio}},\
  }\href {\doibase 10.1103/PhysRevD.104.094512} {\bibfield  {journal} {\bibinfo
   {journal} {Phys. Rev. D}\ }\textbf {\bibinfo {volume} {104}},\ \bibinfo
  {pages} {094512} (\bibinfo {year} {2021}{\natexlab{a}})},\ \Eprint
  {http://arxiv.org/abs/2105.07851} {arXiv:2105.07851 [hep-lat]} \BibitemShut
  {NoStop}%
\bibitem [{\citenamefont {Flynn}\ \emph {et~al.}(2015)\citenamefont {Flynn},
  \citenamefont {Izubuchi}, \citenamefont {Kawanai}, \citenamefont {Lehner},
  \citenamefont {Soni}, \citenamefont {Van~de Water},\ and\ \citenamefont
  {Witzel}}]{Flynn:2015mha}%
  \BibitemOpen
  \bibfield  {author} {\bibinfo {author} {\bibfnamefont {J.~M.}\ \bibnamefont
  {Flynn}}, \bibinfo {author} {\bibfnamefont {T.}~\bibnamefont {Izubuchi}},
  \bibinfo {author} {\bibfnamefont {T.}~\bibnamefont {Kawanai}}, \bibinfo
  {author} {\bibfnamefont {C.}~\bibnamefont {Lehner}}, \bibinfo {author}
  {\bibfnamefont {A.}~\bibnamefont {Soni}}, \bibinfo {author} {\bibfnamefont
  {R.~S.}\ \bibnamefont {Van~de Water}}, \ and\ \bibinfo {author}
  {\bibfnamefont {O.}~\bibnamefont {Witzel}},\ }\href {\doibase
  10.1103/PhysRevD.91.074510} {\bibfield  {journal} {\bibinfo  {journal} {Phys.
  Rev. D}\ }\textbf {\bibinfo {volume} {91}},\ \bibinfo {pages} {074510}
  (\bibinfo {year} {2015})},\ \Eprint {http://arxiv.org/abs/1501.05373}
  {arXiv:1501.05373 [hep-lat]} \BibitemShut {NoStop}%
\bibitem [{\citenamefont {Bailey}\ \emph {et~al.}(2015)\citenamefont {Bailey}
  \emph {et~al.}}]{Lattice:2015tia}%
  \BibitemOpen
  \bibfield  {author} {\bibinfo {author} {\bibfnamefont {J.~A.}\ \bibnamefont
  {Bailey}} \emph {et~al.} (\bibinfo {collaboration} {Fermilab Lattice,
  MILC}),\ }\href {\doibase 10.1103/PhysRevD.92.014024} {\bibfield  {journal}
  {\bibinfo  {journal} {Phys. Rev. D}\ }\textbf {\bibinfo {volume} {92}},\
  \bibinfo {pages} {014024} (\bibinfo {year} {2015})},\ \Eprint
  {http://arxiv.org/abs/1503.07839} {arXiv:1503.07839 [hep-lat]} \BibitemShut
  {NoStop}%
\bibitem [{\citenamefont {Bouchard}\ \emph {et~al.}(2014)\citenamefont
  {Bouchard}, \citenamefont {Lepage}, \citenamefont {Monahan}, \citenamefont
  {Na},\ and\ \citenamefont {Shigemitsu}}]{Bouchard:2014ypa}%
  \BibitemOpen
  \bibfield  {author} {\bibinfo {author} {\bibfnamefont {C.~M.}\ \bibnamefont
  {Bouchard}}, \bibinfo {author} {\bibfnamefont {G.~P.}\ \bibnamefont
  {Lepage}}, \bibinfo {author} {\bibfnamefont {C.}~\bibnamefont {Monahan}},
  \bibinfo {author} {\bibfnamefont {H.}~\bibnamefont {Na}}, \ and\ \bibinfo
  {author} {\bibfnamefont {J.}~\bibnamefont {Shigemitsu}},\ }\href {\doibase
  10.1103/PhysRevD.90.054506} {\bibfield  {journal} {\bibinfo  {journal} {Phys.
  Rev. D}\ }\textbf {\bibinfo {volume} {90}},\ \bibinfo {pages} {054506}
  (\bibinfo {year} {2014})},\ \Eprint {http://arxiv.org/abs/1406.2279}
  {arXiv:1406.2279 [hep-lat]} \BibitemShut {NoStop}%
\bibitem [{\citenamefont {Bazavov}\ \emph {et~al.}(2019)\citenamefont {Bazavov}
  \emph {et~al.}}]{Bazavov:2019aom}%
  \BibitemOpen
  \bibfield  {author} {\bibinfo {author} {\bibfnamefont {A.}~\bibnamefont
  {Bazavov}} \emph {et~al.} (\bibinfo {collaboration} {Fermilab Lattice,
  MILC}),\ }\href {\doibase 10.1103/PhysRevD.100.034501} {\bibfield  {journal}
  {\bibinfo  {journal} {Phys. Rev. D}\ }\textbf {\bibinfo {volume} {100}},\
  \bibinfo {pages} {034501} (\bibinfo {year} {2019})},\ \Eprint
  {http://arxiv.org/abs/1901.02561} {arXiv:1901.02561 [hep-lat]} \BibitemShut
  {NoStop}%
\bibitem [{\citenamefont {Martinelli}\ \emph {et~al.}(2022)\citenamefont
  {Martinelli}, \citenamefont {Simula},\ and\ \citenamefont
  {Vittorio}}]{Martinelli:2021onb}%
  \BibitemOpen
  \bibfield  {author} {\bibinfo {author} {\bibfnamefont {G.}~\bibnamefont
  {Martinelli}}, \bibinfo {author} {\bibfnamefont {S.}~\bibnamefont {Simula}},
  \ and\ \bibinfo {author} {\bibfnamefont {L.}~\bibnamefont {Vittorio}},\
  }\href {\doibase 10.1103/PhysRevD.105.034503} {\bibfield  {journal} {\bibinfo
   {journal} {Phys. Rev. D}\ }\textbf {\bibinfo {volume} {105}},\ \bibinfo
  {pages} {034503} (\bibinfo {year} {2022})},\ \Eprint
  {http://arxiv.org/abs/2105.08674} {arXiv:2105.08674 [hep-ph]} \BibitemShut
  {NoStop}%
\bibitem [{\citenamefont {Martinelli}\ \emph
  {et~al.}(2021{\natexlab{b}})\citenamefont {Martinelli}, \citenamefont
  {Simula},\ and\ \citenamefont {Vittorio}}]{Martinelli:2021myh}%
  \BibitemOpen
  \bibfield  {author} {\bibinfo {author} {\bibfnamefont {G.}~\bibnamefont
  {Martinelli}}, \bibinfo {author} {\bibfnamefont {S.}~\bibnamefont {Simula}},
  \ and\ \bibinfo {author} {\bibfnamefont {L.}~\bibnamefont {Vittorio}},\
  }\href@noop {} {\  (\bibinfo {year} {2021}{\natexlab{b}})},\ \Eprint
  {http://arxiv.org/abs/2109.15248} {arXiv:2109.15248 [hep-ph]} \BibitemShut
  {NoStop}%
\bibitem [{\citenamefont {Boyd}\ \emph {et~al.}(1995)\citenamefont {Boyd},
  \citenamefont {Grinstein},\ and\ \citenamefont {Lebed}}]{Boyd:1994tt}%
  \BibitemOpen
  \bibfield  {author} {\bibinfo {author} {\bibfnamefont {C.~G.}\ \bibnamefont
  {Boyd}}, \bibinfo {author} {\bibfnamefont {B.}~\bibnamefont {Grinstein}}, \
  and\ \bibinfo {author} {\bibfnamefont {R.~F.}\ \bibnamefont {Lebed}},\ }\href
  {\doibase 10.1103/PhysRevLett.74.4603} {\bibfield  {journal} {\bibinfo
  {journal} {Phys. Rev. Lett.}\ }\textbf {\bibinfo {volume} {74}},\ \bibinfo
  {pages} {4603} (\bibinfo {year} {1995})},\ \Eprint
  {http://arxiv.org/abs/hep-ph/9412324} {arXiv:hep-ph/9412324} \BibitemShut
  {NoStop}%
\bibitem [{\citenamefont {Boyd}\ \emph {et~al.}(1997)\citenamefont {Boyd},
  \citenamefont {Grinstein},\ and\ \citenamefont {Lebed}}]{Boyd:1997kz}%
  \BibitemOpen
  \bibfield  {author} {\bibinfo {author} {\bibfnamefont {C.}~\bibnamefont
  {Boyd}}, \bibinfo {author} {\bibfnamefont {B.}~\bibnamefont {Grinstein}}, \
  and\ \bibinfo {author} {\bibfnamefont {R.~F.}\ \bibnamefont {Lebed}},\ }\href
  {\doibase 10.1103/PhysRevD.56.6895} {\bibfield  {journal} {\bibinfo
  {journal} {Phys. Rev. D}\ }\textbf {\bibinfo {volume} {56}},\ \bibinfo
  {pages} {6895} (\bibinfo {year} {1997})},\ \Eprint
  {http://arxiv.org/abs/hep-ph/9705252} {arXiv:hep-ph/9705252} \BibitemShut
  {NoStop}%
\bibitem [{\citenamefont {Caprini}\ \emph {et~al.}(1998)\citenamefont
  {Caprini}, \citenamefont {Lellouch},\ and\ \citenamefont
  {Neubert}}]{Caprini:1997mu}%
  \BibitemOpen
  \bibfield  {author} {\bibinfo {author} {\bibfnamefont {I.}~\bibnamefont
  {Caprini}}, \bibinfo {author} {\bibfnamefont {L.}~\bibnamefont {Lellouch}}, \
  and\ \bibinfo {author} {\bibfnamefont {M.}~\bibnamefont {Neubert}},\ }\href
  {\doibase 10.1016/S0550-3213(98)00350-2} {\bibfield  {journal} {\bibinfo
  {journal} {Nucl. Phys. B}\ }\textbf {\bibinfo {volume} {530}},\ \bibinfo
  {pages} {153} (\bibinfo {year} {1998})},\ \Eprint
  {http://arxiv.org/abs/hep-ph/9712417} {arXiv:hep-ph/9712417} \BibitemShut
  {NoStop}%
\bibitem [{\citenamefont {Bourrely}\ \emph {et~al.}(1981)\citenamefont
  {Bourrely}, \citenamefont {Machet},\ and\ \citenamefont
  {de~Rafael}}]{Bourrely:1980gp}%
  \BibitemOpen
  \bibfield  {author} {\bibinfo {author} {\bibfnamefont {C.}~\bibnamefont
  {Bourrely}}, \bibinfo {author} {\bibfnamefont {B.}~\bibnamefont {Machet}}, \
  and\ \bibinfo {author} {\bibfnamefont {E.}~\bibnamefont {de~Rafael}},\ }\href
  {\doibase 10.1016/0550-3213(81)90086-9} {\bibfield  {journal} {\bibinfo
  {journal} {Nucl. Phys. B}\ }\textbf {\bibinfo {volume} {189}},\ \bibinfo
  {pages} {157} (\bibinfo {year} {1981})}\BibitemShut {NoStop}%
\bibitem [{\citenamefont {Gelzer}\ \emph {et~al.}(2019)\citenamefont {Gelzer}
  \emph {et~al.}}]{Gelzer:2019zwx}%
  \BibitemOpen
  \bibfield  {author} {\bibinfo {author} {\bibfnamefont {Z.}~\bibnamefont
  {Gelzer}} \emph {et~al.} (\bibinfo {collaboration} {Fermilab Lattice,
  MILC}),\ }\href {\doibase 10.22323/1.363.0236} {\bibfield  {journal}
  {\bibinfo  {journal} {PoS}\ }\textbf {\bibinfo {volume} {LATTICE2019}},\
  \bibinfo {pages} {236} (\bibinfo {year} {2019})},\ \Eprint
  {http://arxiv.org/abs/1912.13358} {arXiv:1912.13358 [hep-lat]} \BibitemShut
  {NoStop}%
\bibitem [{\citenamefont {Flynn}\ \emph {et~al.}(2021)\citenamefont {Flynn},
  \citenamefont {Hill}, \citenamefont {J\"uttner}, \citenamefont {Soni},
  \citenamefont {Tsang},\ and\ \citenamefont {Witzel}}]{Flynn:2020nmk}%
  \BibitemOpen
  \bibfield  {author} {\bibinfo {author} {\bibfnamefont {J.~M.}\ \bibnamefont
  {Flynn}}, \bibinfo {author} {\bibfnamefont {R.~C.}\ \bibnamefont {Hill}},
  \bibinfo {author} {\bibfnamefont {A.}~\bibnamefont {J\"uttner}}, \bibinfo
  {author} {\bibfnamefont {A.}~\bibnamefont {Soni}}, \bibinfo {author}
  {\bibfnamefont {J.~T.}\ \bibnamefont {Tsang}}, \ and\ \bibinfo {author}
  {\bibfnamefont {O.}~\bibnamefont {Witzel}} (\bibinfo {collaboration} {RBC,
  UKQCD}),\ }\href {\doibase 10.22323/1.390.0436} {\bibfield  {journal}
  {\bibinfo  {journal} {PoS}\ }\textbf {\bibinfo {volume} {ICHEP2020}},\
  \bibinfo {pages} {436} (\bibinfo {year} {2021})},\ \Eprint
  {http://arxiv.org/abs/2012.04323} {arXiv:2012.04323 [hep-ph]} \BibitemShut
  {NoStop}%
\bibitem [{\citenamefont {Carrasco}\ \emph {et~al.}(2014)\citenamefont
  {Carrasco} \emph {et~al.}}]{Carrasco:2014cwa}%
  \BibitemOpen
  \bibfield  {author} {\bibinfo {author} {\bibfnamefont {N.}~\bibnamefont
  {Carrasco}} \emph {et~al.} (\bibinfo {collaboration} {European Twisted
  Mass}),\ }\href {\doibase 10.1016/j.nuclphysb.2014.07.025} {\bibfield
  {journal} {\bibinfo  {journal} {Nucl. Phys. B}\ }\textbf {\bibinfo {volume}
  {887}},\ \bibinfo {pages} {19} (\bibinfo {year} {2014})},\ \Eprint
  {http://arxiv.org/abs/1403.4504} {arXiv:1403.4504 [hep-lat]} \BibitemShut
  {NoStop}%
\bibitem [{\citenamefont {Gregory}\ \emph {et~al.}(2011)\citenamefont {Gregory}
  \emph {et~al.}}]{Gregory:2010gm}%
  \BibitemOpen
  \bibfield  {author} {\bibinfo {author} {\bibfnamefont {E.~B.}\ \bibnamefont
  {Gregory}} \emph {et~al.},\ }\href {\doibase 10.1103/PhysRevD.83.014506}
  {\bibfield  {journal} {\bibinfo  {journal} {Phys. Rev. D}\ }\textbf {\bibinfo
  {volume} {83}},\ \bibinfo {pages} {014506} (\bibinfo {year} {2011})},\
  \Eprint {http://arxiv.org/abs/1010.3848} {arXiv:1010.3848 [hep-lat]}
  \BibitemShut {NoStop}%
\bibitem [{\citenamefont {Khodjamirian}\ and\ \citenamefont
  {Rusov}(2017)}]{Khodjamirian:2017fxg}%
  \BibitemOpen
  \bibfield  {author} {\bibinfo {author} {\bibfnamefont {A.}~\bibnamefont
  {Khodjamirian}}\ and\ \bibinfo {author} {\bibfnamefont {A.~V.}\ \bibnamefont
  {Rusov}},\ }\href {\doibase 10.1007/JHEP08(2017)112} {\bibfield  {journal}
  {\bibinfo  {journal} {JHEP}\ }\textbf {\bibinfo {volume} {08}},\ \bibinfo
  {pages} {112} (\bibinfo {year} {2017})},\ \Eprint
  {http://arxiv.org/abs/1703.04765} {arXiv:1703.04765 [hep-ph]} \BibitemShut
  {NoStop}%
\bibitem [{\citenamefont {del Amo~Sanchez}\ \emph {et~al.}(2011)\citenamefont
  {del Amo~Sanchez} \emph {et~al.}}]{delAmoSanchez:2010af}%
  \BibitemOpen
  \bibfield  {author} {\bibinfo {author} {\bibfnamefont {P.}~\bibnamefont {del
  Amo~Sanchez}} \emph {et~al.} (\bibinfo {collaboration} {BaBar}),\ }\href
  {\doibase 10.1103/PhysRevD.83.032007} {\bibfield  {journal} {\bibinfo
  {journal} {Phys. Rev. D}\ }\textbf {\bibinfo {volume} {83}},\ \bibinfo
  {pages} {032007} (\bibinfo {year} {2011})},\ \Eprint
  {http://arxiv.org/abs/1005.3288} {arXiv:1005.3288 [hep-ex]} \BibitemShut
  {NoStop}%
\bibitem [{\citenamefont {Ha}\ \emph {et~al.}(2011)\citenamefont {Ha} \emph
  {et~al.}}]{Ha:2010rf}%
  \BibitemOpen
  \bibfield  {author} {\bibinfo {author} {\bibfnamefont {H.}~\bibnamefont {Ha}}
  \emph {et~al.} (\bibinfo {collaboration} {Belle}),\ }\href {\doibase
  10.1103/PhysRevD.83.071101} {\bibfield  {journal} {\bibinfo  {journal} {Phys.
  Rev. D}\ }\textbf {\bibinfo {volume} {83}},\ \bibinfo {pages} {071101}
  (\bibinfo {year} {2011})},\ \Eprint {http://arxiv.org/abs/1012.0090}
  {arXiv:1012.0090 [hep-ex]} \BibitemShut {NoStop}%
\bibitem [{\citenamefont {Lees}\ \emph {et~al.}(2012)\citenamefont {Lees} \emph
  {et~al.}}]{Lees:2012vv}%
  \BibitemOpen
  \bibfield  {author} {\bibinfo {author} {\bibfnamefont {J.~P.}\ \bibnamefont
  {Lees}} \emph {et~al.} (\bibinfo {collaboration} {BaBar}),\ }\href {\doibase
  10.1103/PhysRevD.86.092004} {\bibfield  {journal} {\bibinfo  {journal} {Phys.
  Rev. D}\ }\textbf {\bibinfo {volume} {86}},\ \bibinfo {pages} {092004}
  (\bibinfo {year} {2012})},\ \Eprint {http://arxiv.org/abs/1208.1253}
  {arXiv:1208.1253 [hep-ex]} \BibitemShut {NoStop}%
\bibitem [{\citenamefont {Sibidanov}\ \emph {et~al.}(2013)\citenamefont
  {Sibidanov} \emph {et~al.}}]{Sibidanov:2013rkk}%
  \BibitemOpen
  \bibfield  {author} {\bibinfo {author} {\bibfnamefont {A.}~\bibnamefont
  {Sibidanov}} \emph {et~al.} (\bibinfo {collaboration} {Belle}),\ }\href
  {\doibase 10.1103/PhysRevD.88.032005} {\bibfield  {journal} {\bibinfo
  {journal} {Phys. Rev. D}\ }\textbf {\bibinfo {volume} {88}},\ \bibinfo
  {pages} {032005} (\bibinfo {year} {2013})},\ \Eprint
  {http://arxiv.org/abs/1306.2781} {arXiv:1306.2781 [hep-ex]} \BibitemShut
  {NoStop}%
\bibitem [{\citenamefont {Aaij}\ \emph {et~al.}(2021)\citenamefont {Aaij} \emph
  {et~al.}}]{Aaij:2020nvo}%
  \BibitemOpen
  \bibfield  {author} {\bibinfo {author} {\bibfnamefont {R.}~\bibnamefont
  {Aaij}} \emph {et~al.} (\bibinfo {collaboration} {LHCb}),\ }\href {\doibase
  10.1103/PhysRevLett.126.081804} {\bibfield  {journal} {\bibinfo  {journal}
  {Phys. Rev. Lett.}\ }\textbf {\bibinfo {volume} {126}},\ \bibinfo {pages}
  {081804} (\bibinfo {year} {2021})},\ \Eprint
  {http://arxiv.org/abs/2012.05143} {arXiv:2012.05143 [hep-ex]} \BibitemShut
  {NoStop}%
\bibitem [{\citenamefont {Riggio}\ \emph {et~al.}(2018)\citenamefont {Riggio},
  \citenamefont {Salerno},\ and\ \citenamefont {Simula}}]{Riggio:2017zwh}%
  \BibitemOpen
  \bibfield  {author} {\bibinfo {author} {\bibfnamefont {L.}~\bibnamefont
  {Riggio}}, \bibinfo {author} {\bibfnamefont {G.}~\bibnamefont {Salerno}}, \
  and\ \bibinfo {author} {\bibfnamefont {S.}~\bibnamefont {Simula}},\ }\href
  {\doibase 10.1140/epjc/s10052-018-5943-5} {\bibfield  {journal} {\bibinfo
  {journal} {Eur. Phys. J. C}\ }\textbf {\bibinfo {volume} {78}},\ \bibinfo
  {pages} {501} (\bibinfo {year} {2018})},\ \Eprint
  {http://arxiv.org/abs/1706.03657} {arXiv:1706.03657 [hep-lat]} \BibitemShut
  {NoStop}%
\bibitem [{\citenamefont {Aaij}\ \emph {et~al.}(2020)\citenamefont {Aaij} \emph
  {et~al.}}]{Aaij:2020hsi}%
  \BibitemOpen
  \bibfield  {author} {\bibinfo {author} {\bibfnamefont {R.}~\bibnamefont
  {Aaij}} \emph {et~al.} (\bibinfo {collaboration} {LHCb}),\ }\href {\doibase
  10.1103/PhysRevD.101.072004} {\bibfield  {journal} {\bibinfo  {journal}
  {Phys. Rev. D}\ }\textbf {\bibinfo {volume} {101}},\ \bibinfo {pages}
  {072004} (\bibinfo {year} {2020})},\ \Eprint
  {http://arxiv.org/abs/2001.03225} {arXiv:2001.03225 [hep-ex]} \BibitemShut
  {NoStop}%
\bibitem [{\citenamefont {Mei\ss{}ner}\ and\ \citenamefont
  {Wang}(2014)}]{Meissner:2013pba}%
  \BibitemOpen
  \bibfield  {author} {\bibinfo {author} {\bibfnamefont {U.-G.}\ \bibnamefont
  {Mei\ss{}ner}}\ and\ \bibinfo {author} {\bibfnamefont {W.}~\bibnamefont
  {Wang}},\ }\href {\doibase 10.1007/JHEP01(2014)107} {\bibfield  {journal}
  {\bibinfo  {journal} {JHEP}\ }\textbf {\bibinfo {volume} {01}},\ \bibinfo
  {pages} {107} (\bibinfo {year} {2014})},\ \Eprint
  {http://arxiv.org/abs/1311.5420} {arXiv:1311.5420 [hep-ph]} \BibitemShut
  {NoStop}%
\bibitem [{\citenamefont {Rajeev}\ and\ \citenamefont
  {Dutta}(2018)}]{Rajeev:2018txm}%
  \BibitemOpen
  \bibfield  {author} {\bibinfo {author} {\bibfnamefont {N.}~\bibnamefont
  {Rajeev}}\ and\ \bibinfo {author} {\bibfnamefont {R.}~\bibnamefont {Dutta}},\
  }\href {\doibase 10.1103/PhysRevD.98.055024} {\bibfield  {journal} {\bibinfo
  {journal} {Phys. Rev. D}\ }\textbf {\bibinfo {volume} {98}},\ \bibinfo
  {pages} {055024} (\bibinfo {year} {2018})},\ \Eprint
  {http://arxiv.org/abs/1808.03790} {arXiv:1808.03790 [hep-ph]} \BibitemShut
  {NoStop}%
\bibitem [{\citenamefont {Biswas}\ and\ \citenamefont
  {Nandi}(2021)}]{Biswas:2021cyd}%
  \BibitemOpen
  \bibfield  {author} {\bibinfo {author} {\bibfnamefont {A.}~\bibnamefont
  {Biswas}}\ and\ \bibinfo {author} {\bibfnamefont {S.}~\bibnamefont {Nandi}},\
  }\href {\doibase 10.1007/JHEP09(2021)127} {\bibfield  {journal} {\bibinfo
  {journal} {JHEP}\ }\textbf {\bibinfo {volume} {09}},\ \bibinfo {pages} {127}
  (\bibinfo {year} {2021})},\ \Eprint {http://arxiv.org/abs/2105.01732}
  {arXiv:2105.01732 [hep-ph]} \BibitemShut {NoStop}%
\bibitem [{\citenamefont {Hamer}\ \emph {et~al.}(2016)\citenamefont {Hamer}
  \emph {et~al.}}]{Hamer:2015jsa}%
  \BibitemOpen
  \bibfield  {author} {\bibinfo {author} {\bibfnamefont {P.}~\bibnamefont
  {Hamer}} \emph {et~al.} (\bibinfo {collaboration} {Belle}),\ }\href {\doibase
  10.1103/PhysRevD.93.032007} {\bibfield  {journal} {\bibinfo  {journal} {Phys.
  Rev. D}\ }\textbf {\bibinfo {volume} {93}},\ \bibinfo {pages} {032007}
  (\bibinfo {year} {2016})},\ \Eprint {http://arxiv.org/abs/1509.06521}
  {arXiv:1509.06521 [hep-ex]} \BibitemShut {NoStop}%
\bibitem [{\citenamefont {Altmannshofer}\ \emph {et~al.}(2019)\citenamefont
  {Altmannshofer} \emph {et~al.}}]{Kou:2018nap}%
  \BibitemOpen
  \bibfield  {author} {\bibinfo {author} {\bibfnamefont {W.}~\bibnamefont
  {Altmannshofer}} \emph {et~al.} (\bibinfo {collaboration} {Belle-II}),\
  }\href {\doibase 10.1093/ptep/ptz106} {\bibfield  {journal} {\bibinfo
  {journal} {PTEP}\ }\textbf {\bibinfo {volume} {2019}},\ \bibinfo {pages}
  {123C01} (\bibinfo {year} {2019})},\ \bibinfo {note} {[Erratum: PTEP 2020,
  029201 (2020)]},\ \Eprint {http://arxiv.org/abs/1808.10567} {arXiv:1808.10567
  [hep-ex]} \BibitemShut {NoStop}%
\bibitem [{\citenamefont {Blossier}\ \emph {et~al.}(2010)\citenamefont
  {Blossier} \emph {et~al.}}]{ETM:2009sed}%
  \BibitemOpen
  \bibfield  {author} {\bibinfo {author} {\bibfnamefont {B.}~\bibnamefont
  {Blossier}} \emph {et~al.} (\bibinfo {collaboration} {ETM}),\ }\href
  {\doibase 10.1007/JHEP04(2010)049} {\bibfield  {journal} {\bibinfo  {journal}
  {JHEP}\ }\textbf {\bibinfo {volume} {04}},\ \bibinfo {pages} {049} (\bibinfo
  {year} {2010})},\ \Eprint {http://arxiv.org/abs/0909.3187} {arXiv:0909.3187
  [hep-lat]} \BibitemShut {NoStop}%
\bibitem [{\citenamefont {Bussone}\ \emph {et~al.}(2016)\citenamefont {Bussone}
  \emph {et~al.}}]{ETM:2016nbo}%
  \BibitemOpen
  \bibfield  {author} {\bibinfo {author} {\bibfnamefont {A.}~\bibnamefont
  {Bussone}} \emph {et~al.} (\bibinfo {collaboration} {ETM}),\ }\href {\doibase
  10.1103/PhysRevD.93.114505} {\bibfield  {journal} {\bibinfo  {journal} {Phys.
  Rev. D}\ }\textbf {\bibinfo {volume} {93}},\ \bibinfo {pages} {114505}
  (\bibinfo {year} {2016})},\ \Eprint {http://arxiv.org/abs/1603.04306}
  {arXiv:1603.04306 [hep-lat]} \BibitemShut {NoStop}%
\bibitem [{\citenamefont {Bigi}\ and\ \citenamefont
  {Gambino}(2016)}]{Bigi:2016mdz}%
  \BibitemOpen
  \bibfield  {author} {\bibinfo {author} {\bibfnamefont {D.}~\bibnamefont
  {Bigi}}\ and\ \bibinfo {author} {\bibfnamefont {P.}~\bibnamefont {Gambino}},\
  }\href {\doibase 10.1103/PhysRevD.94.094008} {\bibfield  {journal} {\bibinfo
  {journal} {Phys. Rev. D}\ }\textbf {\bibinfo {volume} {94}},\ \bibinfo
  {pages} {094008} (\bibinfo {year} {2016})},\ \Eprint
  {http://arxiv.org/abs/1606.08030} {arXiv:1606.08030 [hep-ph]} \BibitemShut
  {NoStop}%
\bibitem [{\citenamefont {Bigi}\ \emph
  {et~al.}(2017{\natexlab{a}})\citenamefont {Bigi}, \citenamefont {Gambino},\
  and\ \citenamefont {Schacht}}]{Bigi:2017njr}%
  \BibitemOpen
  \bibfield  {author} {\bibinfo {author} {\bibfnamefont {D.}~\bibnamefont
  {Bigi}}, \bibinfo {author} {\bibfnamefont {P.}~\bibnamefont {Gambino}}, \
  and\ \bibinfo {author} {\bibfnamefont {S.}~\bibnamefont {Schacht}},\ }\href
  {\doibase 10.1016/j.physletb.2017.04.022} {\bibfield  {journal} {\bibinfo
  {journal} {Phys. Lett. B}\ }\textbf {\bibinfo {volume} {769}},\ \bibinfo
  {pages} {441} (\bibinfo {year} {2017}{\natexlab{a}})},\ \Eprint
  {http://arxiv.org/abs/1703.06124} {arXiv:1703.06124 [hep-ph]} \BibitemShut
  {NoStop}%
\bibitem [{\citenamefont {Bigi}\ \emph
  {et~al.}(2017{\natexlab{b}})\citenamefont {Bigi}, \citenamefont {Gambino},\
  and\ \citenamefont {Schacht}}]{Bigi:2017jbd}%
  \BibitemOpen
  \bibfield  {author} {\bibinfo {author} {\bibfnamefont {D.}~\bibnamefont
  {Bigi}}, \bibinfo {author} {\bibfnamefont {P.}~\bibnamefont {Gambino}}, \
  and\ \bibinfo {author} {\bibfnamefont {S.}~\bibnamefont {Schacht}},\ }\href
  {\doibase 10.1007/JHEP11(2017)061} {\bibfield  {journal} {\bibinfo  {journal}
  {JHEP}\ }\textbf {\bibinfo {volume} {11}},\ \bibinfo {pages} {061} (\bibinfo
  {year} {2017}{\natexlab{b}})},\ \Eprint {http://arxiv.org/abs/1707.09509}
  {arXiv:1707.09509 [hep-ph]} \BibitemShut {NoStop}%
\bibitem [{\citenamefont {Baron}\ \emph
  {et~al.}(2010{\natexlab{a}})\citenamefont {Baron} \emph
  {et~al.}}]{Baron:2010bv}%
  \BibitemOpen
  \bibfield  {author} {\bibinfo {author} {\bibfnamefont {R.}~\bibnamefont
  {Baron}} \emph {et~al.},\ }\href {\doibase 10.1007/JHEP06(2010)111}
  {\bibfield  {journal} {\bibinfo  {journal} {JHEP}\ }\textbf {\bibinfo
  {volume} {06}},\ \bibinfo {pages} {111} (\bibinfo {year}
  {2010}{\natexlab{a}})},\ \Eprint {http://arxiv.org/abs/1004.5284}
  {arXiv:1004.5284 [hep-lat]} \BibitemShut {NoStop}%
\bibitem [{\citenamefont {Baron}\ \emph
  {et~al.}(2010{\natexlab{b}})\citenamefont {Baron} \emph
  {et~al.}}]{ETM:2010cqp}%
  \BibitemOpen
  \bibfield  {author} {\bibinfo {author} {\bibfnamefont {R.}~\bibnamefont
  {Baron}} \emph {et~al.} (\bibinfo {collaboration} {ETM}),\ }\href {\doibase
  10.22323/1.105.0123} {\bibfield  {journal} {\bibinfo  {journal} {PoS}\
  }\textbf {\bibinfo {volume} {LATTICE2010}},\ \bibinfo {pages} {123} (\bibinfo
  {year} {2010}{\natexlab{b}})},\ \Eprint {http://arxiv.org/abs/1101.0518}
  {arXiv:1101.0518 [hep-lat]} \BibitemShut {NoStop}%
\bibitem [{\citenamefont {Chetyrkin}\ and\ \citenamefont
  {Retey}(2000)}]{Chetyrkin:1999pq}%
  \BibitemOpen
  \bibfield  {author} {\bibinfo {author} {\bibfnamefont {K.~G.}\ \bibnamefont
  {Chetyrkin}}\ and\ \bibinfo {author} {\bibfnamefont {A.}~\bibnamefont
  {Retey}},\ }\href {\doibase 10.1016/S0550-3213(00)00331-X} {\bibfield
  {journal} {\bibinfo  {journal} {Nucl. Phys. B}\ }\textbf {\bibinfo {volume}
  {583}},\ \bibinfo {pages} {3} (\bibinfo {year} {2000})},\ \Eprint
  {http://arxiv.org/abs/hep-ph/9910332} {arXiv:hep-ph/9910332} \BibitemShut
  {NoStop}%
\bibitem [{\citenamefont {Aoki}\ \emph {et~al.}(2020)\citenamefont {Aoki} \emph
  {et~al.}}]{Aoki:2019cca}%
  \BibitemOpen
  \bibfield  {author} {\bibinfo {author} {\bibfnamefont {S.}~\bibnamefont
  {Aoki}} \emph {et~al.} (\bibinfo {collaboration} {Flavour Lattice Averaging
  Group}),\ }\href {\doibase 10.1140/epjc/s10052-019-7354-7} {\bibfield
  {journal} {\bibinfo  {journal} {Eur. Phys. J. C}\ }\textbf {\bibinfo {volume}
  {80}},\ \bibinfo {pages} {113} (\bibinfo {year} {2020})},\ \Eprint
  {http://arxiv.org/abs/1902.08191} {arXiv:1902.08191 [hep-lat]} \BibitemShut
  {NoStop}%
\bibitem [{\citenamefont {Chetyrkin}\ and\ \citenamefont
  {Steinhauser}(2000)}]{Chetyrkin:1999qi}%
  \BibitemOpen
  \bibfield  {author} {\bibinfo {author} {\bibfnamefont {K.~G.}\ \bibnamefont
  {Chetyrkin}}\ and\ \bibinfo {author} {\bibfnamefont {M.}~\bibnamefont
  {Steinhauser}},\ }\href {\doibase 10.1016/S0550-3213(99)00784-1} {\bibfield
  {journal} {\bibinfo  {journal} {Nucl. Phys. B}\ }\textbf {\bibinfo {volume}
  {573}},\ \bibinfo {pages} {617} (\bibinfo {year} {2000})},\ \Eprint
  {http://arxiv.org/abs/hep-ph/9911434} {arXiv:hep-ph/9911434} \BibitemShut
  {NoStop}%
\bibitem [{\citenamefont {Melnikov}\ and\ \citenamefont
  {Ritbergen}(2000)}]{Melnikov:2000qh}%
  \BibitemOpen
  \bibfield  {author} {\bibinfo {author} {\bibfnamefont {K.}~\bibnamefont
  {Melnikov}}\ and\ \bibinfo {author} {\bibfnamefont {T.~v.}\ \bibnamefont
  {Ritbergen}},\ }\href {\doibase 10.1016/S0370-2693(00)00507-4} {\bibfield
  {journal} {\bibinfo  {journal} {Phys. Lett. B}\ }\textbf {\bibinfo {volume}
  {482}},\ \bibinfo {pages} {99} (\bibinfo {year} {2000})},\ \Eprint
  {http://arxiv.org/abs/hep-ph/9912391} {arXiv:hep-ph/9912391} \BibitemShut
  {NoStop}%
\bibitem [{\citenamefont {Lubicz}\ \emph {et~al.}(2017)\citenamefont {Lubicz},
  \citenamefont {Melis},\ and\ \citenamefont {Simula}}]{Lubicz:2017asp}%
  \BibitemOpen
  \bibfield  {author} {\bibinfo {author} {\bibfnamefont {V.}~\bibnamefont
  {Lubicz}}, \bibinfo {author} {\bibfnamefont {A.}~\bibnamefont {Melis}}, \
  and\ \bibinfo {author} {\bibfnamefont {S.}~\bibnamefont {Simula}} (\bibinfo
  {collaboration} {ETM}),\ }\href {\doibase 10.1103/PhysRevD.96.034524}
  {\bibfield  {journal} {\bibinfo  {journal} {Phys. Rev. D}\ }\textbf {\bibinfo
  {volume} {96}},\ \bibinfo {pages} {034524} (\bibinfo {year} {2017})},\
  \Eprint {http://arxiv.org/abs/1707.04529} {arXiv:1707.04529 [hep-lat]}
  \BibitemShut {NoStop}%
\end{thebibliography}%

\end{document}